\documentclass[aps,pra,twocolumn,groupedaddress,showpacs]{revtex4-1}

\usepackage{amsmath}
\usepackage{amsfonts}
\usepackage{color}
\usepackage{commath}
\usepackage{graphicx}
\usepackage{subfigure}
\usepackage[utf8]{inputenc}

\newcommand{\nn}{\nonumber}
\newcommand{\Nmodes}{n_\text{m}}

\newcommand{\tr}{\text{tr}}
\newcommand{\Log}{\text{Log}}

\newcommand{\delnospace}[1]{\mathopen{}\del{#1}}

\newcommand{\hs}{\hat{\sigma}}
\newcommand{\hE}{\hat{\mathcal{E}}}
\newcommand{\nohs}{\sigma}
\newcommand{\nohE}{\mathcal{E}}

\newcommand{\rddots}{\scalebox{-1}[1]{$\ddots$}}

\begin{document}

\title{Dispersion relations for stationary light in one-dimensional atomic ensembles}

\author{Ivan~Iakoupov$^{1}$}
\author{Johan~R.~\surname{Ott}$^{1}$}
\author{Darrick~E.~Chang$^{2}$}
\author{Anders~S.~S{\o}rensen$^{1}$}
\affiliation{$^1$QUANTOP, The Niels Bohr Institute, University of Copenhagen, Blegdamsvej 17, DK-2100 Copenhagen \O, Denmark\\
$^2$ICFO-Institut de Ciencies Fotoniques, The Barcelona Institute of Science and Technology, 08860 Castelldefels (Barcelona), Spain}
\date{\today}

\begin{abstract}
We investigate the dispersion relations for light coupled to one-dimensional 
ensembles of atoms with different level schemes. The unifying feature of all 
the considered setups is that the forward and backward propagating quantum 
fields are coupled by the applied classical drives such that the group 
velocity can vanish in an effect known as ``stationary light''. We derive the 
dispersion relations for all the considered schemes, highlighting the 
important differences between them. Furthermore, we show that additional 
control of stationary light can be obtained by treating atoms as discrete 
scatterers and placing them at well defined positions. For the latter purpose, 
a multi-mode transfer matrix theory for light is developed.
\end{abstract}


\maketitle

\section{Introduction}
\begin{figure*}[htbp]
\includegraphics{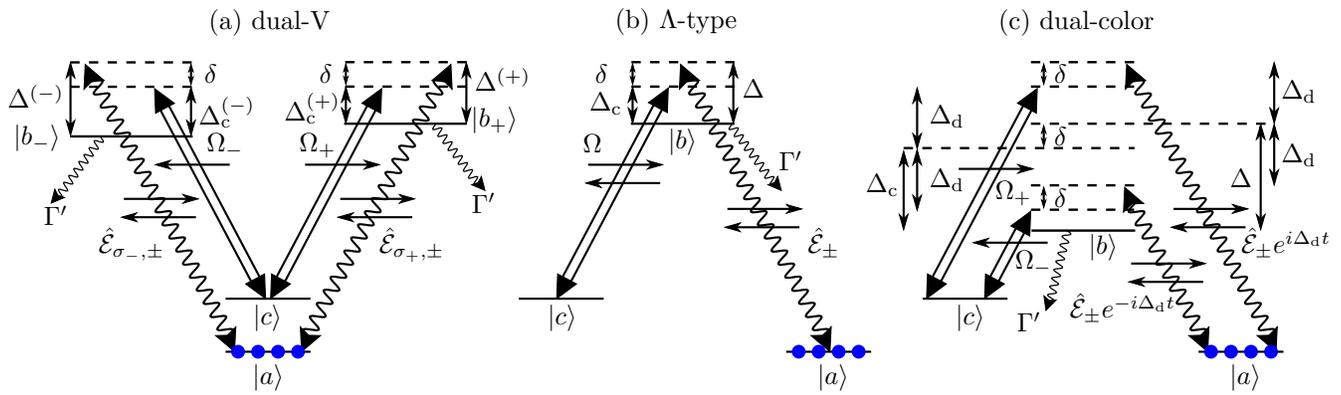}
\caption{(Color online) Level diagrams of the three schemes that we consider. 
The blue circles on state $|a\rangle$ indicate that the atoms are assumed to 
be initialized in this state. The arrows with small wiggly lines originating 
on the excited states $|b_\pm\rangle$ and $|b\rangle$ indicate spontaneous 
emission with a decay rate $\Gamma'$. The arrows between different states indicate 
either quantum fields (wiggly lines), or classical drives (double straight 
lines). The small horizontal arrows on each of these coupling arrows indicate 
the propagation direction. All the transitions are assumed to couple equally 
to both the right-moving and left-moving fields, but the arrows pointing only in a single 
direction on the classical drives for dual\mbox{-}V and dual-color schemes instead 
indicate that the externally applied drives propagate in the shown direction. 
The excited level $|b\rangle$ for the dual-color scheme is shifted vertically
in order to be able to clearly show all the different detunings.
\label{fig:level_diagrams}}
\end{figure*}

A major quest within modern quantum optics is to obtain full control over 
light at the single photon level. Light is, however, highly elusive since it 
travels at great speed making it essential to couple light to matter to 
control it.
A particularly promising system in that respect is an 
ensemble of atoms with the three-level $\Lambda$\mbox{-}type configuration
sketched in Fig.~\ref{fig:level_diagrams}(b). By applying a co-propagating 
classical electric field on one of the 
transitions, the group velocity of a quantized field resonant with the other 
transition can be greatly reduced compared to free-space through the process 
of electromagnetically induced 
transparency~(EIT)~\cite{hau_nature1999,lukin_rmp03}. By further reducing the 
group velocity, EIT even permits the storage of light as long lived 
excitations of the atoms \cite{liu_nature2001,phillips_prl01}.

EIT by itself is a linear optical effect. However, since EIT enables one to 
propagate electric fields near atomic resonance with low absorption, 
variations of EIT also constitute a popular choice for creating non-linear 
optical interactions at low photon numbers 
\cite{harris_prl99,bajcsy_prl09,gorshkov_prl11,chen_science13}. In these 
setups, the relevant figure of merit is the interaction time of the photons, 
which is proportional to the inverse of the group velocity. For EIT, 
decreasing the group velocity of the polaritons (coupled light-matter 
excitations) simultaneously makes them increasingly atomic and less photonic in 
character~\cite{fleischhauer_prl2000} thus also decreasing the optical 
non-linearity. These two effects cancel each other, which results in no 
enhancement of the effective non-linear interaction strength. In this context, 
proposals for ``stationary light'' have emerged as a way of creating 
polaritons with very small (or even vanishing) group velocities within the 
atomic medium, while retaining a non-zero photonic 
component~\cite{andre_prl02a,bajcsy_nature03a}. Building upon the enhanced 
non-linear interactions, it is in principle possible to observe the rich 
physics of non-linear optics at the level of a few 
photons~\cite{chang_naturephys08a,hafezi_pra12a}.

To use stationary light for enhancement of the non-linear interaction 
strength, it is essential to first understand the linear properties, which is 
the focus of this article. We will consider the dispersion relation for 
three different 
stationary light schemes (see Fig.~\ref{fig:level_diagrams}). The dispersion 
relation gives the frequency (two-photon detuning) $\delta$ in terms of the Bloch 
vector $q$. From the dispersion relation, the group velocity 
$v_\text{g}=\pd{\delta}{q}$ can be readily obtained, and by the discussion 
above, it can therefore provide an intuition about how strong the non-linear 
interaction strength is expected to be. For the analysis, we will use two 
different theoretical models. The first is the continuum model, in which the 
atomic operators are defined for any real position coordinate $z$ between $0$ 
and $L$ (the total length of the ensemble). The second is the discrete model, 
where each atom is a linear point scatterer. The latter model is motivated by 
a growing interest in considering systems, where the number of atoms is 
relatively small, while the coupling strength and control over placement of 
the individual atoms are greatly improved. Examples are tapered optical 
fibers~\cite{le_kien_pra04,vetsch_prl10,goban_prl12} and photonic crystal 
waveguides~\cite{yu_apl14,goban_ncomms2014}. In the discrete model, we find 
that placing the atoms in a particular way provides an additional handle for 
controlling the dispersion relation~\cite{witthaut_njp2010,chang_njp11a}.

The dispersion relations for the continuum model have already been 
derived elsewhere~\cite{moiseev_pra2006,zimmer_pra08,moiseev_pra2014}. 
However, as we will show below, the results of the discrete model can be 
understood better, if they are set in context by rederiving the results of the 
continuum model in a different way compared to the previous publications. 
Additionally, even when restricted to the continuum model, treating every 
stationary light scheme in the same framework allows for a much easier 
comparison of the schemes and also for tracking the various (physically 
motivated) approximations that are employed in the derivations. By doing 
numerical calculations with the discrete model afterwards, we can test the 
validity of some of these approximations. We will show that for some of the 
stationary light schemes, the dispersion relations derived analytically using 
the continuum model, can also be obtained numerically as limiting cases of the 
discrete model with randomly placed atoms.

\section{Overview}

We will consider one-dimensional ensembles of atoms with three different level 
and coupling schemes where stationary light can be observed (see 
Fig.~\ref{fig:level_diagrams}). We will focus on the case of cold atoms, 
although for completeness we will also briefly discuss hot $\Lambda$\mbox{-}type 
atoms, which was the scheme used for the first prediction and observation of 
stationary light~\cite{andre_prl02a,bajcsy_nature03a}. Common to all 
stationary light schemes is the presence of two counter-propagating classical 
drives which couple the right-moving and left-moving modes of the quantum 
field through four-wave mixing~\cite{moiseev_pra2006}. One way to 
explain the origin of the four-wave mixing is that an incident photon of the 
quantum field will be temporarily mapped to the meta-stable state $|c\rangle$ 
by the classical drive propagating in the same direction. The other classical 
drive can then retrieve this temporary excitation into a photon of the quantum 
field propagating in the opposite direction. In this picture, stationary light 
can be viewed as simultaneous EIT storage and retrieval in both the forward 
and backward directions~\cite{gorshkov_pra07_2}.

For the $\Lambda$\mbox{-}type scheme (Fig.~\ref{fig:level_diagrams}(b)), a 
different intuitive explanation of stationary light can be given in terms of 
Bragg scattering. In this scheme, the two counter-propagating classical drives 
produce a standing wave, which modulates the refractive index of the ensemble 
such that it behaves as a Bragg grating. Hence, the coupling of the right-moving 
and left-moving modes of the quantum field happens due to the reflection of one into 
the other by the Bragg grating. The dynamics of the cold $\Lambda$-type 
scheme is, however, more complicated, which can be illustrated in terms of the 
allowed processes. Since
both counter-propagating drives are applied on the same transition 
$|b\rangle\leftrightarrow|c\rangle$, it is possible for the atom to be excited 
by one of the classical drives and de-excited by the other. 
This leads to the build up of higher order Fourier components of the atomic 
coherence resulting in a rich and complicated physics of stationary light for 
cold $\Lambda$-type 
atoms~\cite{hansen_pra07a,nikoghosyan_pra09a,lin_prl09,wu_pra10,wu_pra10_2,
peters_pra12}. We will show that depending on the precise details of the 
system and the approximations used, it is possible to get dispersion relations 
with three different scalings close to the two-photon resonance ($\delta=0$): 
$\delta\propto q^2$ (quadratic dispersion relation of stationary light), 
$\delta\propto \pm q$ (EIT-like linear dispersion relation), and 
$\delta \propto \pm |q|^{4/3}$ \cite{moiseev_pra2014}. In the derivations below, 
the first two cases will arise in the continuum model due to different 
truncations of the set of higher order modes of the atomic coherence. 
Afterwards, in the discrete model, we will show that these truncations can 
actually be realized physically by positioning the atoms in certain ways. The 
dispersion relation $\delta \propto |q|^{4/3}$ is obtained in the continuum 
model, when all the higher order Fourier components of the atomic coherence 
are summed to infinite order~\cite{moiseev_pra2014}. In the discrete model, 
such a scaling can be reproduced in the limit of an infinite number of 
randomly placed atoms.

A common trait of the two other schemes for stationary light, 
dual\mbox{-}V~\cite{zimmer_pra08} and dual-color~\cite{moiseev_pra2006} (Figs. 
\ref{fig:level_diagrams}(a) and \ref{fig:level_diagrams}(c) respectively) is 
the separation of the right-moving and left-moving fields (both classical and 
quantum) into different modes, either with different polarizations for 
dual\mbox{-}V or with different frequencies for dual-color. The main purpose 
of this separation is to suppress the higher order Fourier components of the 
atomic coherence since excitation and de-excitation with two different 
classical fields are no longer allowed. The end result of this, is that both 
the dual-V and dual-color schemes have quadratic dispersion relations 
$\delta\propto q^2$, just like stationary light in hot $\Lambda$-type atoms, 
where the higher order Fourier components of the atomic coherence are washed 
away by the thermal motion of the atoms~\cite{andre_prl02a,bajcsy_nature03a}.

Before going into the detailed derivations, we will first outline how the 
different scalings of the dispersion relations can arise in the continuum 
model for the $\Lambda$\mbox{-}type scheme. In the derivations below, the 
equations for the atoms are solved first and then substituted into the 
equations for the electric field. The result has the form
\begin{gather}
\del{\frac{1}{c}\dpd{}{t}\pm \dpd{}{z}}\nohE_{\pm}
=-i n_0 \alpha_1(\delta)\nohE_\pm-i n_0 \alpha_2(\delta)\nohE_\mp,
\end{gather}
where $\mathcal{E}_\pm$ are slowly varying (both in time and space) electric 
fields moving either to the right~($+$) or the left~($-$), and $\alpha_1$ and 
$\alpha_2$ describe the (frequency dependent) atomic polarizability. In the 
equation above, we have separated out the density $n_0$ and a factor of $-i$ 
for consistency with the notation below. Due to the steep dispersion of the 
light field, i.e., since the speed of light $c$ is large, we will omit the 
time derivative of the field. If we look for solutions of the form
$\mathcal{E}_\pm(z)=\mathcal{E}_\pm(0){\rm e}^{iqz}$, we arrive at the coupled 
equations
\begin{gather}
\begin{pmatrix}
\alpha_1(\delta)+\frac{q}{n_0} & \alpha_2 (\delta)   \\
\alpha_2(\delta) & \alpha_1(\delta)-\frac{q}{n_0}
\end{pmatrix}
\begin{pmatrix}
\nohE_+\\
\nohE_-  
\end{pmatrix}
=\begin{pmatrix}
0\\
0
\end{pmatrix}.
\label{eq:genericmatrix}
\end{gather}
To have non-trivial solutions, the determinant 
of the matrix in this equation must vanish, which results in the equation 
\begin{gather}
\alpha_1^2(\delta)-\alpha_2^2(\delta)-\frac{q^2}{n_0^2}=0,
\label{eq:genericcondition}
\end{gather}
which relates $q$ and $\delta$, and thereby gives the dispersion relation.

All the situations that we consider are chosen such that they have a solution at 
$q=0$ and $\delta=0$ (a dark state~\cite{fleischhauer_prl2000}). To fulfill 
Eq.~\eqref{eq:genericcondition} with $q=0$ and $\delta=0$, we must have 
$\alpha_1^2(0)=\alpha_2^2(0)$, and in the detailed derivation below, we will 
choose phases of the classical drives such that $\alpha_1(0)=-\alpha_2(0)$. We 
will then encounter three cases that give different scalings of the dispersion 
relation.

If the coefficients $\alpha_1(0)$ and $\alpha_2(0)$ are non-zero, 
Eq.~\eqref{eq:genericmatrix} with $q=0$ will only have a single solution 
($\mathcal{E}_+=\mathcal{E}_-$ for $\alpha_1(0)=-\alpha_2(0)$). In this 
case we can expand $\alpha_1(\delta)$ and $\alpha_2(\delta)$ to first order in 
the detuning to obtain
\begin{gather}
2\alpha_1(0)(\alpha_1'+\alpha_2')\delta=\frac{q^2}{n_0^2},
\end{gather}
where we have chosen $\alpha_2(0)=-\alpha_1(0)$ and 
$\alpha_k'$ denote the derivatives of $\alpha_k$ ($k=1,2$) at $\delta=0$. 
Assuming $\alpha_1'+\alpha_2'\neq 0$ we obtain a single solution with a 
quadratic dispersion relation
\begin{gather}
\delta\propto q^2.
\end{gather}
This is the original stationary light dispersion 
relation~\cite{andre_prl02a,bajcsy_nature03a} and is the typical situation 
encountered near an extremum of a single dispersion band. This is also the 
situation we will encounter for the dual-V and dual-color schemes.

If $\alpha_1(0)=\alpha_2(0)=0$ and $q=0$, the matrix in Eq.~\eqref{eq:genericmatrix} has 
all elements equal to zero. Hence, any vector is an eigenvector 
of this matrix, and we can pick two orthogonal ones, which can be 
interpreted as two degenerate solutions. 
After expanding $\alpha_1$ and $\alpha_2$ in $\delta$, the lowest order 
contributions to $\alpha_1^2(\delta)-\alpha_2^2(\delta)$ in Eq.~\eqref{eq:genericcondition} is 
then quadratic resulting in
\begin{gather}
\del{{\alpha'_1}^2-{\alpha'_2}^2}\delta^2=\frac{q^2}{n_0^2}.
\end{gather}
Assuming $\del{{\alpha'_1}^2-{\alpha'_2}^2}\neq 0$ we then obtain 
two solutions with a linear dispersion relation
\begin{gather}
\delta\propto \pm q.
\end{gather}
This result reflects the fact that if two dispersion bands cross, they tend to 
be linear around the crossing.

The $\Lambda$\mbox{-}type scheme also provides an example of a different scaling 
of two crossing dispersion bands. It is obtained when 
$\alpha_1(0)=\alpha_2(0)=0$, but the functions $\alpha_1(\delta)$ and 
$\alpha_2(\delta)$ can not be expanded at $\delta=0$ (not analytic). This will 
be the case for the solution for the $\Lambda$\mbox{-}type scheme, when all 
the higher order Fourier components of the atomic coherence are accounted for. 
We obtain the scalings $\alpha_1,\alpha_2\propto \sqrt{\delta}$, but according 
to Eq.~\eqref{eq:genericcondition} the dispersion relation is determined 
by the difference of the squares, which has the scaling
$\alpha_1^2(\delta)-\alpha_2^2(\delta)\propto \delta^{3/2}$ such that we 
obtain
\begin{gather}
\delta\propto \pm|q|^{4/3}.
\end{gather}

The structure of the article is as follows. In the continuum model, the 
dispersion relations for the three stationary light schemes are derived in 
Sec.~\ref{Sec:dual_v_continuum_model} (dual\mbox{-}V), 
Sec.~\ref{Sec:lambda_continuum_model} ($\Lambda$\mbox{-}type), and 
Sec.~\ref{Sec:dual_color_continuum_model} (dual-color). For completeness, in 
Sec.~\ref{Sec:hot_Lambda_continuum_model} we include a brief discussion of how 
the results for the dual-color scheme can provide another confirmation that 
the dispersion relation for hot (i.e. moving) $\Lambda$-type atoms is 
quadratic~\cite{peters_pra12}. In Sec.~\ref{Sec:discrete_model}, the discrete 
model is discussed (Sec.~\ref{Sec:discrete_model_dual_V} for dual\mbox{-}V and 
Sec.~\ref{Sec:discrete_model_Lambda} for $\Lambda$\mbox{-}type). In 
Sec.~\ref{sec_scattering_properties}, we look 
at the connection between the dispersion relations and scattering properties of 
the atomic ensembles under the conditions of stationary light.

\section{Continuum model}\label{Sec:continuum_model}
\subsection{Dispersion relations for cold Dual-V atoms}
\label{Sec:dual_v_continuum_model}
The dual\mbox{-}V scheme as shown in Fig.~\ref{fig:level_diagrams}(a) has already 
been studied in Ref.~\cite{zimmer_pra08} and was shown to have a quadratic 
dispersion relation. Here we do a different derivation of this result to serve 
as the context for the discussion of the other stationary light schemes. We 
take the dual\mbox{-}V scheme as the starting point, because the derivation of 
the dispersion relation is more straightforward, even if the additional atomic 
energy level and two different polarizations of the electric fields make the 
setup of the problem more complicated. In the course of the derivation we will 
introduce most of the definitions that we will also use for the other schemes 
($\Lambda$\mbox{-}type and dual-color).

The atomic ensemble is assumed to be a one-dimensional medium of length $L$ 
consisting of $N$ atoms. In the continuum model, the atomic density 
$n_0=N/L$ is assumed to be constant throughout the length of the ensemble. The 
atoms are described by the collective operators
\begin{align}\label{sigma_collective_definition}
\hs_{\alpha\beta}(z)=\frac{1}{n_0}\sum_{j}\delta(z-z_j)\hs_{\alpha\beta,j}
\end{align}
where $\hs_{\alpha\beta,j}=|\alpha_j\rangle\langle\beta_j|$ is the atomic coherence 
($\alpha\neq\beta$) or population ($\alpha=\beta$) of atom $j$. These 
collective operators have the equal time commutation relations
\begin{align}
[\hs_{\alpha\beta}(z),\hs_{\alpha'\beta'}(z')]
&=\frac{1}{n_0}\delta(z-z')
(\delta_{\beta,\alpha'}\hs_{\alpha\beta'}
-\delta_{\beta',\alpha}\hs_{\alpha'\beta}).
\end{align}
Throughout this paper, all operators are defined to be slowly-varying in time, 
since we work in the interaction picture relative to the carrier frequencies 
of the fields.

The dual\mbox{-}V scheme has two excited states, $|b_+\rangle$ and $|b_-\rangle$, 
which both couple to the ground state $|a\rangle$ but with the different 
polarization modes, $\sigma_+$ and $\sigma_-$, of the quantum field. The 
$\sigma_+$ mode only couples the $|a\rangle\leftrightarrow|b_+\rangle$ 
transition, and the $\sigma_-$ mode only couples the 
$|a\rangle\leftrightarrow|b_-\rangle$ transition. The operator for the total 
quantum field $\hat{\mathcal{E}}_{\sigma_{\pm}}$ for the different 
polarizations can be decomposed as
\begin{gather}
\hat{\mathcal{E}}_{\sigma_\pm}(z)
=\hat{\mathcal{E}}_{\sigma_\pm,+}(z)e^{ik_0 z}
+\hat{\mathcal{E}}_{\sigma_\pm,-}(z)e^{-ik_0 z},
\end{gather}
where $k_0$ is the wave vector corresponding to the carrier frequency of the 
quantum fields $\omega_0$, i.e. $k_0=\omega_0/c$. For the $\sigma_+$ fields, 
$\hat{\mathcal{E}}_{\sigma_+,+}(z)$ is the spatially slowly-varying 
annihilation operator at position $z$ for the field moving to the right (positive direction), and $\hat{\mathcal{E}}_{\sigma_+,-}(z)$ is the operator for the 
field moving to the left (negative direction). Analogous definitions hold for the 
$\sigma_-$ fields. We will be concerned with the dynamics within a frequency 
interval around atomic resonances that is much smaller than the carrier 
frequencies of the fields. Therefore, the right-moving and left-moving quantum 
fields (for each polarization mode) can be regarded as being completely 
separate~\cite{shen_ol05} with the equal time commutation relations
\begin{gather}
[\hat{\mathcal{E}}_\alpha(z),\hat{\mathcal{E}}^\dagger_\beta(z')]
=\delta_{\alpha\beta}\delta(z-z'),
\end{gather}
where $\alpha$ and $\beta$ each denote one of the four possible combinations 
of polarization ($\sigma_\pm$) and propagation direction~($\pm$). 

The transition frequencies between the atomic energy levels $|\alpha\rangle$ 
and $|\beta\rangle$ will be denoted by $\omega_{\alpha \beta}$. The quantum 
fields are detuned from the atomic transition frequencies by 
$\Delta_0^{(\pm)}=\omega_0-\omega_{ab_\pm}$. The excited states $|b_+\rangle$ 
and $|b_-\rangle$ are assumed to have the same incoherent decay rate $\Gamma'$ 
to modes other than the forward and backward propagating ones. We account for 
$\Gamma'$ by making the detunings complex: 
$\tilde{\Delta}_0^{(\pm)}=\Delta_0^{(\pm)}+i\Gamma'/2$. In the calculations 
below, we will employ Fourier transformation, where the Fourier frequencies 
$\omega$ will be defined relative to the carrier frequency $\omega_0$. For 
ease of notation we therefore define the detunings 
$\Delta^{(\pm)}=\Delta^{(\pm)}_0+\omega$. As opposed to the detunings of the 
carrier frequency $\Delta^{(\pm)}_0$, the detunings $\Delta^{(\pm)}$ 
additionally include the shift due to the finite bandwidth of the quantum 
field.

The two counter-propagating classical drives are in the two different 
polarization modes. Here, the polarization and the propagation direction are 
chosen such that $\Omega_+$ is the Rabi frequency of the $\sigma_+$ classical 
drive propagating in the positive direction that couples the transition 
$|b_+\rangle\leftrightarrow|c\rangle$, and $\Omega_-$ is the Rabi frequency of 
the $\sigma_-$ classical drive propagating in the negative direction that 
couples the transition $|b_-\rangle\leftrightarrow|c\rangle$. The classical 
drives have frequency $\omega_\text{c}$ and are detuned from the respective 
transitions by $\Delta_\text{c}^{(\pm)}=\omega_\text{c}-\omega_{b_\pm c}$. 
Furthermore, we define the two-photon detuning 
$\delta_0=\omega_0-\omega_\text{c}-\omega_{ac}$, which has a unique 
definition, since the quantum fields have the same carrier frequency 
($\omega_0$) for both polarizations, and the classical drives have the same 
frequency ($\omega_\text{c}$) for both polarizations. In terms of 
$\Delta_0^{(\pm)}$ and $\Delta_\text{c}^{(\pm)}$ above, we also have
$\delta_0=\Delta_0^{(+)}-\Delta_\text{c}^{(+)}
=\Delta_0^{(-)}-\Delta_\text{c}^{(-)}$. Similar to $\Delta^{(\pm)}$ above, there is a 
complementary definition of the two-photon detning $\delta=\delta_0+\omega$ 
that takes into account the finite bandwidth of the quantum field. The 
wave vector of the classical drive is $k_\text{c}=\omega_\text{c}/c$, but 
throughout our calculations we are going to assume $k_\text{c}\approx k_0$.

The Hamiltonian for the dual\mbox{-}V scheme can be decomposed as 
$\hat{H}_\text{V}=\hat{H}_\text{V,a}+\hat{H}_\text{V,i}+\hat{H}_\text{V,p}$, 
where $\hat{H}_\text{V,a}$ describes the atoms, $\hat{H}_\text{V,p}$ describes 
the photons, and $\hat{H}_\text{V,i}$ describes the light-matter interactions.
In the interaction picture and the rotating wave approximation, the parts are
\begin{subequations}
\begin{align}
&\begin{aligned}
\hat{H}_\text{V,a}=-\hbar n_0\int\left[
\sum_{\alpha\in\cbr{+,-}}\tilde{\Delta}_0^{(\alpha)}
\hat{\sigma}_{b_\alpha b_\alpha}(z)
+\delta_0\hat{\sigma}_{cc}(z)
\right]\dif z,
\end{aligned}\displaybreak[0]\\
&\begin{aligned}
\hat{H}_\text{V,i}=-\hbar n_0\int&
\sum_{\alpha\in\cbr{+,-}}\Bigg\{\left[\hat{\sigma}_{b_\alpha c}(z)\Omega_\alpha 
e^{\alpha ik_\text{c}z}
+\text{H.c.}\right]\\
&+g\sqrt{2\pi}
\left[\hat{\sigma}_{b_\alpha a}(z)\hE_{\sigma_\alpha}(z)
+\text{H.c.}\right]\Bigg\}\dif z,
\end{aligned}\displaybreak[0]\\
&\begin{aligned}
\hat{H}_\text{V,p}=-i\hbar c\int\sum_{\alpha\in\{+,-\}}\Bigg[&
\hE_{\sigma_\alpha,+}^{\dagger}(z)\dpd{\hE_{\sigma_\alpha,+}(z)}{z}\\
&-\hE_{\sigma_\alpha,-}^{\dagger}(z)\dpd{\hE_{\sigma_\alpha,-}(z)}{z}\Bigg]\dif z,
\end{aligned}
\end{align}
\label{dualV_Hamiltonian}
\end{subequations}
where $g=\mu\sqrt{\omega_{ab_\pm}/(4\pi\hbar\epsilon_0 A)}$ (in this constant, 
we assume that $\omega_{ab_+}\approx\omega_{ab_-}$), $\mu$ is the matrix 
element of the atomic dipole, and $A$ is the effective area of the electric field 
mode.

The Heisenberg equations of motion for the electric field operators are given 
by
\begin{subequations}
\begin{align}
\del{\dpd{}{t}\pm c\dpd{}{z}}\nohE_{\sigma_+,\pm}
&=ig\sqrt{2\pi}n_0\nohs_{ab_+}e^{\mp ik_0z}.\label{eqn:hE_sigma_p_pm}\\
\del{\dpd{}{t}\pm c\dpd{}{z}}\nohE_{\sigma_-,\pm}
&=ig\sqrt{2\pi}n_0\nohs_{ab_-}e^{\mp ik_0z}.\label{eqn:hE_sigma_m_pm}
\end{align}\label{eqn:hE_sigma_pm_pm}
\end{subequations}
Here and in the following we will omit the hats above the operators as 
soon as the Heisenberg equations of motion are found, since we will be 
considering linear effects for which the operator character does not play any 
role. The noise operators, normally included in the Heisenberg equations of 
motion whenever incoherent losses are present ($\Gamma'>0$), are also 
omitted, since they can be shown to not have any
effect~\cite{zimmer_pra08,gorshkov_pra07_1,gorshkov_pra07_2}. The equations of 
motion for the atoms are found under the assumption that the probe field is 
weak and that the ensemble is initially prepared in the ground state. Hence, 
we set $\hs_{aa}\approx 1$, $\hs_{b_\pm b_\pm}\approx \hs_{b_\pm b_\mp}\approx
\hs_{cc}\approx\hs_{b_\pm c}\approx 0$, and get the equations
\begin{subequations}
\begin{align}
\label{dual_v_sigma_ab}
&\dpd{\nohs_{ab_\pm}}{t}
=i\tilde{\Delta}_0^{(\pm)}\nohs_{ab_\pm}
+i\Omega_\pm\nohs_{ac}e^{\pm ik_\text{c}z}
+ig\sqrt{2\pi}\nohE_{\sigma_\pm},\\
&\dpd{\nohs_{ac}}{t}
=i\delta_0\nohs_{ac}
+i\Omega_+^*\nohs_{ab_+}e^{-ik_\text{c}z}
+i\Omega_-^*\nohs_{ab_-}e^{ik_\text{c}z}.
\end{align}\label{eqs:coher_dualV}
\end{subequations}
We note that it is in Eqs.~\eqref{eqs:coher_dualV} that the continuum 
approximation is first applied, since both the 
Hamiltonian~\eqref{dualV_Hamiltonian} and Eqs.~\eqref{eqn:hE_sigma_pm_pm} in 
principle retain the discrete nature of the atoms due to the 
definition~\eqref{sigma_collective_definition}. Eqs.~\eqref{eqs:coher_dualV} 
are derived under the approximation $\sigma_{aa}\approx 1$, which can be 
viewed as two separate approximations. The first is that 
$\sigma_{aa,j}\approx 1$ for all the individual atoms $j$. Together with the 
definition~\eqref{sigma_collective_definition}, we see that 
$\sigma_{aa}\approx 1$ also means approximating 
$\sum_j\delta(z-z_j)\approx n_0$, and this is what we mean by the continuum 
approximation. In the analysis done in Ref.~\cite{soerensen_mw_pra08} it was 
shown in a perturbative calculation that this is a good approximation for 
randomly placed atoms. Using the discrete model in 
Sec.~\ref{Sec:discrete_model} below, we will verify it explicitly without any 
perturbative assumptions.

We make two assumptions for simplicity and to be able to relate this 
derivation to the secular approximation for $\Lambda$\mbox{-}type atoms, which we 
discuss below. First, we assume equal atomic transition frequencies, 
$\omega_{b_+c}=\omega_{b_-c}$, so that 
${\Delta_0^{(+)}=\Delta_0^{(-)}=\Delta_0}$, and 
${\Delta_\text{c}^{(+)}=\Delta_\text{c}^{(-)}=\Delta_\text{c}}$. Second, we 
assume equal classical drive strengths, $\Omega_+=\Omega_-=\Omega_0/2$.

\begin{figure}[tbp]
\centering
\includegraphics{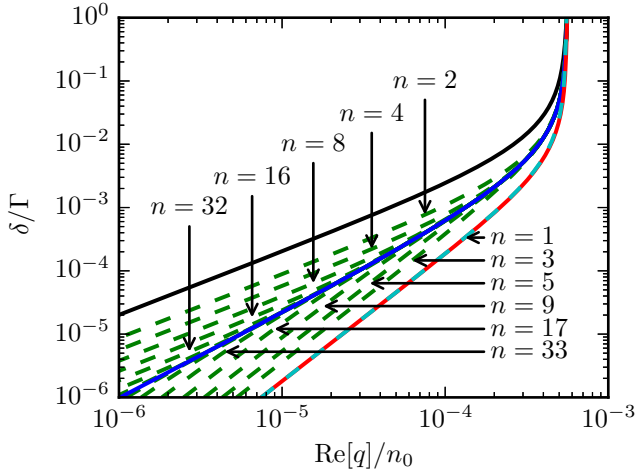}
\caption{(Color online) Log-log plot of the dispersion relations for the 
different setups. The upper solid black curve is for EIT (see 
Eq.~\eqref{Lambda_EIT_dispersion_relation}). The lower solid red curve is the 
quadratic dispersion relation for the dual\mbox{-}V setup (or the secular 
approximation for the $\Lambda$\mbox{-}type scheme) given by 
Eq.~\eqref{dual_v_dispersion_relation_q}. The dashed green curves are the 
dispersion relations for the truncations of 
Eqs.~\eqref{eqs:ss_coher_fourier_components} with increasing number of the 
Fourier components of $\nohs_{ab}$ and $\nohs_{ac}$. The dispersion relations 
for small $\text{Re}[q]/n_0$ alternate between linear and quadratic depending 
on the truncation. The solid blue curve is the analytical limit of these 
dispersion relations given by Eq.~\eqref{Lambda_cold_dispersion_relation}. The 
lower dashed cyan curve is for the dual-color scheme with 
$\Delta_\text{d}/\Gamma=1$. It overlaps the quadratic dual\mbox{-}V curve, so 
that the difference is not visible. The common parameters for all the curves 
are $\Gamma_\text{1D}/\Gamma=0.1$, $\Delta_\text{c}/\Gamma=-90$, and
$\Omega_{0}/\Gamma=1$. The curves are obtained by using a real $\delta$, 
calculating complex $q$ and then plotting $\delta/\Gamma$ as a function of 
$\text{Re}[q]/n_0$. The alternative approach: using real $q$, calculating 
complex $\delta$ and then plotting $\text{Re}[\delta]/\Gamma$ as a function of 
$q/n_0$ will produce results that are indistinguishable for this parameter 
regime (big $\Delta_\text{c}/\Gamma$ and $\Delta_\text{d}/|\delta_\text{S}|$). 
For all the dispersion relations we 
pick the branches such that $\text{Re}[q]/n_0>0$.}
\label{fig:continuum_dispersion_relation_truncations}
\end{figure}

With the above assumptions and defining the slowly-varying versions of 
$\nohs_{ab\pm}$ by
\begin{gather}\label{dual_v_slowly_varying_sigma_ab}
\nohs_{ab}^{\pm}=\nohs_{ab\pm}e^{\mp ik_0 z},
\end{gather}
the equations of motion become
\begin{subequations}
\begin{align}
\label{dual_v_sigma_ab_approx}
&\dpd{\nohs_{ab}^{\pm}}{t}
=i\tilde{\Delta}_0\nohs_{ab}^\pm
+i\frac{\Omega_0}{2}\nohs_{ac}+ig\sqrt{2\pi}\nohE_{\sigma_\pm}e^{\mp ik_0 z},\\
&\dpd{\nohs_{ac}}{t}
=i\delta_0\nohs_{ac}
+i\frac{\Omega_0^*}{2}(\nohs_{ab}^{+}+\nohs_{ab}^{-}),
\end{align}
\label{eqs:coher_dualV_approx}
\end{subequations}
and after the Fourier transform in time,
\begin{subequations}
\begin{align}
\label{dual_v_sigma_ab_approx_ft}
&0
=i\tilde{\Delta}\nohs_{ab}^{\pm}+i\frac{\Omega_0}{2}\nohs_{ac}
+ig\sqrt{2\pi}\nohE_{\sigma_\pm}e^{\mp ik_0 z},\\
\label{dual_v_sigma_ab_prime_approx_ft}
&0
=i\delta\nohs_{ac}
+i\frac{\Omega_0^*}{2}(\nohs_{ab}^{+}+\nohs_{ab}^{-}).
\end{align}
\label{eqs:coher_dualV_approx_ft}
\end{subequations}
Here, we have absorbed the Fourier frequency variable $\omega$ into the 
detunings by defining $\tilde{\Delta}=\tilde{\Delta}_0+\omega$ and 
$\delta=\delta_0+\omega$. Isolating $\nohs_{ac}$ from 
Eq.~\eqref{dual_v_sigma_ab_prime_approx_ft} and inserting into 
Eqs.~\eqref{dual_v_sigma_ab_approx_ft} gives two coupled equations
\begin{gather}\label{dual_v_sigma_ab_approx_ft_ac_eliminated}
0=\left(1-\frac{\delta_\text{S}}{2\delta}\right)\nohs_{ab}^{\pm}
-\frac{\delta_\text{S}}{2\delta}\nohs_{ab}^{\mp}
+\frac{g\sqrt{2\pi}}{\tilde\Delta}\nohE_{\sigma_\pm}e^{\mp ik_0 z}.
\end{gather}
Here, we have introduced 
\begin{gather}
\delta_\text{S}=\frac{|\Omega_0|^2}{2\tilde{\Delta}}.
\end{gather}
For $\delta,\Gamma'\ll \Delta_\text{c}$, 
$\delta_\text{S}\approx |\Omega_0|^2/(2\Delta_\text{c})$ is the total AC Stark 
shift induced by the 
classical drives on the state $|c\rangle$. We will focus on the case when 
$|\delta|\ll |\delta_\text{S}|$. For ${|\delta|\gtrsim |\delta_\text{S}|}$, 
the frequency is outside the scale of the strongest effect induced by the 
classical drives. Therefore, the dispersion relations for the different schemes 
all cross over to the dispersion relation corresponding to a two-level atom, 
as can be seen in Fig.~\ref{fig:continuum_dispersion_relation_truncations}.

Solving Eqs.~\eqref{dual_v_sigma_ab_approx_ft_ac_eliminated}, we find
\begin{gather}\label{dual_v_sigma_ab_approx_ft_ac_eliminated_solved}
\begin{aligned}
\nohs_{ab}^{\pm}
=-\frac{g\sqrt{2\pi}}{\tilde\Delta}
\Bigg[&\frac{\delta-\delta_\text{S}/2}{\delta-\delta_\text{S}}\nohE_{\sigma_\pm}e^{\mp ik_0 z}\\
&+\frac{\delta_\text{S}/2}{\delta-\delta_\text{S}}\nohE_{\sigma_\mp}e^{\pm ik_0 z}\Bigg].
\end{aligned}
\end{gather}
We insert Eqs.~\eqref{dual_v_sigma_ab_approx_ft_ac_eliminated_solved} 
into the Fourier transformed versions of Eqs.~\eqref{eqn:hE_sigma_pm_pm} 
and remove terms with rapid spatial variation, i.e. terms 
containing factors $e^{ink_0 z}$ with the integer $n$ fulfilling $|n|>0$. 
As a consequence, $\nohE_{\sigma_+,+}$ and $\nohE_{\sigma_-,-}$ form a closed 
set of equations, separate from $\nohE_{\sigma_+,-}$ and $\nohE_{\sigma_-,+}$. 
We therefore find
\begin{subequations}
\begin{align}\label{eqn:hE_sigma_pm_pm_ft_ab_inserted}
&\begin{aligned}
&\del{-i\frac{\omega}{cn_0} \pm \frac{1}{n_0}\dpd{}{z}}\nohE_{\sigma_\pm,\pm}\\
&=-i\frac{\Gamma_\text{1D}}{2\tilde\Delta}
\left[\frac{\delta-\delta_\text{S}/2}{\delta-\delta_\text{S}}\nohE_{\sigma_\pm,\pm}
+\frac{\delta_\text{S}/2}{\delta-\delta_\text{S}}\nohE_{\sigma_\mp,\mp}\right],
\end{aligned}\\
\label{eqn:hE_sigma_pm_mp_ft_ab_inserted}
&\begin{aligned}
&\del{-i\frac{\omega}{cn_0} \mp \frac{1}{n_0}\dpd{}{z}}\nohE_{\sigma_\pm,\mp}
=-i\frac{\Gamma_\text{1D}}{2\tilde\Delta}
\frac{\delta-\delta_\text{S}/2}{\delta-\delta_\text{S}}\nohE_{\sigma_\pm,\mp},
\end{aligned}
\end{align}
\label{eqn:hE_sigma_pm_pm_and_mp_ft_ab_inserted}
\end{subequations}
where we have introduced the decay rate $\Gamma_\text{1D}=4\pi g^2/c$ which 
describes the photon emission rate into the one-dimensional modes (the sum of 
right-moving and left-moving) from the atoms. The total decay rate of an excited 
atom is then $\Gamma=\Gamma'+\Gamma_\text{1D}$. In the absence of 
inhomogeneous broadening, the decay rate $\Gamma_\text{1D}$ is related to the 
resonant optical depth $d_\text{opt}$ through
$d_\text{opt}=2N\Gamma_\text{1D}/\Gamma$.

Since the Hamiltonian \eqref{dualV_Hamiltonian} is periodic in space with 
period $2\pi/k_0$ we can invoke Bloch's theorem and look for solutions to 
Eqs.~\eqref{eqn:hE_sigma_pm_pm_and_mp_ft_ab_inserted} of the form
\begin{gather}\label{eqn:hE_sigma_pm_pm_ft_ab_inserted_sought_sol_total_field}
\nohE_{\sigma_\pm}(z,\omega)
=\del{\nohE_{\sigma_\pm,+}(0,\omega)e^{ik_0 z}
+\nohE_{\sigma_\pm,-}(0,\omega)e^{-ik_0 z}} e^{iqz}.
\end{gather}
In general by Bloch's theorem, 
Eq.~\eqref{eqn:hE_sigma_pm_pm_ft_ab_inserted_sought_sol_total_field} should 
have been a product of a periodic function and the factor $e^{iqz}$, where $q$ 
is the Bloch vector. In 
Eq.~\eqref{eqn:hE_sigma_pm_pm_ft_ab_inserted_sought_sol_total_field} we have 
effectively written the periodic function as a Fourier series and kept only 
the $\pm 1$ terms, which were then identified with the components 
$\nohE_{\sigma_\pm,+}$ and $\nohE_{\sigma_\pm,-}$ at $z=0$. Removing higher 
order modes is justified, since we are interested in the dynamics, for which 
${|k_0|=|\omega_0/c|\gg |q|}$. Effectively, after applying the derivative 
$\partial/\partial z$ in 
Eqs.~\eqref{eqn:hE_sigma_pm_pm_and_mp_ft_ab_inserted}, the higher order modes 
will have an energy difference that is multiple of $c k_0$, which corresponds 
to a multiple of the optical 
frequency of the atomic transition.

On the other hand, the frequency $\omega$ 
in Eqs.~\eqref{eqn:hE_sigma_pm_pm_and_mp_ft_ab_inserted} is relative to the 
carrier frequency $\omega_0$ and is assumed to fulfill ${|\omega/c|\ll |q|}$, 
i.e. within the narrow frequency range of interest the stationary light 
dispersion is the dominant contribution to the dispersion relation and the 
vacuum dispersion relation can be neglected. Therefore, we remove the terms 
$\omega/(cn_0)$ in the following.

The form of 
Eq.~\eqref{eqn:hE_sigma_pm_pm_ft_ab_inserted_sought_sol_total_field} implies 
that we should insert
\begin{gather}\label{eqn:hE_sigma_pm_pm_ft_ab_inserted_sought_sol}
\nohE_{\sigma_\pm,\pm}(z,\omega)=\nohE_{\sigma_\pm,\pm}(0,\omega)e^{iqz},
\end{gather}
into Eqs.~\eqref{eqn:hE_sigma_pm_pm_ft_ab_inserted} and 
\begin{gather}\label{eqn:hE_sigma_pm_mp_ft_ab_inserted_sought_sol}
\nohE_{\sigma_\pm,\mp}(z,\omega)=\nohE_{\sigma_\pm,\mp}(0,\omega)e^{iqz},
\end{gather}
into Eqs.~\eqref{eqn:hE_sigma_pm_mp_ft_ab_inserted}. After removing terms with 
rapid spatial variation, this gives
\begin{subequations}
\begin{align}
\label{eqn:hE_sigma_pm_pm_ft_ab_inserted_bloch}
&\pm\frac{q}{n_0}\nohE_{\sigma_\pm,\pm}
=-\frac{\Gamma_\text{1D}}{2\tilde\Delta}
\left[\frac{\delta-\delta_\text{S}/2}{\delta-\delta_\text{S}}\nohE_{\sigma_\pm,\pm}
+\frac{\delta_\text{S}/2}{\delta-\delta_\text{S}}\nohE_{\sigma_\mp,\mp}\right],
\\
\label{eqn:hE_sigma_pm_mp_ft_ab_inserted_bloch}
&\mp\frac{q}{n_0}\nohE_{\sigma_\pm,\mp}
=-\frac{\Gamma_\text{1D}}{2\tilde\Delta}
\frac{\delta-\delta_\text{S}/2}{\delta-\delta_\text{S}}\nohE_{\sigma_\pm,\mp}.
\end{align}
\end{subequations}

The equations above describe coupling between the different electric field 
modes. We first solve for the field modes moving in the opposite direction 
compared to the classical fields of the same polarization 
($\nohE_{\sigma_\pm,\mp}$). Due to momentum conservation (or equivalently the 
lack of mode matching), these do not couple to any other field modes. As a 
consequence, we essentially have two separate $\Lambda$\mbox{-}systems. One of 
them involves the states $|a\rangle$, $|b_+\rangle$, and $|c\rangle$, which are coupled 
by the fields $\nohE_{\sigma_+,-}$ and $\Omega_+$. The other one involves the states
$|a\rangle$, $|b_-\rangle$, and $|c\rangle$, which are coupled by the fields 
$\nohE_{\sigma_-,+}$ and $\Omega_-$. From 
Eqs.~\eqref{eqn:hE_sigma_pm_mp_ft_ab_inserted_bloch} we immediately find the 
dispersion relations
\begin{gather}\label{dual_v_dispersion_relation_linear_q}
\frac{q}{n_0}=\pm\frac{\Gamma_\text{1D}}{2\tilde\Delta}
\frac{\delta-\delta_\text{S}/2}{\delta-\delta_\text{S}}.
\end{gather}
Solving these equations for $\delta$ and expanding for small $q/n_0$ gives
\begin{align}\label{dual_v_dispersion_relation_linear_q_approx}
&\delta
\approx\frac{\delta_\text{S}}{2}
\mp\frac{|\Omega_0|^2}{2\Gamma_\text{1D}}\frac{q}{n_0}.
\end{align}
This is the regular EIT dispersion relation (c.f. Eqs. 
\eqref{Lambda_EIT_dispersion_relation} and 
\eqref{Lambda_EIT_dispersion_relation_small_delta} below) only shifted by the 
AC Stark shift of the classical drive not participating in the EIT (since it 
is only shifted by one of the fields, the shift is $\delta_\text{S}/2$).

The quadratic dispersion relation is obtained from 
Eqs.~\eqref{eqn:hE_sigma_pm_pm_ft_ab_inserted_bloch}.
Here, the forward and backward propagation are 
coupled and can be written in matrix form as
\begin{align}\label{dual_v_coupled_equations_E_matrix}
\begin{pmatrix}
\alpha_1+ \frac{q}{n_0} & \alpha_2\\
\alpha_2 & \alpha_1- \frac{q}{n_0}
\end{pmatrix}
\begin{pmatrix}
\nohE_{\sigma_+,+}\\
\nohE_{\sigma_-,-}
\end{pmatrix}
=
\begin{pmatrix}
0\\
0
\end{pmatrix}
\end{align}
with
\begin{gather}\label{alpha_secular_definition}
\alpha_1=\left(\frac{\Gamma_\text{1D}}{2\tilde\Delta}\right)
\frac{\delta-\delta_\text{S}/2}{\delta-\delta_\text{S}},\quad
\alpha_2=\left(\frac{\Gamma_\text{1D}}{2\tilde\Delta}\right)
\frac{\delta_\text{S}/2}{\delta-\delta_\text{S}}.
\end{gather}
In order for Eq.~\eqref{dual_v_coupled_equations_E_matrix} to have non-trivial 
solutions, the determinant of the matrix on the left hand side has to be zero. 
This produces the equation
\begin{align}\label{dual_v_dispersion_relation_q}
\left(\frac{q}{n_0}\right)^2
&=\left(\frac{\Gamma_\text{1D}}{2\tilde\Delta}\right)^2
\frac{\delta}{\delta-\delta_\text{S}},
\end{align}
which determines the dispersion relation.
Solving Eq.~\eqref{dual_v_dispersion_relation_q} for $\delta$, and expanding 
the solution for small $q/n_0$, we get the quadratic dispersion relation
\begin{align}\label{dual_v_dispersion_relation_delta_approx}
\delta\approx\frac{1}{2m} \left(\frac{q}{n_0 }\right)^2
\end{align}
with the effective mass
\begin{gather}\label{dual_v_dispersion_relation_mass}
m=-\frac{\Gamma_\text{1D}^2}{4(\Delta_\text{c}+i\Gamma'/2)|\Omega_0|^2}.
\end{gather}
The quadratic dispersion relation 
\eqref{dual_v_dispersion_relation_delta_approx} is the same as for the original 
stationary light in hot $\Lambda$\mbox{-}type 
atoms~\cite{andre_prl02a,bajcsy_nature03a} (see below for a discussion of the 
connection between cold dual\mbox{-}V and hot $\Lambda$\mbox{-}type schemes). 
We plot the full dispersion relation given by 
Eq.~\eqref{dual_v_dispersion_relation_q} in 
Fig.~\ref{fig:continuum_dispersion_relation_truncations} as the solid red 
curve.

Having gone through the derivation, we now return to highlight some important 
parts, which will be of relevance later. We note that when solving the atomic 
equations (Eqs.~\eqref{eqs:coher_dualV_approx}), the full spatial dependence 
of the field was included, i.e. no attempt was made to remove fast-varying 
terms at this level. Such a procedure was only made after substituting the 
atomic solutions into the Fourier transforms of the field 
equations~\eqref{eqn:hE_sigma_pm_pm}. We will show below, that for the cold 
$\Lambda$\mbox{-}type atoms, it is very important, at which point and how the removal 
of the fast-varying terms is performed.

\subsection{Dispersion relations for cold $\Lambda$\mbox{-}type atoms}
\label{Sec:lambda_continuum_model}

We now turn to the $\Lambda$\mbox{-}type scheme shown in 
Fig.~\ref{fig:level_diagrams}(b). The atoms have fewer energy levels than in 
the dual\mbox{-}V scheme, but the dynamics in the case of cold atoms is 
complicated by presence of higher order Fourier components of the atomic 
coherence~\cite{hansen_pra07a,nikoghosyan_pra09a,lin_prl09,wu_pra10,wu_pra10_2,peters_pra12}. The dispersion relation for 
the cold $\Lambda$\mbox{-}type scheme, that effectively sums all the 
Fourier components to the infinite order, has been found in 
Ref.~\cite{moiseev_pra2014}. However, the result in Ref.~\cite{moiseev_pra2014} does 
not provide much intuition about the underlying physics. Here, we will do a 
different derivation that explicitly tracks the different Fourier components 
of the atomic coherence. This will illustrate the differences from the 
dual\mbox{-}V scheme and lead to the discussion of the ``secular approximation'' for the 
$\Lambda$\mbox{-}type scheme, which makes the two schemes equivalent. This derivation 
will also serve as a connection between the continuum and discrete models of 
the $\Lambda$\mbox{-}type scheme. In short, the different truncations of the infinite 
set of Fourier components that we will discuss in the continuum model can physically be implemented by 
positioning the atoms in the discrete model the certain way (see 
Sec.~\ref{Sec:discrete_model_Lambda}). For completeness, we will also do a 
second derivation of the dispersion relation for the $\Lambda$\mbox{-}type 
scheme that is more similar to Ref.~\cite{moiseev_pra2014}, but with more focus 
on the off-resonant regime ($\Delta_\text{c}\neq 0$, $\delta\neq 0$).

Compared to the dual\mbox{-}V scheme, the $\Lambda$\mbox{-}type atoms have only one 
excited state $|b\rangle$, and there is only one polarization mode for both 
the quantum and the classical fields. The quantum field has detuning $\Delta_0$ 
from the $|a\rangle\leftrightarrow|b\rangle$ transition, and the classical 
drive has detuning $\Delta_\text{c}$ from the 
$|b\rangle\leftrightarrow|c\rangle$ transition. The operator for the total 
quantum field $\hat{\mathcal{E}}$ can be decomposed as
$\hat{\mathcal{E}}(z)=\hat{\mathcal{E}}_+(z)e^{ik_0z}+\hat{\mathcal{E}}_-(z)e^{-ik_0z}$, 
where $\hat{\mathcal{E}}_\pm$ are the spatially slowly-varying components.
The classical drive is given by the sum of the two parts moving in both 
directions, $\Omega(z)=\Omega_0\cos(k_0 z)$ (assuming $k_\text{c}\approx k_0$). 
Similar to the dual-V scheme and using the definitions above, the Hamiltonian is 
$\hat{H}_\text{3}=\hat{H}_\text{3,a}+\hat{H}_\text{3,i}+\hat{H}_\text{3,p}$ 
(sum of the atomic, interaction, and photonic parts), where
\begin{subequations}
\begin{align}
&\begin{aligned}
\hat{H}_\text{3,a}=-\hbar n_0\int\Big[\tilde{\Delta}_0\hat{\sigma}_{bb}(z)
+\delta_0\hat{\sigma}_{cc}(z)\Big]\dif z
\end{aligned}\displaybreak[0]\\
&\begin{aligned}
\hat{H}_\text{3,i}=-\hbar n_0\int\Bigg\{&\left[\hat{\sigma}_{bc}(z)\Omega(z)
+\text{H.c.}\right]\\
&+g\sqrt{2\pi}\left[\hat{\sigma}_{ba}(z)\hat{\mathcal{E}}(z)
+\text{H.c.}\right]\Bigg\}\dif z
\end{aligned}\displaybreak[0]\\
&\begin{aligned}
\hat{H}_\text{3,p}=-i\hbar c\int\Bigg[&\hE_{+}^{\dagger}(z)\dpd{\hE_{+}(z)}{z}
-\hE_{-}^{\dagger}(z)\dpd{\hE_{-}(z)}{z}\Bigg]\dif z.
\end{aligned}
\end{align}
\label{three_level_H}
\end{subequations}
With this Hamiltonian, the Heisenberg equations of motion for the electric 
field operators and the atomic operators are given by
\begin{align}
\label{eqn:hE_pm}
\del{\dpd{}{t}\pm c\dpd{}{z}}\nohE_{\pm}&=ig\sqrt{2\pi}n_0\nohs_{ab}e^{\mp ik_0z},
\end{align}
and
\begin{subequations}
\begin{align}
&\dpd{\nohs_{ab}}{t}
=i\tilde\Delta_0\nohs_{ab}+i\Omega\nohs_{ac}+ig\sqrt{2\pi}\nohE,\\
&\dpd{\nohs_{ac}}{t}
=i\delta_0\nohs_{ac}
+i\Omega^*\nohs_{ab}.
\end{align}\label{eqs:coher}
\end{subequations}

If $\Omega$ were independent of position ($\Omega(z)=\Omega_0$), 
Eqs.~\eqref{eqs:coher} would describe the usual EIT system, which can be shown 
to have the dispersion relation
\begin{gather}\label{Lambda_EIT_dispersion_relation}
\frac{q}{n_0}=\pm \frac{\Gamma_\text{1D}}{2\tilde{\Delta}}
\frac{\delta}{\delta-2\delta_\text{S}},
\end{gather}
or for small $q/n_0$,
\begin{gather}\label{Lambda_EIT_dispersion_relation_small_delta}
\delta\approx \pm \frac{2|\Omega_0|^2}{\Gamma_\text{1D}}\frac{q}{n_0}.
\end{gather}
We note that for the $\Lambda$\mbox{-}type scheme, $\delta_\text{S}$ has a 
different meaning. For EIT with $\Omega(z)=\Omega_0$, it is a half of the AC 
Stark shift induced by the field. For $\Omega(z)=\Omega_0\cos(k_0 z)$ below, 
it is the average of the AC Stark shift. Also note that the group velocity 
(factor in front of $q$) in 
Eq.~\eqref{Lambda_EIT_dispersion_relation_small_delta} differs by a factor of 
4 from the group velocity
in Eq.~\eqref{dual_v_dispersion_relation_linear_q_approx}. This difference 
arises from the fact that the strength of the field participating in the EIT in 
that case is given by $\Omega_\pm=\Omega_0/2$.

We now calculate the dispersion relation for the case when $\Omega$ is a 
standing wave ($\Omega(z)=\Omega_0\cos(k_0 z)$). The Fourier transform in time 
of Eqs.~\eqref{eqs:coher} gives
\begin{subequations} 
\begin{align}
\label{eqs:ss_coher_app_a}
&0=i\tilde{\Delta}\nohs_{ab}+i\Omega\nohs_{ac}+ig\sqrt{2\pi}\nohE,\\
\label{eqs:ss_coher_app_b}
&0=i\delta\nohs_{ac}+i\Omega^*\nohs_{ab},
\end{align}
\label{eqs:ss_coher_app}
\end{subequations}
where, as before, we have absorbed the Fourier frequency variable $\omega$ 
into the detunings by defining $\tilde{\Delta}=\tilde{\Delta}_0+\omega$ and 
$\delta=\delta_0+\omega$.

By Bloch's theorem, $\nohs_{ab}$, $\nohs_{ac}$ and $\nohE$ need to be periodic 
functions in space multiplied by the factor $e^{iqz}$, with $q$ being the 
Bloch vector. The periodic parts have the same periodicity as $\Omega(z)$, and 
we write each one of them as a Fourier series
\begin{subequations} 
\begin{align}
\label{sigma_ab_fourier_sum_def}
&\nohs_{ab}(z,\omega)=\sum_{n=-\infty}^\infty
\nohs_{ab}^{(n)}(\omega) e^{ink_0 z} e^{iqz},\\
\label{sigma_ac_fourier_sum_def}
&\nohs_{ac}(z,\omega)=\sum_{n=-\infty}^\infty
\nohs_{ac}^{(n)}(\omega) e^{ink_0 z} e^{iqz},\\
\label{E_fourier_sum_def}
&\nohE(z,\omega)
=\del{\nohE_+(0,\omega)e^{ik_0 z}+\nohE_-(0,\omega)e^{-ik_0 z}}e^{iqz},
\end{align}
\label{sigma_ab_ac_and_E_fourier_sum_def}
\end{subequations}
where we have kept only the lowest order terms in the Fourier series for the 
field, similar to 
Eq.~\eqref{eqn:hE_sigma_pm_pm_ft_ab_inserted_sought_sol_total_field}.

After inserting Eqs. \eqref{sigma_ab_fourier_sum_def} and 
\eqref{sigma_ac_fourier_sum_def} into Eqs.~\eqref{eqs:ss_coher_app} and 
collecting the terms with equal exponents of $ink_0 z$, we obtain an infinite 
set of coupled equations
\begin{subequations} 
\begin{align}
0=\;&i\tilde{\Delta}\nohs_{ab}^{(n)}+i\frac{\Omega_0}{2}\del{\nohs_{ac}^{(n+1)}+\nohs_{ac}^{(n-1)}}\nn\\
&+ig\sqrt{2\pi}\del{\nohE_+\delta_{n,1}+\nohE_-\delta_{n,-1}},\\
0=\;&i\delta\nohs_{ac}^{(n)}+i\frac{\Omega_0^*}{2}\del{\nohs_{ab}^{(n+1)}+\nohs_{ab}^{(n-1)}},
\end{align}
\label{eqs:ss_coher_fourier_components}
\end{subequations}
where $\delta_{j,j'}$ is the Kronecker delta.

From the above equations, we see the crucial difference between the 
dual\mbox{-}V scheme and the cold $\Lambda$\mbox{-}type scheme. In the 
dual\mbox{-}V scheme, described in Eqs.~\eqref{eqs:coher_dualV_approx_ft}, 
there are only two components of the atomic coherence for the excited states 
($\nohs_{ab\pm}$).  For the cold $\Lambda$\mbox{-}type atoms, by writing 
$\nohs_{ab}$ as a Fourier series, we have obtained an infinite set of coupled 
components. This can be explained by the fact that a dual\mbox{-}V atom in 
state $|c\rangle$ can transition to state $|b_+\rangle$ (i.e. be excited) by 
absorbing a photon of the classical drive propagating in the positive 
direction, and can only transition back to state $|c\rangle$ (i.e. be 
de-excited) by emitting a photon in the same direction. On the other hand, a 
cold $\Lambda$\mbox{-}type atom in state $|c\rangle$ can transition to state 
$|b\rangle$ by a photon of the clasical drive coming from one direction and 
transition back to state $|c\rangle$ by emitting a photon in the opposite 
direction. This couples a Fourier component $\nohs_{ab}^{(n)}$ with a certain 
wave number $n$ to components differing by two wave numbers, i.e.  
$\nohs_{ab}^{(n\pm2)}$ (through $\nohs_{ac}^{(n\pm 1)}$), and leads to an 
infinite set of coupled equations.

To obtain any results from Eqs.~\eqref{eqs:ss_coher_fourier_components}, 
truncation of the Fourier components of $\nohs_{ab}$ and $\nohs_{ac}$ is 
needed. The smallest non-trivial truncated set of equations involves 
$\sigma_{ab}^{(\pm 1)}$ and $\sigma_{ac}^{(0)}$ and can be written
\begin{subequations} 
\begin{align}
0=\;&i\tilde{\Delta}\nohs_{ab}^{(\pm 1)}+i\frac{\Omega_0}{2}\nohs_{ac}^{(0)}
+ig\sqrt{2\pi}\nohE_\pm,\\
0=\;&i\delta\nohs_{ac}^{(0)}+i\frac{\Omega_0^*}{2}\del{\nohs_{ab}^{(+1)}+\nohs_{ab}^{(-1)}}.
\end{align}
\label{eqs:ss_coher_fourier_components_truncated1}
\end{subequations}
This particular truncation is also known as the ``secular approximation'' in 
the literature \cite{zimmer_pra08, nikoghosyan_pra09a}. If we had approximated 
$\nohE_{\sigma_\pm}e^{\mp ik_0 z}\approx \nohE_{\sigma_\pm,\pm}$ in 
Eqs.~\eqref{eqs:coher_dualV_approx_ft} (which would not have changed the 
quadratic dispersion relation for the dual\mbox{-}V scheme), then 
Eqs.~\eqref{eqs:ss_coher_fourier_components_truncated1} would have had exactly 
the same form as Eqs.~\eqref{eqs:coher_dualV_approx_ft}.

The equations for the electric field \eqref{eqn:hE_pm}, in principle, 
contain all the Fourier components $\nohs_{ab}^{(n)}$, but, as for the dual\mbox{-}V 
scheme, we will make the approximation, where we remove terms with rapid 
spatial variation. This effectively means that we approximate 
$\nohs_{ab}e^{\mp ik_0z}\approx\nohs_{ab}^{(\pm 1)}e^{iqz}$ in Eqs.~\eqref{eqn:hE_pm}. 
Fourier transforming these equations, we end up with 
\begin{align}\label{eqn:hE_pm_fourier} \del{-i\omega\pm 
c\dpd{}{z}}\nohE_{\pm}&=ig\sqrt{2\pi}n_0\nohs_{ab}^{(\pm 1)}e^{iqz}.
\end{align}
Proceeding as for the dual\mbox{-}V case, 
Eqs.~\eqref{eqs:ss_coher_fourier_components_truncated1} and 
Eqs.~\eqref{eqn:hE_pm_fourier} together with the sought form of the Bloch 
solutions
\begin{gather}\label{eqn:hE_pm_Lambda_sought_sol}
\nohE_{\pm}(z,\omega)=\nohE_{\pm}(0,\omega)e^{iqz},
\end{gather}
which is similar to Eqs. \eqref{eqn:hE_sigma_pm_pm_ft_ab_inserted_sought_sol} 
and \eqref{eqn:hE_sigma_pm_mp_ft_ab_inserted_sought_sol} for the 
dual\mbox{-}V scheme, result in the coupled equations for the fields
\begin{align}\label{Lambda_coupled_equations_E_matrix}
\begin{pmatrix}
\alpha_1+ \frac{q}{n_0} & \alpha_2\\
\alpha_2 & \alpha_1- \frac{q}{n_0}
\end{pmatrix}
\begin{pmatrix}
\nohE_{+}\\
\nohE_{-}
\end{pmatrix}
=
\begin{pmatrix}
0\\
0
\end{pmatrix}.
\end{align}
This is the same as Eq.~\eqref{dual_v_coupled_equations_E_matrix} with the 
same $\alpha_1$ and $\alpha_2$ (but with different definitions of the 
electric fields). Hence, exactly the same quadratic dispersion 
relation~\eqref{dual_v_dispersion_relation_delta_approx} is obtained.

A completely different dispersion relation can be found by considering the 
next smallest truncated set of equations. That set additionally involves 
$\sigma_{ac}^{(\pm 2)}$, so that the system of equations is
\begin{subequations}
\begin{align}
0=\;&i\tilde{\Delta}\nohs_{ab}^{(\pm 1)}
+i\frac{\Omega_0}{2}\del{\nohs_{ac}^{(0)}+\nohs_{ac}^{(\pm 2)}}
+ig\sqrt{2\pi}\nohE_\pm,\\
0=\;&i\delta\nohs_{ac}^{(0)}+i\frac{\Omega_0^*}{2}\del{\nohs_{ab}^{(+1)}+\nohs_{ab}^{(-1)}},\\
0=\;&i\delta\nohs_{ac}^{(\pm 2)}+i\frac{\Omega_0^*}{2}\nohs_{ab}^{(\pm 1)}.
\end{align}
\label{eqs:ss_coher_fourier_components_truncated2}
\end{subequations}
Following the same procedure as above, we get the dispersion relation
\begin{align}
\del{\frac{q}{n_0}}^2
&=\del{\frac{\Gamma_\text{1D}}{2\tilde\Delta}}^2
\frac{\delta^2}{(\delta-\delta_\text{S}/2)(\delta-3\delta_\text{S}/2)},
\end{align}
which for small $q/n_0$ can be approximated by
\begin{gather}\label{Lambda_second_truncation_dispersion_relation}
\delta\approx \pm \frac{\sqrt{3}|\Omega_0|^2}{2\Gamma_\text{1D}}\frac{q}{n_0}.
\end{gather}
This dispersion relation is linear instead of 
quadratic. Comparing it with the dispersion relation for 
EIT~\eqref{Lambda_EIT_dispersion_relation_small_delta}, 
we observe that Eq.~\eqref{Lambda_second_truncation_dispersion_relation} only 
differs by a constant factor.

One could continue calculating dispersion relations for even higher order 
truncations. As the analytical calculations quickly become complicated, we 
only do it numerically, as described in App.~\ref{App:Continuum_numerical}. 
The resulting dispersion relations are shown in 
Fig.~\ref{fig:continuum_dispersion_relation_truncations}. We find that 
truncations which contain Fourier components up to and including 
$\nohs_{ab}^{(\pm n)}$ with odd $n$, result in a quadratic dispersion relation 
for small $q/n_0$. On the other hand, truncations that contain Fourier 
components up to and including $\nohs_{ac}^{(\pm n)}$ with even $n$, result in 
a linear dispersion relation for small~$q/n_0$.

It is possible to find the limiting dispersion relation ($n\rightarrow\infty$) 
analytically~\cite{moiseev_pra2014}. To derive it, we will not use 
the Fourier series representation in Eqs.~\eqref{sigma_ab_fourier_sum_def} and 
\eqref{sigma_ac_fourier_sum_def}, but instead solve 
Eqs.~\eqref{eqs:ss_coher_app} directly. Isolating $\nohs_{ac}$ from 
Eq.~\eqref{eqs:ss_coher_app_b} and inserting in Eq.~\eqref{eqs:ss_coher_app_a} 
gives
\begin{align}\label{eqn:hs_ab_app}
\nohs_{ab}(z,\omega)&=-\frac{g\sqrt{2\pi}}{\tilde\Delta}\gamma(z)\nohE(z,\omega),
\end{align}
where we have defined the dimensionless position dependent coupling parameter
\begin{align}
\gamma(z)
=\frac{1}{1-(2\delta_\text{S}/\delta)\cos^2(k_0 z)}.
\end{align}
We then introduce the Fourier series of $\gamma$, i.e.
\begin{align}\label{gamma_Fourier_series}
\gamma(z)=\sum_{\ell=-\infty}^{\infty}\gamma^{(\ell)}e^{2i\ell k_0 z},
\end{align}
with
\begin{gather}
\begin{aligned}
\gamma^{(\ell)}&=\frac{1}{\pi}\int_{-\frac{\pi}{2}}^{\frac{\pi}{2}} 
\gamma(z) e^{-2i\ell k_0 z}\dif\,(k_0 z)\\
&={}_3\tilde{F}_2\sbr{\cbr{\frac{1}{2},1,1},\{1-\ell,1+\ell\},(2\delta_\text{S}/\delta)},
\end{aligned}
\end{gather}
where ${}_p\tilde{F}{}_q$ is the regularized generalized hypergeometric 
function. The terms with 
$\ell=0$ and $\ell=\pm1$ are
\begin{subequations}
\begin{align}
\gamma^{(0)}
&=\frac{1}{\sqrt{1-(2\delta_\text{S}/\delta)}},\\
\gamma^{(\pm1)}
&=-\frac{2 \sqrt{1-(2\delta_\text{S}/\delta)}+(2\delta_\text{S}/\delta)-2}
{(2\delta_\text{S}/\delta)\sqrt{1-(2\delta_\text{S}/\delta)}}.
\end{align}
\end{subequations}
Inserting Eq.~\eqref{gamma_Fourier_series} into Eq.~\eqref{eqn:hs_ab_app} we can write
\begin{align}\label{sigma_ab_fourier_sum}
\nohs_{ab}&=-\frac{g\sqrt{2\pi}}{\tilde\Delta}
\sum_{\ell=-\infty}^{\infty}
[\gamma^{(\ell)}\nohE_{+}+\gamma^{(\ell+1)}\nohE_{-}]e^{i(2\ell+1)k_0 z}.
\end{align}
In Eqs.~\eqref{eqn:hE_pm_fourier}, we only need the terms from 
Eq.~\eqref{sigma_ab_fourier_sum} that have the factors $e^{\pm ik_0z}$, i.e. 
the terms corresponding to $\ell=0$ and $\ell=-1$. Inserting those terms into 
Eqs.~\eqref{eqn:hE_pm_fourier} and proceeding as for the previous calculations 
we get two coupled equations for the fields as in 
Eqs.~\eqref{Lambda_coupled_equations_E_matrix}, but with the different 
$\alpha_1$ and $\alpha_2$:
\begin{gather}\label{alpha_cold_Lambda_definition}
\alpha_1=\frac{\Gamma_\text{1D}}{2\tilde\Delta}\gamma^{(0)},\quad
\alpha_2=\frac{\Gamma_\text{1D}}{2\tilde\Delta}\gamma^{(\pm1)}.
\end{gather}
Finally, we get the dispersion relation
\begin{align}\label{Lambda_cold_dispersion_relation}
\del{\frac{q}{n_0 }}^2
&=\del{\frac{\Gamma_\text{1D}}{2\tilde\Delta}}^2
\del{\frac{4\del{-1 + \sqrt{1 - (2\delta_\text{S}/\delta)}}^2}
{\sqrt{1 - (2\delta_\text{S}/\delta)}(2\delta_\text{S}/\delta)^2}}
\end{align}
which for small real $q/n_0$ can be approximated by 
\begin{align}\label{Lambda_cold_dispersion_relation_small_delta}
\delta
\approx \frac{(-\Delta_\text{c}-i\Gamma'/2)^{1/3}|\Omega_0|^2}
{\Gamma_\text{1D}^{4/3}}
\envert{\frac{q}{n_0}}^{4/3},
\end{align}
where we have restricted the solution to the branch with $\text{Re}[\delta]$ 
having the same sign as $-\Delta_\text{c}$ ($\Delta_\text{c}\neq 0$), and 
${(-\Delta_\text{c}-i\Gamma'/2)^{1/3}}$ means third root of 
${-\Delta_\text{c}-i\Gamma'/2}$ such that 
${\text{Re}[(-\Delta_\text{c}-i\Gamma'/2)^{1/3}]}$ has the same sign as 
$-\Delta_\text{c}$.

We see that for small $q/n_0$, the dispersion relation is neither quadratic, 
nor linear, but goes as $\delta \propto |q|^{4/3}$. The dispersion 
relation~\eqref{Lambda_cold_dispersion_relation} is shown by the solid blue 
curve in Fig.~\ref{fig:continuum_dispersion_relation_truncations}. It is seen 
to lie in between the curves for the EIT and dual\mbox{-}V results and is the 
limiting case as we increase the number of Fourier components for the atomic 
coherence.

\subsection{Dispersion relations for cold dual-color atoms}
\label{Sec:dual_color_continuum_model}
We now consider the dual-color scheme shown in 
Fig.~\ref{fig:level_diagrams}(c). The dispersion for this scheme has been 
originally derived in Ref.~\cite{moiseev_pra2006} under the secular 
approximation. Using the secular approximation for this scheme makes the 
dual-color scheme equivalent to the dual-V scheme. However, in the analysis 
below, we want to illustrate the fact that the dynamics of the dual-color 
scheme can potentially be much more complex compared to the dual-V and 
$\Lambda$\mbox{-}type schemes.

The atomic level structure of the dual-color scheme is the same as for 
the $\Lambda$-type scheme, but the two counter-propagating classical drives 
are at two different frequencies instead of only one. The detuning 
$\Delta_\text{c}$ now has a different meaning---it is relative to the mean of 
the two frequencies. Hence, if $\omega_{\text{c}\pm}$ are the frequencies of 
the two classical drives, then
$\Delta_\text{c}=(\omega_{\text{c}+}+\omega_{\text{c}-})/2-\omega_{bc}$. We 
also define the detuning 
$\Delta_\text{d}=|\omega_{\text{c}+}-\omega_{\text{c}-}|/2$, which measures 
how far the two frequencies are separated from each other. With the modified 
definition of $\Delta_\text{c}$, the Hamiltonian is the same as for the 
$\Lambda$\mbox{-}type atom, i.e. it is given by Eq.~\eqref{three_level_H}, but 
with $\Omega(z,t)=\Omega_{0}\cos(\Delta_\text{d}t+k_\text{c}z)$. The 
Heisenberg equations of motion are also the same as for the $\Lambda$\mbox{-}type 
scheme (Eqs.~\eqref{eqn:hE_pm} and Eqs.~\eqref{eqs:coher}), just with the 
different definition of $\Omega(z,t)$.

Compared to the $\Lambda$\mbox{-}type scheme, the dual-color scheme has a 
time-dependent Hamiltonian, but since it is periodic in time, it allows us to 
use Floquet's theorem in addition to Bloch's 
theorem~\cite{holthaus_prb2011,holthaus_floquet_tutorial}. According to the 
two theorems, $\nohs_{ab}$, $\nohs_{ac}$ and $\nohE$ need to be periodic 
functions in space and time multiplied by the factor $e^{iqz-i\omega_q t}$, 
with $q$ being the Bloch vector, and $\omega_q$ being the Floquet quasi-energy 
divided by $\hbar$. The periodic parts have the same periodicity as 
$\Omega(z,t)$, and we write each one of them as a Fourier series
\begin{subequations} 
\begin{align}
\label{sigma_ab_fourier_z_t_sum_def}
&\nohs_{ab}(z,t)=\sum_{n_1=-\infty}^\infty \sum_{n_2=-\infty}^\infty
\nohs_{ab}^{(n_1,n_2)} e^{in_1 k_0 z} e^{in_2\Delta_\text{d}t}
e^{iqz-i\omega_q t},\\
\label{sigma_ac_fourier_z_t_sum_def}
&\nohs_{ac}(z,t)=\sum_{n_1=-\infty}^\infty \sum_{n_2=-\infty}^\infty
\nohs_{ac}^{(n_1,n_2)} e^{in_1 k_0 z} e^{in_2\Delta_\text{d}t}
e^{iqz-i\omega_q t},\\
\label{E_fourier_z_t_sum_def}
&\begin{aligned}
\nohE(z,t) = \Big(&\nohE_+(0,0)e^{ik_0 z+i\Delta_\text{d} t}\\
&+\nohE_-(0,0)e^{-ik_0 z-i\Delta_\text{d} t}\Big) e^{iqz-i\omega_q t},
\end{aligned}
\end{align}
\label{sigma_ab_ac_and_E_fourier_z_t_sum_def}
\end{subequations}
where we have kept only two terms in the Fourier series for the electric field 
and removed all other terms. The justification for removing the terms with 
$e^{\mp ik_0 z\pm i\Delta_\text{d}t}$ is that we expect them to only add 
separate linear dispersion bands, similar to the linear bands for 
$\mathcal{E}_{\sigma_\pm,\mp}$ for the dual\mbox{-}V scheme. Also, we have not 
included other $n\Delta_\text{d}$ terms except the ones for $n=\pm 1$, since 
the other terms will not contribute to the dynamics for 
$\Delta_\text{d}\gg |\delta_\text{S}|$.

Inserting Eqs.~\eqref{sigma_ab_ac_and_E_fourier_z_t_sum_def} into 
Eqs.~\eqref{eqs:coher}, and collecting terms of equal exponents gives
\begin{subequations} 
\begin{align}
&0
=i(\tilde\Delta-n\Delta_\text{d})\nohs_{ab}^{(n)}
+i\frac{\Omega_{0}}{2}\del{\nohs_{ac}^{(n+1)}+\nohs_{ac}^{(n-1)}}\nn\\
&\quad\qquad+ig\sqrt{2\pi}\del{\nohE_{+}\delta_{n,1}+\nohE_{-}\delta_{n,-1}},\\
&0=i(\delta-n\Delta_\text{d})\nohs_{ac}^{(n)}
+i\frac{\Omega_{0}^*}{2}\del{\nohs_{ab}^{(n+1)}+\nohs_{ab}^{(n-1)}},
\end{align}
\label{eqs:dual_color_coher_fourier_components_ft}
\end{subequations}
where by $\sigma_{ab}^{(n)}$ and $\sigma_{ac}^{(n)}$ we mean 
$\sigma_{ab}^{(n,n)}$ and $\sigma_{ac}^{(n,n)}$ respectively. The absence of 
$\sigma_{ab}^{(n_1,n_2)}$ and $\sigma_{ac}^{(n_1,n_2)}$ for $n_1\neq n_2$ in 
this system of equations is a consequence of the classical drive only coupling 
the Fourier terms to the ones with both an increased (decreased) wave vector 
and increased (decreased) detuning, combined with only considering the lowest 
order quantum field components in Eq.~\eqref{E_fourier_z_t_sum_def}. We have 
absorbed $\omega_\text{q}$ into the detunings by defining 
$\tilde{\Delta}=\tilde{\Delta}_0+\omega_\text{q}$ and 
$\delta=\delta_0+\omega_\text{q}$. The only but important difference from 
Eqs.~\eqref{eqs:ss_coher_fourier_components} is that the frequencies of the 
different Fourier components are shifted by $n\Delta_\text{d}$ in 
Eqs.~\eqref{eqs:dual_color_coher_fourier_components_ft}. The result of this 
difference is that the higher order Fourier components of the atomic coherence 
contribute little for $\Delta_\text{d}\gg |\delta_\text{S}|$ and therefore can 
be neglected, thus giving the same effect as in the secular approximation. 
Hence, the dispersion relation will be the same as the quadratic dispersion 
relation of the dual\mbox{-}V scheme. We verify numerically (see 
Fig.~\ref{fig:continuum_dispersion_relation_truncations} and 
App.~\ref{App:Continuum_numerical}) that this is the case for 
$\Delta_\text{d}/\Gamma=1$ and $|\delta_\text{S}|/\Gamma\approx 10^{-2}$.

The summary of the discussion in Sec.~\ref{Sec:lambda_continuum_model} is that 
the reason for the difference in the dispersion relation between dual\mbox{-}V 
and the cold $\Lambda$\mbox{-}type schemes is that the cold $\Lambda$\mbox{-}type 
scheme allows excitations and de-excitations by the classical fields from 
different directions, whereas the dual\mbox{-}V does not due to separation of 
the different directions into different polarization modes. For the dual-color 
scheme, such mismatched excitations and de-excitations are suppressed by the 
frequency difference between the right-moving and left-moving fields.

\subsection{Dispersion relations for hot $\Lambda$\mbox{-}type atoms}
\label{Sec:hot_Lambda_continuum_model}
Stationary light was first considered for hot $\Lambda$\mbox{-}type atoms, 
where a quadratic dispersion relation was 
predicted~\cite{andre_prl02a,bajcsy_nature03a}. For completeness, we will 
briefly discuss how this result arises from the results of the dual-color 
scheme~\cite{peters_pra12}. The main difference between the cold atoms and the 
hot atoms is that the latter ones move and hence have an associated Doppler 
shift in the transition frequency. In the one-dimensional approximation this 
amounts to having the right propagating fields being shifted by 
$\omega_\text{D}$, and the left propagating fields being shifted by 
$-\omega_\text{D}$, where $\omega_\text{D}$ is the Doppler shift that is 
determined by the velocity of the atoms. For each individual velocity class 
with the same $\omega_\text{D}$, the dynamics will be completely equivalent to 
the dual-color setup, where instead of $\Delta_\text{d}$ we now have 
$\omega_\text{D}$. That is, the system is described by Eqs.~\eqref{eqn:hE_pm} 
and Eqs.~\eqref{eqs:coher} with
$\Omega(z,t)=\Omega_{0}\cos(\omega_\text{D}t+k_\text{c} z)$. Hence, for
$\omega_\text{D}\gg |\delta_\text{S}|$, the quadratic dispersion relation is 
valid. If the width of the distribution of $\omega_\text{D}$ is much bigger 
than $|\delta_\text{S}|$, then the contribution of the velocity classes, where 
$\omega_\text{D}\gg |\delta_\text{S}|$ is not fulfilled, is small, and the 
quadratic dispersion relation~\eqref{dual_v_dispersion_relation_delta_approx} 
should be true for the ensemble as a whole.

In the original derivations of stationary 
light~\cite{andre_prl02a,bajcsy_nature03a} the dispersion relation was 
obtained by arguing that the thermal motion of the atoms washes out any 
spatial coherences with Fourier components $|n|\geq 2$. This argument is 
essentially equivalent to the Doppler shift argument above, except that it is 
formulated in time rather than frequency. As originally noted in 
Ref.~\cite{zimmer_pra08} the level structure of the dual\mbox{-}V scheme does 
not allow these higher order Fourier components. Hence, the dispersion 
relation~\eqref{dual_v_dispersion_relation_delta_approx} originally derived 
for hot $\Lambda$\mbox{-}type atoms also applies for the dual\mbox{-}V system 
regardless of the temperature.

\section{Discrete model}\label{Sec:discrete_model}
\subsection{Transfer matrix formalism}\label{Sec:discrete_model_transfer_matrix}
To support some of the conclusions reached above and to provide additional 
possibilities for how the dispersion relation can be controlled, we will now 
consider a model where we account for the individual atoms instead of using the 
continuum model. To do this we will use the transfer matrix formalism (see 
App.~\ref{App:Multi_mode_transfer_matrices} for details) to describe 
stationary light. In the transfer matrix formalism, the electric field at the 
position $z$ is represented by the vector
\begin{gather}\label{multimode_E_field_vector}
\mathbf{E}(z)=\begin{pmatrix}
\mathbf{E}_+(z)\\
\mathbf{E}_-(z)
\end{pmatrix}.
\end{gather}
The two parts $\mathbf{E}_\pm(z)$ (right-moving and left-moving fields) are, 
in general, vectors with $\Nmodes$ elements---one for each of $\Nmodes$ 
different modes of the electric field. For the $\Lambda$\mbox{-}type scheme 
(see Fig.~\ref{fig:level_diagrams}(b)), only a single polarization mode of the 
field is necessary, so that we have $\Nmodes=1$, and $E_\pm(z)$ are scalars 
(omitting the bold script). In terms of the definitions of the fields for the 
continuum model we have
\begin{gather}
\label{Lambda_E_vector_definition}
E_\pm(z)=\mathcal{E}_\pm(z) e^{\pm ik_0 z},
\end{gather}
i.e. contrary to $\mathcal{E}_\pm$, these fields are not slowly-varying in 
space. For the dual\mbox{-}V scheme (see Fig.~\ref{fig:level_diagrams}(a)), we 
have $\Nmodes=2$ (for the $\sigma_+$ and $\sigma_-$ polarization modes), and 
the vectors are similarly related to the continuum model definitions by
\begin{gather}
\label{dualV_E_vector_definition}
\mathbf{E}_\pm(z)=\begin{pmatrix}
\mathcal{E}_{\sigma_+,\pm}(z)\\
\mathcal{E}_{\sigma_-,\pm}(z)
\end{pmatrix}e^{\pm ik_0z}.
\end{gather}

The atoms are assumed to be linear scatterers, hence both the atoms and the 
space between atoms are represented by matrices that relate the vector of 
electric fields at one position to the vector at another position. The 
transfer matrix $T_{\text{a},j}$ of atom $j$ at position $z_j$ is such that it 
fulfills the relation
$\mathbf{E}(z_j^+)=T_{\text{a},j}\mathbf{E}(z_j^-)$, where $z_j^+=z_j+\epsilon$ 
and $z_j^-=z_j-\epsilon$ in the limit $\epsilon\rightarrow 0$. This limit 
expresses the fact that the atoms are assumed to be point scatterers with no 
spatial extent. The transfer matrix $T_{\text{f},j}$ of the free propagation 
between the atoms $j$ and $j+1$ at the positions $z_j$ and $z_{j+1}$ 
respectively is such that it fulfills the relation 
$\mathbf{E}(z_{j+1}^-)=T_{\text{f},j}\mathbf{E}(z_j^+)$. For the last transfer 
matrix $T_{\text{f},N}$, we define $\mathbf{E}(z_{N+1}^-)=\mathbf{E}(L)$, 
where $N$ is the total number of atoms, and $L$ is the total length of the 
ensemble. The transfer matrix of the whole ensemble is the product of the 
transfer matrices of each atom in the ensemble and the free propagation 
between them.

We will consider two types of placement of the atoms: periodic with respect 
to the classical drives and completely random. The former will allow us to 
tailor the properties of the stationary light, and the latter is used to 
reproduce the results of the continuum model investigated above.
If the arrangement of the atoms is periodic, then studying the repeated unit cell is 
sufficient to obtain full information about the system. If the 
arrangement of the atoms is random, then we need to do statistical averaging 
over placement of the atoms inside a single period of the classical drives. 

For the random placement of the atoms, the starting point is the observation 
(shown in App.~\ref{App:scattering_matrix_Lambda} and 
App.~\ref{App:scattering_matrix_dualV}) that the scattering matrix for both 
the $\Lambda$\mbox{-}type and dual\mbox{-}V atoms with applied 
counter-propagating classical drives is invariant under shift of the atomic 
position by $\pi/k_0$ (assuming $k_\text{c}\approx k_0$) and not $2\pi/k_0$, 
which is the periodicity of each of the classical drives. For the 
$\Lambda$\mbox{-}type atoms, this is due to the fact that the two classical 
drives form a standing wave, which has half the period of the individual 
running waves. For the dual\mbox{-}V atoms, it is also true, even though there 
is no obvious standing wave pattern due to the two classical drives.

Having identified $\pi/k_0$ as the period of the effective potential due to 
the two classical drives, we can now explain the statistical averaging 
procedure. The basic idea is to take an integer number of periods as 
the length $L_\text{u}$ of the unit cell and randomly place $N_\text{u}$ atoms
within this unit cell with a uniform distribution. Then this unit cell is used 
to find the dispersion relation in the same way as the unit cells for the periodic 
placement of the atoms (with one technical difference as discussed below). To 
obtain a better statistical averaging, we increase the number of periods in 
$L_\text{u}$, while simultaneously increasing the number of atoms $N_\text{u}$, 
such that the density $n_0=N_\text{u}/L_\text{u}$ is held fixed.

In the transfer matrix theory, Bloch's theorem is a statement about the 
eigenvalues and eigenvectors of the transfer matrix for the unit cell 
$T_\text{cell}$. If $\mathbf{E}_\lambda$ is an eigenvector of $T_\text{cell}$ 
with the eigenvalue $\lambda$, then $\mathbf{E}_\lambda$ is the periodic part 
of the Bloch wave (that spatially varies in discrete steps by successively 
applying transfer matrices whose product is equal to $T_\text{cell}$), and
the eigenvalue $\lambda$ is related to the Bloch vector. One natural relation is 
\begin{gather}\label{transfer_matrix_eigenvalue_in_terms_of_bloch_fast}
\lambda=\exp\delnospace{i\tilde{q}L_\text{u}},
\end{gather}
where we denote the Bloch vector with $\tilde{q}$ to make it distinct from the 
Bloch vector $q$ that we used in the continuum model. The difference is 
entirely due to defining the electric fields either slowly varying in space 
(continuum model) or not (discrete model).

For consistency with the continuum model, we will also use a slightly modified 
relation. Since the elements of the electric field vectors (Eqs. 
\eqref{Lambda_E_vector_definition} and \eqref{dualV_E_vector_definition}) are 
defined not to be slowly varying in space, the length of the unit cell 
$L_\text{u}=n_\text{u}\pi/k_0$ with integer $n_\text{u}$ results in free 
propagation factors $e^{\pm ik_0L_\text{u}}=(-1)^{n_\text{u}}$ being 
multiplied onto the vectors. Therefore, we take the relation between the 
eigenvalue and the Bloch vector to be
\begin{gather}\label{transfer_matrix_eigenvalue_in_terms_of_bloch}
\lambda=(-1)^{n_\text{u}}\exp\delnospace{iqL_\text{u}}.
\end{gather}
which is equivalent to a constant shift of $q$ compared to $\tilde{q}$.

For the $\Lambda$\mbox{-}type atoms and dual\mbox{-}V 
atoms with $\Omega_+=\Omega_-$ and $\omega_{b_+c}=\omega_{b_-c}$, the four 
blocks of the scattering matrix that represent reflection and transmission are 
either scalars or symmetric matrices (see 
App.~\ref{App:scattering_matrix_Lambda} and 
App.~\ref{App:scattering_matrix_dualV}). Using this property, one can show 
that the transfer matrix $T_\text{cell}$ is symplectic, which implies that its 
eigenvalues come in reciprocal pairs (see 
App.~\ref{App:Multi_mode_transfer_matrices}), i.e. if $\lambda$ is an 
eigenvalue, then $1/\lambda$ is also an eigenvalue. Hence, if $q$ is a Bloch 
vector, then $-q$ is also a Bloch vector. Inverting 
Eq.~\eqref{transfer_matrix_eigenvalue_in_terms_of_bloch}, we can find the 
Bloch vector from the eigenvalue through
\begin{gather}\label{bloch_eigenvalue_equation}
\frac{q}{n_0}=-\frac{i}{N_\text{u}}\Log\delnospace{(-1)^{n_\text{u}}\lambda},
\end{gather}
where $\Log$ is the complex logarithm.

When using Eq.~\eqref{bloch_eigenvalue_equation} to determine the Bloch 
vector, care is required in selecting the right 
branch of the complex logarithm, when $q$ is calculated as a function of 
$\delta$. If the principal branch of the complex logarithm is always used, then 
$\text{Im}[\Log(\lambda)]$ is constrained to the interval $(-\pi,\pi]$, so that 
Eq.~\eqref{bloch_eigenvalue_equation} will result in $\text{Re}[q]/n_0$ being 
constrained to the interval $(-\pi/N_\text{u},\pi/N_\text{u}]$. As we let 
$N_\text{u}$ go to infinity to obtain good statistical averaging, this 
interval becomes arbitrarily small. In practice, this means that as $\delta$ 
is increased, and if $\text{Re}[q]/n_0$ increases and reaches $\pi/N_\text{u}$, 
all the subsequent values of $\text{Re}[q]/n_0$ will be shifted by 
$-2\pi/N_\text{u}$. In the numerical evaluation of the dispersion relations 
with statistical averaging we thus need to undo these shifts, which is 
equivalent to selecting different branches of the complex logarithm.

\subsection{Dispersion relations for cold dual\mbox{-}V atoms}
\label{Sec:discrete_model_dual_V}
\begin{figure}[tbp]
\centering
\includegraphics{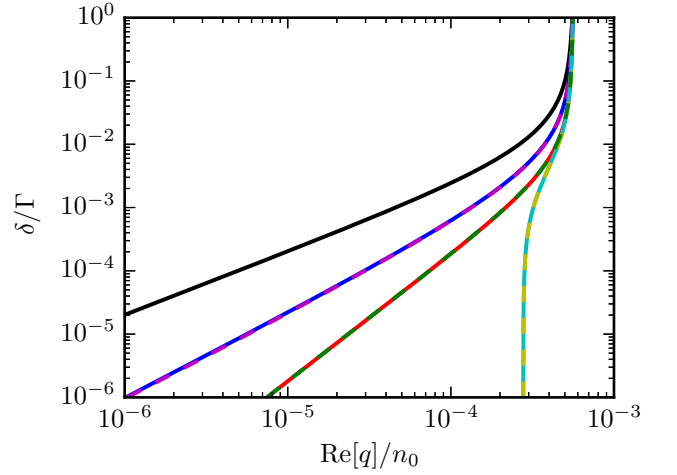}
\caption{(Color online) Log-log plot of the dispersion relations calculated 
analytically with the continuum model and numerically with the discrete model 
for randomly placed atoms. The solid black (upper), red (lower) and blue (in 
between) curves are as in 
Fig.~\ref{fig:continuum_dispersion_relation_truncations} and are shown for 
reference. The middle dashed magenta curve is for $\Lambda$\mbox{-}type scheme 
computed numerically with the discrete model. It overlaps with the solid blue 
curve (the same dispersion relation computed analytically), so that the 
difference is not visible. The lower dashed green curve is the quadratic 
dispersion relation for the dual\mbox{-}V scheme found numerically with the 
discrete model. The solid cyan and dashed yellow curves that are almost 
vertical for small $\delta/\Gamma$ show the linear dispersion relation for the 
dual-V scheme. The solid cyan curve is the analytical result given by 
Eq.~\eqref{dual_v_dispersion_relation_linear_q}, while the dashed yellow curve 
is computed numerically with the discrete model. Both the numerical curves for 
the two dispersion relations for the dual\mbox{-}V scheme (linear and 
quadratic) overlap with the respective analytical solutions, so that the 
difference in not visible. The common parameters are: 
$\Gamma_\text{1D}/\Gamma=0.1$, $\Delta_\text{c}/\Gamma=-90$, 
$\Omega_0/\Gamma=1$, $k_0/n_0=\pi/2$ and $N_\text{u}=10^4$ (i.e. 
$L_\text{u}=(10^4/2)\pi/k_0$).
\label{fig:dispersion_relation_plot_dual_v}}
\end{figure}

We first use the transfer matrix formalism to find the dispersion 
relations for ensembles of randomly and regularly placed dual\mbox{-}V atoms. 
In Fig.~\ref{fig:dispersion_relation_plot_dual_v} we plot the dispersion 
relations for the randomly placed atoms. The dashed yellow curve is the linear 
dispersion relation and the dashed green curve is the quadratic dispersion 
relation. They have an excellent agreement with the analytical solutions given 
by Eqs.~\eqref{dual_v_dispersion_relation_linear_q} and 
\eqref{dual_v_dispersion_relation_q}, which are shown by the solid cyan and 
red curves respectively. The curves showing the linear dispersion relation for 
the dual\mbox{-}V scheme have a non-zero $\text{Re}[q]$ for $\delta=0$ 
(see Eq.~\eqref{dual_v_dispersion_relation_linear_q}), and hence look vertical 
for small $\delta/\Gamma$ on the log-log plot.

If the dual\mbox{-}V atoms are placed regularly, the only noticeable 
difference we have found between the continuum and discrete theory is when the 
atoms in the discrete model are spaced with either $d=\pi/k_0$ or 
$d=\pi/(2k_0)$. The former is equivalent to the atomic 
mirror~\cite{chang_njp12}, and since we neglect the vacuum dispersion 
relation, for $d=\pi/k_0$ we find the constant Bloch vector $q$ independent of 
$\delta$. The latter, $d=\pi/(2k_0)$, changes the linear dispersion 
relation~\eqref{dual_v_dispersion_relation_linear_q}. The reason for this is 
that in the derivation of Eq.~\eqref{dual_v_dispersion_relation_linear_q}, we have 
neglected the terms with $e^{\pm 2ik_0z}$ and $e^{\pm 4ik_0z}$. For discrete 
positions $z=jd=j\pi/(2k_0)$ ($j$ is an integer), these factors are 
$e^{\pm 2ik_0z}=e^{\pm i\pi j}$ and $e^{\pm 4ik_0z}=1$. We see that for 
discrete atoms with spacing $d=\pi/(2k_0)$, the factors $e^{\pm 4ik_0z}=1$ 
should not be neglected, since they are constant and not rapidly varying. With this correction, 
Eqs.~\eqref{eqn:hE_sigma_pm_mp_ft_ab_inserted_bloch} become
\begin{align}
\mp\frac{q}{n_0}\nohE_{\sigma_\pm,\mp}
=-\frac{\Gamma_\text{1D}}{2\tilde\Delta}\left[
\frac{\delta-\delta_\text{S}/2}{\delta-\delta_\text{S}}\nohE_{\sigma_\pm,\mp}
+\frac{\delta_\text{S}/2}{\delta-\delta_\text{S}}\nohE_{\sigma_\mp,\pm}
\right],
\end{align}
which makes them of exactly the same coupled form as 
Eqs.~\eqref{eqn:hE_sigma_pm_pm_ft_ab_inserted_bloch} and therefore results in 
the same quadratic dispersion 
relation~\eqref{dual_v_dispersion_relation_q} instead of a linear one. 
This behavior is reproduced by the numerical calculations with the discrete 
model.

\subsection{Dispersion relations for cold $\Lambda$\mbox{-}type atoms}
\label{Sec:discrete_model_Lambda}
As for the dual\mbox{-}V atoms above, we can calculate the dispersion 
relation of an ensemble with randomly placed $\Lambda$\mbox{-}type atoms. As 
shown in Fig.~\ref{fig:dispersion_relation_plot_dual_v}, the dispersion 
relation obtained in this way (dashed magenta) matches the one that was found 
analytically for the continuum model (solid blue).

\begin{figure}[tbp]
\centering
\includegraphics{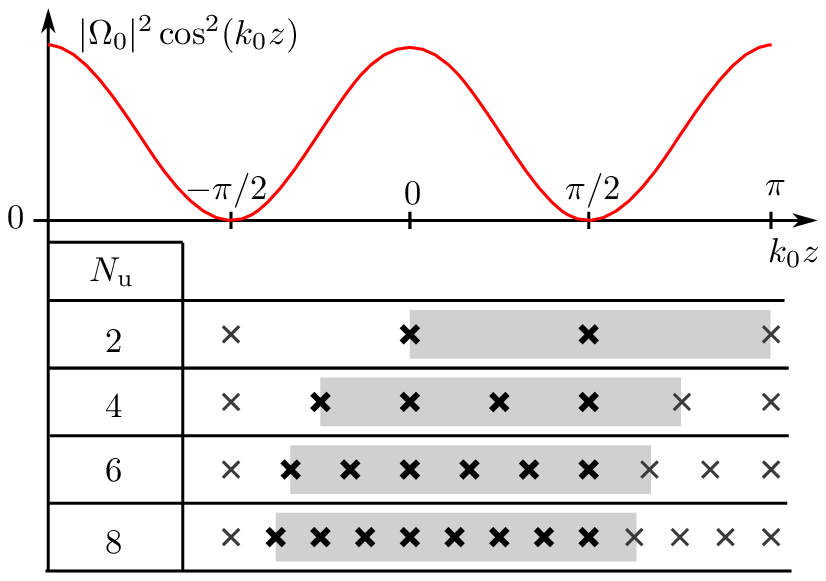}
\caption{(Color online) Placement of atoms for periodic ensembles. At 
the top, the standing wave of the classical drive is plotted. In the table 
below, the crosses indicate the positions of the atoms in the standing wave of 
the classical drive for different values of the number of atoms per unit 
cell $N_\text{u}$. The thick crosses are the atoms in the chosen unit cell, 
and the thin crosses are the other atoms in the ensemble. The particular 
choice of the unit cell (gray) is such that the atoms with the non-zero 
classical drive are taken first (when propagating from the left), and the last 
atom is placed on the node of the classical drive (at $k_0 z=\pi/2$) which 
effectively makes it a two-level atom.
\label{fig:variable_N_u}}
\end{figure}

For the regularly placed $\Lambda$\mbox{-}type atoms, we can also obtain 
dispersion relations, which are different from the predictions of the 
continuum model. To this end we consider the ensembles shown in 
Fig.~\ref{fig:variable_N_u}. The atoms are spaced with a distance 
$d=\pi/(N_\text{u}k_0)$, where we only take even $N_\text{u}$ for simplicity. 
(As explained above, adding integer multiples of $\pi/k_0$ to $d$ does not 
change the results.) A unit cell consists of $N_\text{u}-1$ atoms, which 
experience a non-zero classical drive, and one atom, which is placed such that 
the classical drive is zero, i.e. on the node of the standing wave of the 
classical drive. For such a setup, we show in 
App.~\ref{App:Regular_Spaced_Mass} that the dispersion relation for two-photon 
detunings fulfilling
\begin{gather}\label{regular_spacing_dispersion_relation_condition}
\delta\ll 2|\delta_\text{S}|\cos^2\del{\frac{\pi}{2}-\frac{\pi}{N_\text{u}}}
\approx 2|\delta_\text{S}|\del{\frac{\pi}{N_\text{u}}}^2
\end{gather}
(i.e. if frequency is within the smallest EIT window of the atoms that are not 
placed on the node) is given by
\begin{gather}\label{regular_spacing_dispersion_relation}
\delta\approx\frac{1}{2m}\del{\frac{q}{n_0}}^2,
\end{gather}
where $n_0=1/d$, and
\begin{gather}\label{regularly_spaced_mass}
m=-\frac{(N_\text{u}-1)\Gamma_\text{1D}^2}
{2N_\text{u}^2(\Delta_\text{c}+i\Gamma'/2)|\Omega_0|^2}
\end{gather}
is the effective mass. Note that the quadratic dispersion relation in 
Eq.~\eqref{regular_spacing_dispersion_relation} is of the same form as 
Eq.~\eqref{dual_v_dispersion_relation_delta_approx}, but with the effective 
mass in Eq.~\eqref{regularly_spaced_mass} differing by a factor 
$2(N_\text{u}-1)/N_\text{u}^2$ from the one in 
Eq.~\eqref{dual_v_dispersion_relation_mass}.

\begin{figure}[tbp]
\centering
\includegraphics{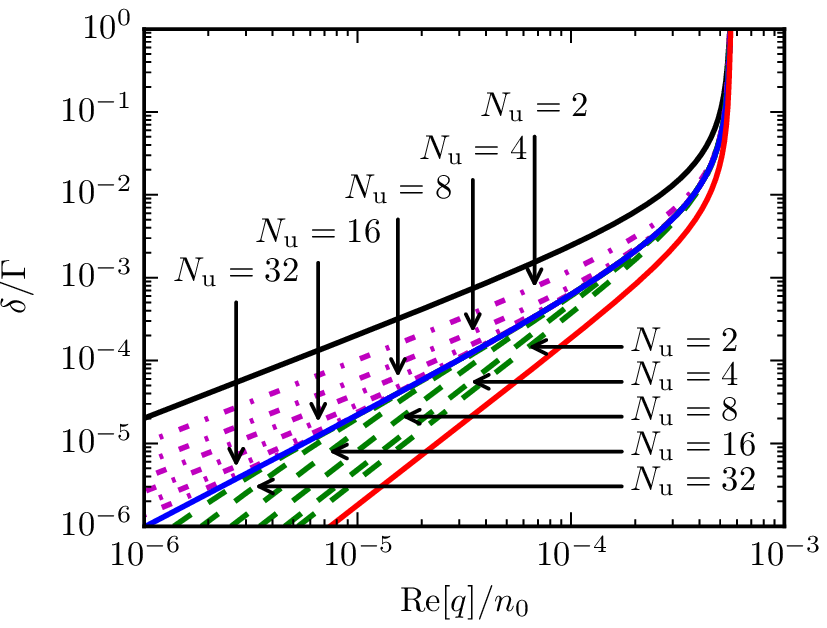}
\caption{(Color online) Log-log plot of the dispersion relations for 
$\Lambda$\mbox{-}type atoms with different placement of the atoms within the 
unit cell. The solid black (upper), red (lower) and blue (in between) curves 
are as in Fig.~\ref{fig:continuum_dispersion_relation_truncations} and are 
shown for reference. The dashed green curves are for ensembles with regularly 
placed atoms (see Fig.~\ref{fig:variable_N_u}) for the period lengths 
$N_\text{u}=2,4,8,16,32$. The dash-dotted magenta curves are for the same 
setups, but with a shifted standing wave of the classical drive: 
$\Omega(z)=\Omega_0\cos(k_0 z+\varphi)$ with $\varphi=\pi/(2N_\text{u})$. The 
common parameters are: $\Gamma_\text{1D}/\Gamma=0.1$, 
$\Delta_\text{c}/\Gamma=-90$, $\Omega_0/\Gamma=1$. The density of the atoms 
$n_0$ is related to the spacing between the atoms $d$ by $n_0=1/d$. The 
distance $d$ depends on the desired period length and is given by 
$d=\pi/(N_\text{u}k_0)$ (plus any integer multiple of $\pi/k_0$).
\label{fig:regularly_spaced_dispersion_relation_plot}}
\end{figure}

The above quadratic dispersion relation is obtained by placing the atoms such 
that one of them coincides exactly with the node of the standing wave of the 
classical drive. The dispersion relation can be completely changed, however, 
by shifting the position of the atoms relative to the drive. This can be 
achieved if the classical drive is given by 
$\Omega(z)=\Omega_0\cos(k_0z +\varphi)$ (with the situation above 
corresponding to $\varphi=0$). By choosing $\varphi=k_0 d/2=\pi/(2N_\text{u})$, 
the node of the standing wave is placed exactly in the middle of the 
free-space separation between two atoms.

We show the numerically calculated dispersion relation for $\varphi=0$ and 
$\varphi=k_0 d/2$ in Fig.~\ref{fig:regularly_spaced_dispersion_relation_plot}. 
For $\varphi=0$ (dashed green curves), the dispersion relation becomes 
quadratic for small $\text{Re}[q]/n_0$ as given by 
Eq.~\eqref{regular_spacing_dispersion_relation}. The range of validity of the 
quadratic approximation becomes smaller for increasing $N_\text{u}$, as 
predicted by the condition in 
Eq.~\eqref{regular_spacing_dispersion_relation_condition}. For 
$\varphi=k_0 d/2$ (dash-dotted magenta curves), the dispersion relation 
becomes linear (parallel to the EIT dispersion relation) instead of quadratic 
for small $\text{Re}[q]/n_0$. As $N_\text{u}$ increases, both for $\varphi=0$ 
and $\varphi=k_0 d/2$, the dispersion relation approaches the one for an
ensemble of cold randomly placed $\Lambda$\mbox{-}type atoms (solid blue). The 
two choices of the phase, $\varphi=0$ and $\varphi=k_0 d/2$, are thus similar 
to respectively the odd and even $n$ truncations in 
Fig.~\ref{fig:continuum_dispersion_relation_truncations}. In essence, having a 
finite number of atoms per unit cell gives a truncation because a finite 
number of atoms can only support a finite number of Fourier components of 
$\nohs_{ab}$ and $\nohs_{ac}$.

The two situations, $\varphi=0$ and $\varphi=k_0 d/2$, considered in 
Fig.~\ref{fig:regularly_spaced_dispersion_relation_plot}, represent the two 
extreme cases with the node of the classical drive either coinciding with an 
atom or being placed as far away from the atoms as possible. In between these 
extremes there is a whole continuum of possibilities. In general, if no atoms 
are placed at the nodes, all atoms will have a finite EIT window and hence the 
dispersion relation will be linear for sufficiently small~$\delta$. This also 
implies that with a finite number of randomly placed stationary $\Lambda$-type 
atoms, it is impossible to realize a $\delta\propto |q|^{4/3}$ dispersion in 
the limit $\delta\rightarrow 0$, as there is formally zero probability for the 
point-like atoms to sit exactly at the nodes, and hence the dispersion 
relation will eventually cross over to the linear one.

\section{Scattering properties}
\label{sec_scattering_properties}
A different way to compare the ensembles with regularly and randomly placed 
$\Lambda$\mbox{-}type atoms is to look at the scattering properties 
(transmission and reflection coefficients) of the whole ensemble. Contrary to 
the dispersion relation, which, in principle, is only valid for an infinite 
ensemble, the total number of atoms does matter for the scattering properties. 
If the number of the atoms is sufficiently large, the dispersion relation is 
still reflected in the behavior of the transmission and reflection 
coefficients. Hence, the scattering properties can also be used to characterize 
the dispersion relation. Below, the transmission coefficients~$t$ and 
reflection coefficients~$r$ will be obtained numerically by multiplying the 
transfer matrices for the atoms and free propagation to obtain the transfer 
matrix for the whole ensemble and afterwards extracting the scattering 
coefficients from this transfer matrix, as described in 
App.~\ref{App:Multi_mode_transfer_matrices}. For regularly placed discrete 
atoms and the continuum model, one can derive closed-form expressions for the 
transfer matrix for the whole ensemble (see 
App.~\ref{App:transfer_matrix_for_uniform_ensemble}).

\begin{figure}[tbp]
\centering
\includegraphics{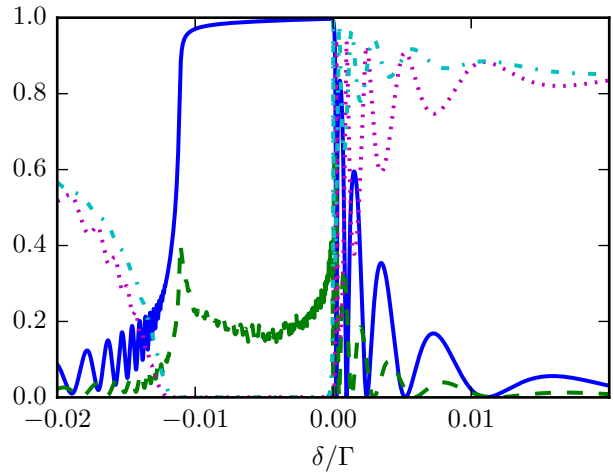}
\caption{(Color online) Plot of transmittance $|t|^2$ and reflectance $|r|^2$ 
of ensembles with $N=4\cdot10^4$ atoms. The dotted magenta and solid blue 
curves are respectively the transmittance and reflectance of an ensemble with 
regularly placed $\Lambda$\mbox{-}type atoms and $N_\text{u}=2$ (with the 
placement shown in Fig.~\ref{fig:variable_N_u}). The dash-dotted cyan and 
dashed green curves are respectively the transmittance and reflectance of an 
ensemble with randomly placed $\Lambda$\mbox{-}type atoms and is averaged over 
100 ensemble realizations. The other parameters are: 
$\Gamma_\text{1D}/\Gamma=0.1$, ${\Delta_\text{c}/\Gamma=-90}$, 
$\Omega_0/\Gamma=1$, $k_0/n_0=\pi/2$. The interval around $\delta=0$ is shown 
in more detail in Fig.~\ref{fig:discrete_model_t_r_2}(a).
\label{fig:discrete_model_t_r_1}}
\end{figure}

\begin{figure}[tbp]
\centering
\includegraphics{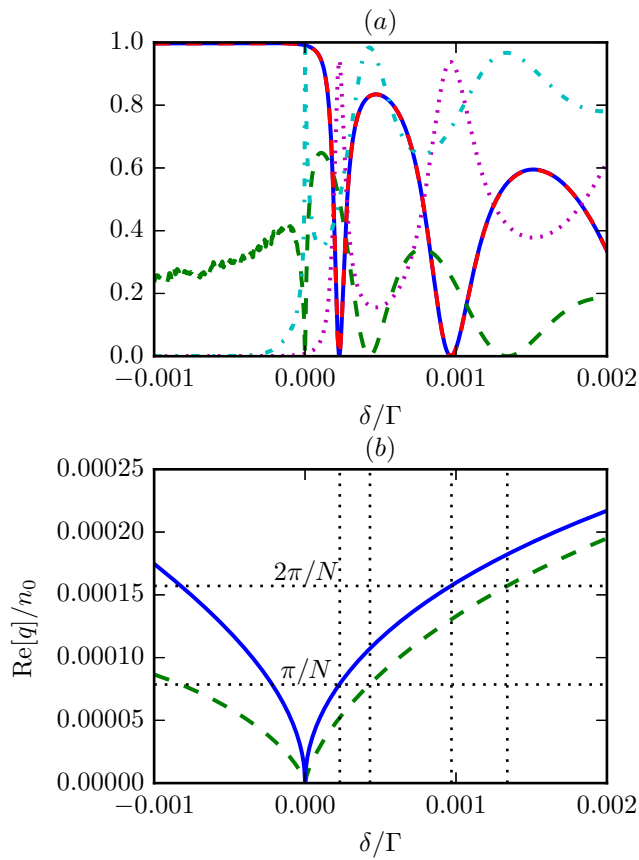}
\caption{(Color online) (a) Same as Fig.~\ref{fig:discrete_model_t_r_1}, but 
zoomed in around $\delta=0$. Additionally, the reflectance for an ensemble 
with randomly placed dual\mbox{-}V atoms is plotted (dashed red), and it 
completely overlaps the reflectance for the regularly placed $\Lambda$\mbox{-}type 
scheme. The dual\mbox{-}V scheme has the same parameters except that $\Omega_0$ 
is multiplied by $\sqrt{2}$ to make the dispersion relation equal to the one 
of the regularly placed $\Lambda$\mbox{-}type scheme. (b)~Dispersion relations 
for $\Lambda$\mbox{-}type scheme: regularly placed (solid blue) and randomly 
placed (dashed green). The dispersion relation was calculated numerically with 
the transfer matrix formalism for the regularly placed ensemble, and using 
Eq.~\eqref{Lambda_cold_dispersion_relation} for the randomly placed ensemble. 
The two horizontal dotted lines at ${\rm Re}[q]/n_0=\pi/N$ and 
${\rm Re}[q]/n_0=2\pi/N$, give the condition for the first and the second high 
transmission resonance. At each intersection (for $\delta>0$) of these 
horizontal lines with the dispersion relation curves, vertical dotted lines 
are drawn, which can be seen to coincide with the high transmission resonances 
in (a).
\label{fig:discrete_model_t_r_2}}
\end{figure}

In Figs. \ref{fig:discrete_model_t_r_1} and \ref{fig:discrete_model_t_r_2}(a), 
we plot transmittance $|t|^2$ and reflectance $|r|^2$ for ensembles with 
regular ($N_\text{u}=2$) and random (average of 100 ensemble realizations) 
placement of $\Lambda$\mbox{-}type atoms. The latter case can also be calculated 
using the continuum model together with 
App.~\ref{App:transfer_matrix_for_uniform_ensemble}. The main visible 
difference between the discrete model with random placement and the continuum 
model is that the former has noise in the region 
$-0.01\lesssim\delta/\Gamma\leq 0$ due to finite number of ensemble realizations.
In Fig.~\ref{fig:discrete_model_t_r_2}(a) we additionally show the reflection 
coefficient for randomly placed dual\mbox{-}V atoms (single ensemble 
realization) with $\sigma_+$ input incident from the left and finding the left-moving $\sigma_-$ field to the left of the ensemble (such that the quadratic 
dispersion relation is valid). The reflection coefficient of the dual\mbox{-}V 
scheme overlaps completely with the reflection coefficient of the regularly 
placed $\Lambda$\mbox{-}type scheme, because we have increased the classical 
drive strength $\Omega_0$ of the dual\mbox{-}V scheme by a factor of $\sqrt{2}$ 
to make the masses in Eq.~\eqref{dual_v_dispersion_relation_mass} and 
Eq.~\eqref{regularly_spaced_mass} equal.

The plots and the chosen parameters are similar to the ones in 
Ref.~\cite{hafezi_pra12a}. As opposed to Ref.~\cite{hafezi_pra12a}, however, 
we do not make the secular approximation for the $\Lambda$\mbox{-}type scheme, 
and this leads to very different results, which depend on how the atoms are 
placed (and whether we use the dual\mbox{-}V scheme instead). For the 
regularly placed $\Lambda$\mbox{-}type atoms, we see a clear signature of a 
photonic band gap in the region $-0.01\lesssim\delta/\Gamma\leq 0$, where there 
is a near unit reflectance and negligible transmittance. For the randomly 
placed $\Lambda$\mbox{-}type atoms, the situation is more complex with a 
similar negligible transmittance but a rather limited reflectance. For 
$\delta>0$, the position of the resonances with low reflection and high 
transmission corresponds to the condition $\sin\delnospace{{\rm Re}[q]L}=0$, 
i.e. there is a standing wave of the Bloch vectors inside the ensemble. 
Specifically, the high transmission resonance occurs each time 
${\rm Re}[q]/n_0$ crosses a multiple of $\pi/N$, as can be seen in 
Fig.~\ref{fig:discrete_model_t_r_2}(b). This behavior can also be seen from 
the analytical results derived in 
App.~\ref{App:transfer_matrix_for_uniform_ensemble}. Due to non-zero 
incoherent decay rate $\Gamma'$, the sum $|t|^2+|r|^2$ is in general not equal 
to unity.

As we have shown above, the regularly and randomly placed $\Lambda$\mbox{-}type 
atoms have very different dispersion relations, and this translates into very 
different positions of the high transmission resonances in Figs. 
\ref{fig:discrete_model_t_r_1} and \ref{fig:discrete_model_t_r_2}(a). For the 
randomly placed $\Lambda$\mbox{-}type scheme, there additionally occurs a high 
transmission resonance at $\delta=0$, since no atoms are placed exactly at the 
node of the standing wave of the classical drive, and hence all the atoms are 
transmissive due to EIT. For the regularly placed setup with $N_\text{u}=2$, 
half of the atoms are placed on the nodes and therefore behave as effective 
two-level atoms. For $\delta=0$, the other half of the atoms becomes 
transparent, and the whole ensemble is exactly equivalent to the atomic mirror 
\cite{chang_njp12}. For the dual\mbox{-}V atoms, as shown in 
App.~\ref{App:scattering_matrix_dualV} the reflection coefficients of a single 
atom do not become zero for $\delta=0$, regardless of how the individual 
atoms are placed.

\section{Conclusion}

We have analyzed a number of different setups for stationary light. These 
setups lead have different behaviors and dispersion relations depending on the 
exact details. For small Bloch vectors $q$, the dispersion relations are 
either linear, quadratic, or in between with $\delta \propto |q|^{4/3}$. For 
higher values of $q$, the dispersion relations either continue to have the 
same behavior, or the linear and quadratic dispersion relations may cross over 
into the $\delta\propto |q|^{4/3}$ results.

Overall, these results demonstrate the rich physics of stationary light. This 
opens the possibility of tailoring the light propagation to meet specific 
desired functionalities. In addition to the strong interest in understanding 
and controlling the propagation of light, another interesting possibility of 
stationary light is to use for non-linear optics. The ability to achieve a 
vanishing group velocity, and the corresponding increase in interaction time 
between photons in an optical pulse, may in principle lead to strong optical 
non-linearities down to the single photon 
level~\cite{andre_prl02a,chang_naturephys08a,hafezi_pra12a}. In order to fully 
assess such possibilities, it is essential to first have a thorough 
understanding of the linear properties of the system as determined in this 
work.

\begin{acknowledgments}
The research leading to these results was funded by the European Union Seventh 
Framework Programme through SIQS (Grant No. 600645), ERC Grant QIOS (Grant 
No. 306576), ERC Starting Grant FOQAL (Grant No. 639643), the MINECO 
Plan Nacional Grant CANS (Grant No. FIS2014-58419-P), the MINECO Severo Ochoa Grant SEV-2015-0522, and 
Fundacio Privada Cellex. II, JO and AS want to thank Ben Buchler, Ping Koy Lam 
and their team for helpful discussions.
\end{acknowledgments}

\appendix

\section{Numerical methods for the continuum model}\label{App:Continuum_numerical}
The truncations of both Eqs.~\eqref{eqs:ss_coher_fourier_components} for the $\Lambda$\mbox{-}type 
scheme and Eqs.~\eqref{eqs:dual_color_coher_fourier_components_ft} for the 
dual-color scheme can be written
\begin{gather}\label{eqs:matrix_form_fourier_components}
0=M\boldsymbol{\sigma}+g\sqrt{2\pi}V\boldsymbol{\mathcal{E}},
\end{gather}
where 
\begin{gather}
\boldsymbol{\sigma}
=
\begin{pmatrix}
\vdots\\
\sigma_{ac}^{(+2)}\\
\sigma_{ab}^{(+1)}\\
\sigma_{ac}^{(0)}\\
\sigma_{ab}^{(-1)}\\
\sigma_{ac}^{(-2)}\\
\vdots
\end{pmatrix},
\quad
V
=
\begin{pmatrix}
\vdots & \vdots\\
0 & 0\\
1 & 0\\
0 & 0\\
0 & 1\\
0 & 0\\
\vdots & \vdots
\end{pmatrix},
\quad
\boldsymbol{\mathcal{E}}
=
\begin{pmatrix}
\mathcal{E}_+\\
\mathcal{E}_-
\end{pmatrix},
\end{gather}
and the definition of the matrix $M$ depends on whether we consider 
Eqs.~\eqref{eqs:ss_coher_fourier_components} or 
Eqs.~\eqref{eqs:dual_color_coher_fourier_components_ft}. For 
Eqs.~\eqref{eqs:ss_coher_fourier_components}, we have
\begin{gather}
M=
\begin{pmatrix}
\ddots & \vdots & \vdots & \vdots & \vdots & \vdots & \rddots\\
\cdots & \delta & \Omega_0^*/2 & 0 & 0 & 0 & \cdots\\
\cdots & \Omega_0/2 & \tilde{\Delta} & \Omega_0/2 & 0 & 0 & \cdots\\
\cdots & 0 & \Omega_0^*/2 & \delta & \Omega_0^*/2 & 0 & \cdots\\
\cdots & 0 & 0 & \Omega_0/2 & \tilde{\Delta} & \Omega_0/2 & \cdots\\ 
\cdots & 0 & 0 & 0 & \Omega_0^*/2 & \delta & \cdots\\
\rddots & \vdots & \vdots & \vdots & \vdots & \vdots & \ddots
\end{pmatrix}.
\end{gather}
For Eqs.~\eqref{eqs:dual_color_coher_fourier_components_ft}, we subtract 
$n\Delta_\text{d}$ ($n$ is the number of the row such that the middle one has $n=0$) 
from the diagonal elements of the above matrix.

We can write the equations for the electric field as
\begin{gather}\label{app_coupled_equations_E_matrix}
\begin{pmatrix}
\frac{q}{n_0} & 0\\
0 & -\frac{q}{n_0}
\end{pmatrix}
\boldsymbol{\mathcal{E}}
=\frac{g\sqrt{2\pi}}{c}V^\text{T}\boldsymbol{\sigma},
\end{gather}
where $V^\text{T}$ is the transpose of the matrix $V$. Using 
Eq.~\eqref{eqs:matrix_form_fourier_components} and defining
$M_\mathcal{E}=(\Gamma_\text{1D}/2)V^\text{T}M^{-1}V$, 
Eqs.~\eqref{app_coupled_equations_E_matrix} become
\begin{gather}\label{continuum_numerical_dispersion_relation_matrix_eq}
\begin{pmatrix}
M_{\mathcal{E},11}+\frac{q}{n_0} & M_{\mathcal{E},12}\\
M_{\mathcal{E},21} & M_{\mathcal{E},22}-\frac{q}{n_0}
\end{pmatrix}
\boldsymbol{\mathcal{E}}
=
\begin{pmatrix}
0\\
0
\end{pmatrix},
\end{gather}
where $M_{\mathcal{E},kl}$ are the elements of $M_{\mathcal{E}}$. This 
equation is the equivalent of Eq.~\eqref{Lambda_coupled_equations_E_matrix}, 
but more general, since it is possible that 
$M_{\mathcal{E},11}\neq M_{\mathcal{E},22}$ (for the dual-color scheme). For 
Eq.~\eqref{continuum_numerical_dispersion_relation_matrix_eq} to have 
non-trivial solutions, the determinant of the matrix on the left hand side 
should be equal to zero. Hence, we get the equation
\begin{gather}\label{continuum_numerical_dispersion_relation_eq}
\del{\frac{q}{n_0}}^2+\frac{q}{n_0}(M_{\mathcal{E},11}-M_{\mathcal{E},22})
-\det(M_{\mathcal{E}})=0,
\end{gather}
where $\det(M_{\mathcal{E}})$ is the determinant of $M_{\mathcal{E}}$. The 
dispersion relation is found by solving 
Eq.~\eqref{continuum_numerical_dispersion_relation_eq}.

\section{Multi-mode transfer matrices}\label{App:Multi_mode_transfer_matrices}
In this appendix, we show how to transform between scattering matrices and 
transfer matrices for the multi-mode electric fields. The approach is very 
similar to the transfer matrix theory used in elastostatics~\cite{stephen_2006}.
This is a more general version of the single-mode transfer matrix 
formalism~\cite{deutsch_pra95a} that is commonly used for calculating 
electric fields in one-dimensional systems.

When one solves the scattering problem for an atom $j$ with position $z_j$, 
the result is the scattering matrix. In terms of the right-moving and 
left-moving parts of the electric field vector defined by 
Eq.~\eqref{multimode_E_field_vector} the relation is of the form
\begin{gather}
\label{scattering_matrix_E_relation}
\begin{pmatrix}
\mathbf{E}_+(z_j^+)\\
\mathbf{E}_-(z_j^-)
\end{pmatrix}
=
\begin{pmatrix}
S_{j,11} & S_{j,12}\\
S_{j,21} & S_{j,22}
\end{pmatrix}
\begin{pmatrix}
\mathbf{E}_+(z_j^-)\\
\mathbf{E}_-(z_j^+)
\end{pmatrix},
\end{gather}
where the blocks $S_{j,kl}$ are in general $\Nmodes$~by~$\Nmodes$ matrices 
describing the mixing of the $\Nmodes$ possible modes propagating in each 
direction. The scattering matrix relates output fields on both sides of the 
scatterer to the inputs. We find the scattering matrices for the $\Lambda$\mbox{-}type 
and dual\mbox{-}V atoms in App.~\ref{App:scattering_matrix_Lambda} and 
App.~\ref{App:scattering_matrix_dualV} respectively. In this appendix we only 
consider the general properties.

A transfer matrix for the atom
\begin{gather}
T_{\text{a},j}
=
\begin{pmatrix}
T_{\text{a},j,{11}} & T_{\text{a},j,{12}}\\
T_{\text{a},j,{21}} & T_{\text{a},j,{22}}
\end{pmatrix}
\end{gather}
is a relation of the form
\begin{gather}
\label{transfer_matrix_E_relation}
\begin{pmatrix}
\mathbf{E}_+(z_j^+)\\
\mathbf{E}_-(z_j^+)
\end{pmatrix}
=
T_{\text{a},j}
\begin{pmatrix}
\mathbf{E}_+(z_j^-)\\
\mathbf{E}_-(z_j^-)
\end{pmatrix},
\end{gather}
i.e. it relates the fields on one side of the atom to the fields on the 
other side. By rearranging Eq.~\eqref{scattering_matrix_E_relation} into the 
form of Eq.~\eqref{transfer_matrix_E_relation} one can show that
\begin{subequations}
\begin{align}
&T_{\text{a},j,{11}}=S_{j,11}-S_{j,12}S_{j,22}^{-1}S_{j,21},\\
&T_{\text{a},j,{12}}=S_{j,12}S_{j,22}^{-1},\\
&T_{\text{a},j,{21}}=-S_{j,22}^{-1}S_{j,21},\\
&T_{\text{a},j,{22}}=S_{j,22}^{-1}.
\end{align}
\label{transfer_matrix_blocks}
\end{subequations}
In App.~\ref{App:scattering_matrix_Lambda} and 
App.~\ref{App:scattering_matrix_dualV} we show that the blocks of the 
scattering matrix for the $\Lambda$\mbox{-}type and dual\mbox{-}V atoms fulfill
\begin{subequations}
\begin{gather}
S_{j,11}=S_{j,22}=S_{j,t},\\
S_{j,12}=S_{j,21}=S_{j,r},
\end{gather}
\label{S_r_S_t_definition}
\end{subequations}
where the matrices $S_{j,r}$ and $S_{j,t}$ are related by
\begin{gather}\label{S_r_S_t_relation}
S_{j,t}=I+S_{j,r},
\end{gather}
with $I$ being the $\Nmodes$~by~$\Nmodes$ identity matrix. From 
Eq.~\eqref{S_r_S_t_relation} we see that the matrices $S_{j,r}$ and $S_{j,t}$ commute. 
By writing
\begin{gather}
S_{j,r}S_{j,t}=S_{j,t}S_{j,r}
\end{gather}
and multiplying both sides by $S_{j,t}^{-1}$ from right and left, we get
\begin{gather}
S_{j,t}^{-1}S_{j,r}=S_{j,r}S_{j,t}^{-1},
\end{gather}
which implies that $S_{j,r}$ and $S_{j,t}^{-1}$ commute. This allows us to write 
Eqs.~\eqref{transfer_matrix_blocks} in terms of a single matrix
\begin{gather}\label{beta_matrix_definition}
\beta_j=-S_{j,t}^{-1}S_{j,r}.
\end{gather}
We obtain
\begin{subequations}
\begin{align}
&T_{\text{a},j,{11}}=S_{j,t}-S_{j,t}^{-1}S_{j,r}^2=I-\beta_j,\\
&T_{\text{a},j,{12}}=S_{j,t}^{-1}S_{j,r}=-\beta_j,\\
&T_{\text{a},j,{21}}=-S_{j,t}^{-1}S_{j,r}=\beta_j,\\
&T_{\text{a},j,{22}}=S_{j,t}^{-1}=I+\beta_j.
\end{align}
\label{transfer_matrix_blocks_B}
\end{subequations}

For the $\Lambda$\mbox{-}type atoms and dual\mbox{-}V 
atoms with ${\Omega_+=\Omega_-}$ and $\omega_{b_+c}=\omega_{b_-c}$, $\beta_j$ is 
symmetric, i.e. $\beta_j=\beta_j^\text{T}$, where $\beta_j^\text{T}$ 
is the transpose of $\beta_j$ (see App.~\ref{App:scattering_matrix_Lambda} and 
App.~\ref{App:scattering_matrix_dualV}).
Using this fact we also see that the transfer matrix $T_{\text{a},j}$ is 
symplectic. This means that if we define a matrix
\begin{gather}
J=\begin{pmatrix}
0 & I\\
-I & 0
\end{pmatrix},
\end{gather}
where zeros mean $\Nmodes$~by~$\Nmodes$ matrices with all elements equal to 
zero, then it holds that
\begin{gather}\label{T_a_is_symplectic}
T_{\text{a},j}^\text{T}JT_{\text{a},j}=J.
\end{gather}
This can be seen from the fact that if $\beta_j$ is symmetric, then so is 
$I\pm\beta_j$, and Eq.~\eqref{T_a_is_symplectic} can be shown by writing out the 
left hand side using Eqs.~\eqref{transfer_matrix_blocks_B}.

Free propagation of the electric field with the wave vector $k_0$ for a 
distance $d$ has the transfer matrix
\begin{gather}\label{free_space_transfer_matrix}
T_\text{f}=
\begin{pmatrix}
e^{ik_0 d}I & 0\\
0 & e^{-ik_0 d}I
\end{pmatrix}.
\end{gather}
The free propagation matrix $T_{\text{f},j}$ between atoms $j$ and $j+1$ at 
positions $z_j$ and $z_{j+1}$ fulfills 
$\mathbf{E}(z_{j+1}^-)=T_{\text{f},j}\mathbf{E}(z_j^+)$ and is given by 
Eq.~\eqref{free_space_transfer_matrix} with $d=z_{j+1}-z_j$.

The free propagation transfer matrices $T_{\text{f},j}$ are always symplectic.  
Therefore, the transfer matrix of a unit cell (or the whole ensemble), which 
is a product of the matrices $T_{\text{a},j}$ and $T_{\text{f},j}$, is 
symplectic if $T_{\text{a},j}$ is symplectic for all~$j$. This can be seen by 
considering a product of two symplectic transfer matrices, $T_1$ and $T_2$. It 
holds that
\begin{gather}
(T_1T_2)^\text{T}J(T_1T_2)
=T_2^\text{T}T_1^\text{T}JT_1T_2
=T_2^\text{T}JT_2
=J,
\end{gather}
hence the matrix $T_1T_2$ is symplectic.

For the purposes of finding the dispersion relation, we need to diagonalize 
the transfer matrix for the unit cell $T_\text{cell}$. Assuming that the unit 
cell has length $L_\text{u}$ and starts at $z=0$, we have the relation
\begin{gather}
\label{transfer_matrix_cell_E_relation}
\begin{pmatrix}
\mathbf{E}_+(L_\text{u}^+)\\
\mathbf{E}_-(L_\text{u}^+)
\end{pmatrix}
=
T_\text{cell}
\begin{pmatrix}
\mathbf{E}_+(0^-)\\
\mathbf{E}_-(0^-)
\end{pmatrix}.
\end{gather}
We note that if $T_\text{cell}$ is symplectic, then it has the property that 
its eigenvalues occur in reciprocal pairs. To see this, assume that 
$\mathbf{E}_\lambda$ is an eigenvector of $T_\text{cell}$ with eigenvalue 
$\lambda$, i.e.
\begin{gather}
T_\text{cell}\mathbf{E}_\lambda=\lambda\mathbf{E}_\lambda.
\end{gather}
Then using the property $T_\text{cell}^\text{T}JT_\text{cell}=J$ we have
\begin{gather}
T_\text{cell}^\text{T}(J\mathbf{E}_\lambda)
=T_\text{cell}^\text{T}JT_\text{cell}(1/\lambda)\mathbf{E}_\lambda
=(1/\lambda)(J\mathbf{E}_\lambda).
\end{gather}
Therefore, $J\mathbf{E}_\lambda$ is an eigenvector of $T_\text{cell}^\text{T}$ 
with the eigenvalue $1/\lambda$. Since $T_\text{cell}$ and 
$T_\text{cell}^\text{T}$ have the same set of eigenvalues, $1/\lambda$ is also 
an eigenvalue of $T_\text{cell}$.

The transmission and reflection coefficients for the whole ensemble can be 
obtained from its transfer matrix~$T_\text{e}$. Assuming that the ensemble has 
length $L$ and starts at $z=0$, we have the relation
\begin{gather}
\label{transfer_matrix_ensemble_E_relation}
\begin{pmatrix}
\mathbf{E}_+(L^+)\\
\mathbf{E}_-(L^+)
\end{pmatrix}
=
\begin{pmatrix}
T_{\text{e},{11}} & T_{\text{e},{12}}\\
T_{\text{e},{21}} & T_{\text{e},{22}}
\end{pmatrix}
\begin{pmatrix}
\mathbf{E}_+(0^-)\\
\mathbf{E}_-(0^-)
\end{pmatrix}.
\end{gather}
For concreteness, we assume a two-mode transfer matrix as is relevant for the 
dual\mbox{-}V scheme. Hence, the vectors $\mathbf{E}_\pm$ have two elements. We adopt 
the convention that the first element is a $\sigma_+$ component, and the second 
element is the $\sigma_-$ component of the field (the same definition as in 
Eq.~\eqref{dualV_E_vector_definition}). As an example, consider a scattering 
problem with the incoming fields
\begin{gather}\label{input_E_for_T}
\mathbf{E}_+(0^-)=
\begin{pmatrix}
1\\
0
\end{pmatrix},\quad
\mathbf{E}_-(L^+)=
\begin{pmatrix}
0\\
0
\end{pmatrix},
\end{gather}
i.e. there is only a $\sigma_+$ input field from the left. We want to find the 
outgoing fields: $\mathbf{E}_+(L^+)$ (the transmitted field) and 
$\mathbf{E}_-(0^-)$ (the reflected field).

After insertion of Eqs.~\eqref{input_E_for_T} into 
Eq.~\eqref{transfer_matrix_ensemble_E_relation} we find
\begin{subequations}
\begin{align}
&\mathbf{E}_-(0^-)=-T_{\text{e},{22}}^{-1}T_{\text{e},{12}}\mathbf{E}_+(0^-),\\
&\mathbf{E}_+(L^+)
=(T_{\text{e},{11}}
-T_{\text{e},{12}}T_{\text{e},{22}}^{-1}T_{\text{e},{21}})\mathbf{E}_+(0^-).
\end{align}
\label{output_E_from_T}
\end{subequations}

For the single-mode transfer matrices, $T_{\text{e},{kl}}$ are scalars. 
Furthermore, from Eqs. \eqref{transfer_matrix_blocks_B} and 
\eqref{free_space_transfer_matrix} we see that the transfer matrices for atoms 
and free propagation have determinants equal to unity. Using the fact that 
$\det(T_1T_2)=\det(T_1)\det(T_2)$ for any two square matrices $T_1$ and $T_2$, 
we have
$T_{\text{e},{11}}T_{\text{e},{22}}-T_{\text{e},{12}}T_{\text{e},{21}}
=\det(T_\text{e})=1$. This leads to a simplification of 
Eqs.~\eqref{output_E_from_T}, so that they become
\begin{subequations}
\begin{align}
&E_-(0^-)=-(T_{\text{e},{12}}/T_{\text{e},{22}})E_+(0^-),\\
&E_+(L^+)=(1/T_{\text{e},{22}})E_+(0^-).
\end{align}
\end{subequations}

\section{Transfer matrix for a uniform ensemble}\label{App:transfer_matrix_for_uniform_ensemble}
In this appendix, we will find closed-form expressions for the transfer matrix 
for the whole ensemble that either consists of $n_\text{e}$ copies of the same unit 
cell with the transfer matrix 
\begin{gather}\label{transfer_matrix_cell_definition}
T_\text{cell}=
\begin{pmatrix}
T_{11} & T_{12} \\
T_{21} & T_{22}
\end{pmatrix}
\end{gather}
in the discrete case, or is governed by the equations of the form
\begin{gather}\label{continuum_transfer_matrix_definition}
\dpd{}{z}
\begin{pmatrix}
\nohE_{+}\\
\nohE_{-}
\end{pmatrix}
=in_0
\begin{pmatrix}
-\alpha_1 & -\alpha_2\\
\alpha_2 & \alpha_1
\end{pmatrix}
\begin{pmatrix}
\nohE_{+}\\
\nohE_{-}
\end{pmatrix}
\end{gather}
in the continuum case. For the continuum case, we note that 
Eqs.~\eqref{eqn:hE_sigma_pm_pm_ft_ab_inserted} for the dual-V scheme and the 
equivalent equations for the $\Lambda$-type scheme can be written in the form 
above (neglecting the vacuum dispersion relation) with $\alpha_1$ and 
$\alpha_2$ being given by either Eqs.~\eqref{alpha_secular_definition} or 
Eqs.~\eqref{alpha_cold_Lambda_definition}, depending on the scheme.

The starting point of the derivation is diagonalizing either the 
transfer matrix in Eq.~\eqref{transfer_matrix_cell_definition} or the 
matrix in Eq.~\eqref{continuum_transfer_matrix_definition}. This gives
\begin{align}\label{transfer_matrix_cell_diagonalization}
&\begin{pmatrix}
T_{11} & T_{12} \\
T_{21} & T_{22}
\end{pmatrix}
=V_\text{cell}D_\text{cell}V_\text{cell}^{-1},\\
&\begin{pmatrix}
-\alpha_1 & -\alpha_2\\
\alpha_2 & \alpha_1
\end{pmatrix}
=V_\alpha D_\alpha V_\alpha^{-1},
\end{align}
where the diagonal matrix $D_\text{cell}$ has elements (eigenvalues) 
$\exp(\pm i\tilde{q}L_\text{u})$, and the diagonal matrix $D_\alpha$ has 
elements $\pm q/n_0$. Here we use the 
relation~\eqref{transfer_matrix_eigenvalue_in_terms_of_bloch_fast} between the 
Bloch vector and the elements of $D_\text{cell}$ for brevity. The 
eigenvector matrices are
\begin{align}
&V_\text{cell}=
\begin{pmatrix}
1 & 1\\
(e^{i\tilde{q}L_\text{u}}-T_{11})/T_{12}
& (e^{-i\tilde{q}L_\text{u}}-T_{11})/T_{12}
\end{pmatrix},\\
&V_\alpha=
\begin{pmatrix}
1 & 1\\
-(q/n_0+\alpha_1)/\alpha_2
& -(-q/n_0+\alpha_1)/\alpha_2
\end{pmatrix}.
\end{align}

In the discrete case, the transfer matrix for the whole ensemble is
$T_\text{e}=T_\text{cell}^{n_\text{e}}$, where $n_\text{e}=L/L_\text{u}$ is an 
integer. This expression can be written as
\begin{gather}
\begin{aligned}
T_\text{e}
&=V_\text{cell}D_\text{cell}^{n_\text{e}}V_\text{cell}^{-1}\\
&=V_\text{cell}\del{\cos(\tilde{q}L)I+i\sin(\tilde{q}L)\sigma_z}V_\text{cell}^{-1}\\
&=\cos(\tilde{q}L)I+i\sin(\tilde{q}L)V_\text{cell}\sigma_zV_\text{cell}^{-1},
\end{aligned}
\end{gather}
where $I$ is the identity matrix and
\begin{gather}
\sigma_z=
\begin{pmatrix}
1 & 0\\
0 & -1
\end{pmatrix}.
\end{gather}
By doing the matrix multiplications and using 
$2\cos(\tilde{q}L_\text{u})=\tr(T_\text{cell})=T_{11}+T_{12}$ and 
$\det(T_\text{cell})=1$, we find
\begin{gather}
V_\text{cell}\sigma_zV_\text{cell}^{-1}
=\frac{1}{\sin(\tilde{q}L_\text{u})}
\begin{pmatrix}
\frac{i}{2}(T_{22}-T_{11}) & -iT_{12}\\
-iT_{21} & -\frac{i}{2}(T_{22}-T_{11})
\end{pmatrix}.
\end{gather}

In the continuum case, the transfer matrix for the whole ensemble is
\begin{gather}
\begin{aligned}
T_\text{e}
&=\exp\del{in_0
\begin{pmatrix}
-\alpha_1 & -\alpha_2\\
\alpha_2 & \alpha_1
\end{pmatrix}
L
}\\
&=\cos(qL)I+i\sin(qL)V_\alpha\sigma_zV_\alpha^{-1},
\end{aligned}
\end{gather}
where (using $\alpha_1^2-(q/n_0)^2-\alpha_2^2=0$)
\begin{gather}
V_\alpha\sigma_zV_\alpha^{-1}
=\frac{1}{q/n_0}
\begin{pmatrix}
-\alpha_1 & -\alpha_2\\
\alpha_2 & \alpha_1
\end{pmatrix}.
\end{gather}

\section{Scattering matrix for $\Lambda$\mbox{-}type atoms}
\label{App:scattering_matrix_Lambda}
In this appendix, we find the scattering matrix (i.e. the reflection and 
transmission coefficients) for a $\Lambda$\mbox{-}type atom (see 
Fig.~\ref{fig:level_diagrams}(b)). The derivation is based on 
Ref.~\cite{chang_njp11a}. The electric field is given by the operator
\begin{gather}\label{E_field_definition_single_atom_lambda}
\hE(z)=\hE_+(z)e^{ik_0(z-z_j)}
+\hE_-(z)e^{-ik_0(z-z_j)}.
\end{gather}
Compared to the continuum model, we have shifted the spatial phases such that 
they vanish at the position of the atom $z_j$ ($j$ is the index of the atom). 
The effects of the propagation phases will be accounted for separately by the 
transfer matrices of free propagation.

The Hamiltonian~\eqref{three_level_H}, which we have used for the continuum 
model, can also be used to describe a single $\Lambda$-type 
atom, since the discrete nature of the atoms is still present due to the 
definition of the atomic operators given by 
Eq.~\eqref{sigma_collective_definition}. Because of considering only a 
single atom, Eq.~\eqref{sigma_collective_definition} becomes 
$\hs_{\alpha\beta}(z)=\frac{1}{n_0}\delta(z-z_j)\hs_{\alpha\beta,j}$, and 
inserting this into Eqs.~\eqref{three_level_H} results in
\begin{subequations}
\begin{align}
&\begin{aligned}
\hat{H}_\text{3,a}=-\hbar\sbr{\tilde{\Delta}_0\hs_{bb,j}+\delta_0\hs_{cc,j}},
\end{aligned}\displaybreak[0]\\
&\begin{aligned}
\hat{H}_\text{3,i}=&-\hbar\sbr{\hs_{bc,j}\Omega(z_j)+\text{H.c.}}\\
&-\hbar g\sqrt{2\pi}
\left[
\hs_{ba,j}\hE(z_j)+\text{H.c.}\right],
\end{aligned}\displaybreak[0]\\
&\begin{aligned}
\hat{H}_\text{3,p}=-i\hbar c\int \sbr{\hE_{+}^\dagger(z)\dpd{\hE_{+}(z)}{z}
-\hE_{-}^\dagger(z)\dpd{\hE_{-}(z)}{z}}\dif z.
\end{aligned}
\end{align}
\end{subequations}
From the Hamiltonian, we get the Heisenberg equations for the atom
\begin{subequations}
\begin{align}\label{Lambda_system_sigma_ab_equation}
&\dpd{\nohs_{ab,j}}{t}
=i\tilde{\Delta}_0\nohs_{ab,j}+i\Omega(z_j)\nohs_{ac,j}+ig\sqrt{2\pi}\nohE(z_j,t),\\
\label{Lambda_system_sigma_ac_equation}
&\dpd{\nohs_{ac,j}}{t}=i\delta_0\nohs_{ac,j}+i\Omega^*(z_j)\nohs_{ab,j}.
\end{align}
\label{Lambda_system_sigma_ab_ac_equations}
\end{subequations}
These equations are similar to Eqs.~\eqref{eqs:coher}, except that here we do 
not make the continuum approximation.

For the electric field we have the equations
\begin{align}\label{Lambda_system_sigma_E_field_eqs}
&\del{\dpd{}{t}\pm c\dpd{}{z}}\nohE_{\pm}(z,t)
=ig\sqrt{2\pi}\delta(z-z_j)\nohs_{ab,j},
\end{align}
which are exactly the same as Eqs.~\eqref{eqn:hE_pm} due the 
definition~\eqref{sigma_collective_definition}. In this form, however, we can 
formally solve them~\cite{chang_njp12}, so that we obtain
\begin{gather}\label{Lambda_system_sigma_E_field_eqs_sols}
\begin{aligned}
\nohE_{\pm}(z,t)
=\;&\nohE_{\pm,\text{in}}(z \mp ct)\\
&+\frac{ig\sqrt{2\pi}}{c}
\theta\delnospace{\pm(z-z_j)}\nohs_{ab,j}\delnospace{t\mp\frac{z-z_j}{c}},
\end{aligned}
\end{gather}
where $\nohE_{\pm,\text{in}}(z \pm ct)$ are the input fields, and $\theta$ is 
the Heaviside theta function.

Since the scattering problem is symmetric, and since the equations are linear, 
we can gain full information about the scattering by setting 
$\nohE_{+,\text{in}}(z - ct)=1$ and $\nohE_{-,\text{in}}(z + ct)=0$ in 
Eqs.~\eqref{Lambda_system_sigma_E_field_eqs_sols}. 
Then we find the 
total electric field~\eqref{E_field_definition_single_atom_lambda} to be
\begin{gather}
\nohE(z_j,t)=1+\frac{ig\sqrt{2\pi}}{c}\nohs_{ab,j}\del{t}
\end{gather}
Upon inserting this expression into \eqref{Lambda_system_sigma_ab_equation}, we obtain
\begin{gather}
\label{Lambda_system_sigma_ab_equation2}
\dpd{\nohs_{ab,j}}{t}
=i\del{\tilde{\Delta}_0+i\frac{\Gamma_\text{1D}}{2}}\nohs_{ab,j}
+i\Omega(z_j)\nohs_{ac,j}+ig\sqrt{2\pi}.
\end{gather}
After Fourier transforming Eqs.~\eqref{Lambda_system_sigma_E_field_eqs_sols} 
we get the reflection and transmission coefficients
\begin{subequations}
\begin{align}
&r_j=\nohE_{-}(z_j^-,\omega)=\frac{ig\sqrt{2\pi}}{c}\nohs_{ab,j}(\omega),\\
&t_j=\nohE_{+}(z_j^+,\omega)=1+r_j.
\end{align}
\label{Lambda_system_t_r_sigma_ab_relations}
\end{subequations}

We also Fourier transform Eq.~\eqref{Lambda_system_sigma_ac_equation} and 
Eq.~\eqref{Lambda_system_sigma_ab_equation2} and get
\begin{subequations}
\begin{align}\label{Lambda_system_sigma_ab_equation_ft}
&0
=i\del{\tilde{\Delta}+i\frac{\Gamma_\text{1D}}{2}}\nohs_{ab,j}
+i\Omega(z_j)\nohs_{ac,j}+ig\sqrt{2\pi},\\
\label{Lambda_system_sigma_ac_equation_ft}
&0=i\delta\nohs_{ac,j}+i\Omega^*(z_j)\nohs_{ab,j},
\end{align}
\label{Lambda_system_sigma_ab_ac_equations_ft}
\end{subequations}
where, as before, we have absorbed the Fourier frequency variable $\omega$ into the 
detunings by defining $\tilde{\Delta}=\tilde{\Delta}_0+\omega$ and 
$\delta=\delta_0+\omega$.

Now we solve Eqs. \eqref{Lambda_system_t_r_sigma_ab_relations} and 
\eqref{Lambda_system_sigma_ab_ac_equations_ft} and find
\begin{subequations}
\begin{align}
&r_j
=-\frac{i(\Gamma_\text{1D}/2)\delta}
{(\tilde{\Delta}+i\Gamma_\text{1D}/2)\delta-|\Omega(z_j)|^2},\\ 
&t_j
=\frac{\tilde{\Delta}\delta-|\Omega(z_j)|^2}
{(\tilde{\Delta}+i\Gamma_\text{1D}/2)\delta-|\Omega(z_j)|^2}.
\end{align}
\end{subequations}
The parameter \eqref{beta_matrix_definition}, in terms of which the blocks of 
the transfer matrix \eqref{transfer_matrix_blocks_B} are written, is a scalar 
in the single-mode case and is given by
\begin{gather}\label{lambda_type_beta}
\beta_j
=-\frac{r_j}{t_j}
=\frac{i(\Gamma_\text{1D}/2)\delta}
{\tilde{\Delta}\delta-|\Omega(z_j)|^2}.
\end{gather}

\section{Scattering matrix for the dual\mbox{-}V atoms.}\label{App:scattering_matrix_dualV}
The derivation of the scattering matrix for the dual\mbox{-}V atoms proceeds 
in a similar manner as the derivation for the $\Lambda$\mbox{-}type atoms in 
App.~\ref{App:scattering_matrix_Lambda}. Similar to 
Eq.~\eqref{E_field_definition_single_atom_lambda} we define the electric field 
operators
\begin{gather}
\hE_{\sigma_\pm}(z)=\hE_{\sigma_\pm,+}(z)e^{ik_0(z-z_j)}
+\hE_{\sigma_\pm,-}(z)e^{-ik_0(z-z_j)}.
\end{gather}
The Hamiltonian for a single dual\mbox{-}V atom interacting with light is 
given by Eqs.~\eqref{dualV_Hamiltonian} with the atomic operators 
$\hs_{\alpha\beta}(z)=\frac{1}{n_0}\delta(z-z_j)\hs_{\alpha\beta,j}$ (special 
case of the definition~\eqref{sigma_collective_definition}). Therefore, 
Eqs.~\eqref{dualV_Hamiltonian} can be written
\begin{subequations}
\begin{align}
&\begin{aligned}
\hat{H}_\text{V,a}=
-\hbar\sbr{\sum_{\alpha\in\{+,-\}}
\tilde{\Delta}_0^{(\alpha)}\hs_{b_\alpha b_\alpha,j}+\delta_0\hs_{cc,j}},
\end{aligned}\displaybreak[0]\\
&\begin{aligned}
\hat{H}_\text{V,i}=&-\hbar\sum_{\alpha\in\{+,-\}}\Bigg\{
\sbr{\hs_{b_\alpha c,j}\Omega_\alpha e^{\alpha ik_\text{c}z_j}+\text{H.c.}}\\
&+g\sqrt{2\pi}
\left[
\hs_{b_\alpha a,j}\hE_{\sigma_\alpha}(z_j)+\text{H.c.}\right]\Bigg\},
\end{aligned}\displaybreak[0]\\
&\begin{aligned}
\hat{H}_\text{V,p}=-i\hbar c\int\sum_{\alpha\in\{+,-\}}
\Bigg[&\hE_{\sigma_\alpha,+}^\dagger(z)\dpd{\hE_{\sigma_\alpha,+}(z)}{z}\\
&-\hE_{\sigma_\alpha,-}^\dagger(z)\dpd{\hE_{\sigma_\alpha,-}(z)}{z}\Bigg]\dif z.
\end{aligned}
\end{align}
\end{subequations}
From the Hamiltonian, the equations for the atom are
\begin{subequations}
\begin{align}\label{dualV_system_sigma_ab_equations}
&\begin{aligned}
\dpd{\nohs_{ab_\pm,j}}{t}
=\;&i\tilde{\Delta}_0^{(\pm)}\nohs_{ab_\pm,j}
+i\Omega_\pm\nohs_{ac,j}e^{\pm ik_\text{c}z_j}\\
&+ig\sqrt{2\pi}\nohE_{\sigma_\pm}(z_j,t),
\end{aligned}\\
\label{dualV_system_sigma_ac_equation}
&\dpd{\nohs_{ac,j}}{t}=i\delta_0\nohs_{ac,j}
+i\Omega_+^*\nohs_{ab_+,j}e^{-ik_\text{c}z_j}
+i\Omega_-^*\nohs_{ab_-,j}e^{ik_\text{c}z_j}.
\end{align}
\label{dualV_system_sigma_ab_ac_equations}
\end{subequations}
The formal solutions to the equations for the field are
\begin{subequations}
\begin{align}
&\begin{aligned}
&\nohE_{\sigma_+,\pm}(z,t)
=\nohE_{\sigma_+,\pm,\text{in}}(z \mp ct)\\
&+\frac{ig\sqrt{2\pi}}{c}
\theta\delnospace{\pm(z-z_j)}\nohs_{ab_+,j}\delnospace{t\mp\frac{z-z_j}{c}},
\end{aligned}\\
&\begin{aligned}
&\nohE_{\sigma_-,\pm}(z,t)
=\nohE_{\sigma_-,\pm,\text{in}}(z \mp ct)\\
&+\frac{ig\sqrt{2\pi}}{c}
\theta\delnospace{\pm(z-z_j)}\nohs_{ab_-,j}\delnospace{t\mp\frac{z-z_j}{c}}.
\end{aligned}
\end{align}
\label{dualV_system_sigma_E_field_eqs_sols}
\end{subequations}

Because of the symmetry of the system, we only need to consider two cases: 
$\nohE_{\sigma_+,+,\text{in}}(z_j - ct)=1$ with the rest of the input fields 
being zero, and $\nohE_{\sigma_-,+,\text{in}}(z_j - ct)=1$ with the rest of 
the input fields being zero.

Starting with the first case ($\nohE_{\sigma_+,+,\text{in}}(z_j - ct)=1$) and 
Fourier transforming, Eqs.~\eqref{dualV_system_sigma_ab_ac_equations} become
\begin{subequations}
\begin{align}\label{dualV_system_sigma_ab_plus_equation2_case1_ft}
&0
=i\tilde{\Delta}^{(+)}_\text{tot}\nohs_{ab_+,j}
+i\Omega_+\nohs_{ac,j}e^{ik_\text{c}z_j}+ig\sqrt{2\pi},\\
\label{dualV_system_sigma_ab_minus_equation2_case1_ft}
&0
=i\tilde{\Delta}^{(-)}_\text{tot}\nohs_{ab_-,j}
+i\Omega_-\nohs_{ac,j}e^{-ik_\text{c}z_j},\\
\label{dualV_system_sigma_ac_equation2_case1_ft}
&0=i\delta\nohs_{ac,j}
+i\Omega_+^*\nohs_{ab_+,j}e^{-ik_\text{c}z_j}
+i\Omega_-^*\nohs_{ab_-,j}e^{ik_\text{c}z_j},
\end{align}
\label{dualV_system_sigma_ab_ac_equations2_case1_ft}
\end{subequations}
with 
$\tilde{\Delta}^{(\pm)}_\text{tot}=\tilde{\Delta}_0^{(\pm)}+i(\Gamma_\text{1D}/2)+\omega$
defined for notational convenience, such that we have now absorbed the 
total decay rate $\Gamma=\Gamma'+\Gamma_\text{1D}$ into 
$\tilde{\Delta}^{(\pm)}_\text{tot}$; and $\delta=\delta_0+\omega$. From 
Eqs.~\eqref{dualV_system_sigma_E_field_eqs_sols} we have the relations
\begin{subequations}
\begin{align}
&r_{j,++}=\nohE_{\sigma_+,-}(z_j^-,\omega)=\frac{ig\sqrt{2\pi}}{c}\nohs_{ab_+,j}(\omega),\displaybreak[0]\\
&t_{j,++}=\nohE_{\sigma_+,+}(z_j^+,\omega)=1+r_{j,++},\displaybreak[0]\\
&r_{j,+-}=\nohE_{\sigma_-,-}(z_j^-,\omega)=\frac{ig\sqrt{2\pi}}{c}\nohs_{ab_-,j}(\omega),\displaybreak[0]\\
&t_{j,+-}=\nohE_{\sigma_-,+}(z_j^+,\omega)=r_{j,+-}.
\end{align}
\label{dualV_system_t_r_sigma_ab_relations_case1}
\end{subequations}
After solving Eqs. \eqref{dualV_system_sigma_ab_ac_equations2_case1_ft} and 
\eqref{dualV_system_t_r_sigma_ab_relations_case1} (with $k_\text{c}\approx k_0$) 
we get
\begin{subequations}
\begin{align}
&r_{j,++}
=-\frac{i(\Gamma_\text{1D}/2)\del{\tilde{\Delta}^{(-)}_\text{tot}\delta
-|\Omega_-|^2}}
{\tilde{\Delta}^{(+)}_\text{tot}\tilde{\Delta}^{(-)}_\text{tot}\delta
-\tilde{\Delta}^{(+)}_\text{tot}|\Omega_-|^2
-\tilde{\Delta}^{(-)}_\text{tot}|\Omega_+|^2},\displaybreak[0]\\
&t_{j,++}=1+r_{j,++},\displaybreak[0]\\
\label{dualV_system_t_r_sigma_ab_relations_case1_sol_rpm_tpm}
&r_{j,+-}=t_{j,+-}
=\frac{\Omega_-\Omega_+^*e^{-2ik_0 z_j}}
{\tilde{\Delta}^{(-)}_\text{tot}\delta-|\Omega_-|^2}
r_{j,++}.
\end{align}
\label{dualV_system_t_r_sigma_ab_relations_case1_sol}
\end{subequations}

For the second case ($\nohE_{\sigma_-,+,\text{in}}(z_j - ct)=1$), instead of 
Eqs.~\eqref{dualV_system_sigma_ab_ac_equations2_case1_ft} we have
\begin{subequations}
\begin{align}\label{dualV_system_sigma_ab_plus_equation2_case2_ft}
&0
=i\tilde{\Delta}^{(+)}_\text{tot}\nohs_{ab_+,j}
+i\Omega_+\nohs_{ac,j}e^{ik_\text{c}z_j},\\
\label{dualV_system_sigma_ab_minus_equation2_case2_ft}
&0
=i\tilde{\Delta}^{(-)}_\text{tot}\nohs_{ab_-,j}
+i\Omega_-\nohs_{ac,j}e^{-ik_\text{c}z_j}+ig\sqrt{2\pi},\\
\label{dualV_system_sigma_ac_equation2_case2_ft}
&0=i\delta\nohs_{ac,j}
+i\Omega_+^*\nohs_{ab_+,j}e^{-ik_\text{c}z_j}
+i\Omega_-^*\nohs_{ab_-,j}e^{ik_\text{c}z_j}.
\end{align}
\label{dualV_system_sigma_ab_ac_equations2_case2_ft}
\end{subequations}
Instead of Eqs.~\eqref{dualV_system_t_r_sigma_ab_relations_case1_sol} we have
\begin{subequations}
\begin{align}
&r_{j,--}=\nohE_{\sigma_-,-}(z_j^-,\omega)=\frac{ig\sqrt{2\pi}}{c}\nohs_{ab_-,j}(\omega),\displaybreak[0]\\
&t_{j,--}=\nohE_{\sigma_-,+}(z_j^+,\omega)=1+r_{j,--},\displaybreak[0]\\
&r_{j,-+}=\nohE_{\sigma_+,-}(z_j^-,\omega)=\frac{ig\sqrt{2\pi}}{c}\nohs_{ab_+,j}(\omega),\displaybreak[0]\\
&t_{j,-+}=\nohE_{\sigma_+,+}(z_j^+,\omega)=r_{j,-+}.
\end{align}
\label{dualV_system_t_r_sigma_ab_relations_case2}
\end{subequations}
After solving Eqs. 
\eqref{dualV_system_sigma_ab_ac_equations2_case2_ft} and 
\eqref{dualV_system_t_r_sigma_ab_relations_case2} we get
\begin{subequations}
\begin{align}
&r_{j,--}
=-\frac{i(\Gamma_\text{1D}/2)\del{\tilde{\Delta}^{(+)}_\text{tot}\delta
-|\Omega_+|^2}}
{\tilde{\Delta}^{(+)}_\text{tot}\tilde{\Delta}^{(-)}_\text{tot}\delta
-\tilde{\Delta}^{(+)}_\text{tot}|\Omega_-|^2
-\tilde{\Delta}^{(-)}_\text{tot}|\Omega_+|^2},\displaybreak[0]\\
&t_{j,--}=1+r_{j,--},\displaybreak[0]\\
\label{dualV_system_t_r_sigma_ab_relations_case2_sol_rmp_tmp}
&r_{j,-+}=t_{j,-+}
=\frac{\Omega_+\Omega_-^*e^{2ik_0 z_j}}
{\tilde{\Delta}^{(+)}_\text{tot}\delta-|\Omega_+|^2}
r_{j,--}.
\end{align}
\label{dualV_system_t_r_sigma_ab_relations_case2_sol}
\end{subequations}

In terms of Eqs. \eqref{dualV_system_t_r_sigma_ab_relations_case1_sol} and 
\eqref{dualV_system_t_r_sigma_ab_relations_case2_sol}, the blocks of the 
scattering matrix in Eq.~\eqref{scattering_matrix_E_relation} with the 
definition of the fields in Eq.~\eqref{dualV_E_vector_definition} are
\begin{subequations}
\begin{gather}
S_{j,11}=S_{j,22}=
\begin{pmatrix}
t_{j,++} & t_{j,-+}\\
t_{j,+-} & t_{j,--}
\end{pmatrix},\\
S_{j,12}=S_{j,21}=
\begin{pmatrix}
r_{j,++} & r_{j,-+}\\
r_{j,+-} & r_{j,--}
\end{pmatrix}.
\end{gather}
\end{subequations}
If we use the definitions of Eq.~\eqref{S_r_S_t_definition}, we further see that 
Eq.~\eqref{S_r_S_t_relation} holds.

As for the calculations using the continuum model in 
Sec.~\ref{Sec:dual_v_continuum_model}, we will also use 
${\omega_{b_+c}=\omega_{b_-c}}$ and ${\Omega_+=\Omega_-=\Omega_0/2}$ in 
the discrete model. This implies that $r_{j,+-}=r_{j,-+}$, and hence that the 
matrices $S_{j,kl}$ are symmetric. Since the product of commuting symmetric 
matrices is symmetric, it also follows that $\beta_j$ given by 
Eq.~\eqref{beta_matrix_definition} is symmetric.

\section{Effective mass for the regularly placed $\Lambda$\mbox{-}type scheme}\label{App:Regular_Spaced_Mass}
In this appendix, we derive the expression for the effective 
mass~\eqref{regularly_spaced_mass}. For the single-mode case, we have that 
\begin{gather}\label{bloch_eigenvalue_equation_single_mode_lambda}
\tr(T_\text{cell})=\lambda+1/\lambda,
\end{gather}
where $\tr(T_\text{cell})$ is the trace of $T_\text{cell}$, and $\lambda$ is 
one of the eigenvalues of $T_\text{cell}$. Since the length of the unit cell 
is $L_\text{u}=\pi/k_0$, 
Eq.~\eqref{bloch_eigenvalue_equation_single_mode_lambda} together with 
Eq.~\eqref{transfer_matrix_eigenvalue_in_terms_of_bloch} implies that
\begin{gather}\label{bloch_eigenvalue_equation_single_mode}
\cos(qL_\text{u})=-\frac{1}{2}\tr(T_\text{cell}),
\end{gather}
The right hand 
side of this equation is a function of $\delta$. We will solve it 
perturbatively to find $\delta$ as a function of $q$. Then the mass is found 
as the coefficient of the second order term in $q$ in the series expansion. 

For small $\delta$ and $\Omega_0\neq 0$, the scattering coefficient $\beta_j$ 
(given by Eq.~\eqref{lambda_type_beta} with 
$\Omega(z_j)=\Omega_0\cos(k_0 z_j)$) can be approximated by
\begin{gather}\label{beta3_approximate_expression}
\beta_j\approx -i\frac{\Gamma_\text{1D}}
{2|\Omega_0|^2\cos^2(k_0 z_j)}\delta.
\end{gather}
The precise condition for this approximation to be valid is that 
\begin{gather}\label{regular_spacing_dispersion_relation_condition_general} 
\delta\ll \frac{|\Omega_0|^2}{|\tilde{\Delta}|}\cos^2(k_0 z_j) \end{gather} 
has to be fulfilled for all the atoms in the unit cell which experience 
non-vanishing classical drive, i.e. the frequency has to be within their EIT 
windows. The right hand side of 
Eq.~\eqref{regular_spacing_dispersion_relation_condition_general} is smallest 
for the atoms placed at $k_0 z_j = \pm(\pi/2-\pi/N_\text{u})$ (see 
Fig.~\ref{fig:variable_N_u}). This leads to the condition given by 
Eq.~\eqref{regular_spacing_dispersion_relation_condition} of the main text.

For the chosen unit cells in Fig.~\ref{fig:variable_N_u} and numbering the 
atoms from the left (such that the leftmost atom in the unit cell has index 
$j=1$), we have within the approximation above that $\beta_j$ for 
${1\leq j \leq N_\text{u}-1}$ is inversely proportional to the classical field 
strength. We define
\begin{gather}\label{beta_c_definition}
\beta_\text{c}=-i\frac{\Gamma_\text{1D}}{2|\Omega_0|^2}\delta,
\end{gather}
so that for the chosen unit cells we have
\begin{gather}\label{beta_j_in_terms_of_beta_c}
\beta_j\approx \frac{\beta_\text{c}}{\cos^2((j-N_\text{u}/2)k_0 d)}
\end{gather}
with $d=\pi/(N_\text{u}k_0)$.

On the other hand, the last atom in the unit cell,
which is positioned at the node of the standing wave of the classical drive, 
will instead be described by Eq.~\eqref{lambda_type_beta} with 
${\Omega(z_j)=0}$, i.e.
\begin{gather}\label{beta2_expression}
\beta_{N_\text{u}}
\approx\frac{\Gamma_\text{1D}}{\Gamma'-2i\Delta_\text{c}},
\end{gather}
where we have approximated $\Delta\approx\Delta_\text{c}$, since we assume 
$\delta \ll \Delta_\text{c}$. This last atom effectively behaves as a 
two-level atom.

The claim now is that in this approximation and for even $N_\text{u}$, we have 
to first order in $\beta_\text{c}$ that
\begin{gather}\label{even_N_u_tr_M_cell}
\frac{1}{2}\tr(T_\text{cell})
\approx -1-2(N_\text{u}-1)\beta_{N_\text{u}}\beta_\text{c}.
\end{gather}
We will prove this claim below, but first we show how it leads to the desired 
expression for the effective mass~\eqref{regularly_spaced_mass}. If we expand 
the left hand side of Eq.~\eqref{bloch_eigenvalue_equation_single_mode} around 
$qd=0$, we find
\begin{gather}\label{bloch_eigenvalue_equation_single_mode_lhs}
\cos(qL_\text{u})
=\cos(N_\text{u}qd)
\approx 1+\frac{1}{2}N_\text{u}^2\del{qd}^2.
\end{gather}
Then, using Eqs. \eqref{even_N_u_tr_M_cell} and 
\eqref{bloch_eigenvalue_equation_single_mode_lhs} for respectively the right 
hand side and the left hand side of 
Eq.~\eqref{bloch_eigenvalue_equation_single_mode} together with Eqs. \eqref{beta2_expression} and \eqref{beta_c_definition}, we
get
\begin{gather}
\frac{1}{2}\del{qd}^2
\approx -\frac{(N_\text{u}-1)\Gamma_\text{1D}^2}
{2N_\text{u}^2(\Delta_\text{c}+i\Gamma'/2)|\Omega_0|^2}\delta.
\end{gather}
Comparing this expression with 
Eq.~\eqref{regular_spacing_dispersion_relation}, we find the mass given by 
Eq.~\eqref{regularly_spaced_mass}.

Now we prove the claim \eqref{even_N_u_tr_M_cell}. The transfer matrices for 
the atoms have elements given by Eqs.~\eqref{transfer_matrix_blocks_B} with the 
scalar $\beta_j$ given by either Eq.~\eqref{beta3_approximate_expression} or 
Eq.~\eqref{beta2_expression}. We first find the product of the 
transfer matrices for the atoms with $1\leq j \leq N_\text{u}-1$ and free 
propagation between them. If $T_{\text{a},j}$ is the transfer matrix for the 
atom with $\beta_j$, and $T_\text{f}$ is the free propagation matrix given by 
Eq.~\eqref{free_space_transfer_matrix} with $d=\pi/(N_\text{u}k_0)$, then we 
can recursively define the partial product by 
$T^{(j)}=T_\text{f}T_{\text{a},j}T^{(j-1)}$ for $2\leq j \leq N_\text{u}-1$, 
and $T^{(1)}=T_\text{f}T_{\text{a},1}$. In terms of the elements of the matrix 
$T^{(j)}$ we have to first order in $\beta_j$ that
\begin{subequations}
\begin{align}
&T^{(j)}_{11}\approx\del{1-\sum_{j'=1}^{j} \beta_{j'}}e^{ijk_0 d}\\
&T^{(j)}_{22}\approx\del{1+\sum_{j'=1}^{j} \beta_{j'}}e^{-ijk_0 d}\\
&T^{(j)}_{21}\approx\sum_{j'=1}^{j} \beta_{j'} e^{i(2j'-j-2)k_0 d}\\
&T^{(j)}_{12}\approx-\sum_{j'=1}^{j} \beta_{j'} e^{-i(2j'-j-2)k_0 d}
\end{align}
\end{subequations}
We can now find
\begin{gather}
\tr(T_\text{cell})=\tr\delnospace{T_\text{f}T_{a,N_\text{u}}T^{(N_\text{u}-1)}}.
\end{gather}
After writing the matrix product out, taking the trace and using 
$\exp(iN_\text{u}k_0 d)=-1$ we get
\begin{gather}
\tr(T_\text{cell})
\approx -2
-4\beta_{N_\text{u}}\sum_{j=1}^{N_\text{u}-1}\beta_j
\cos^2((j-N_\text{u}/2)k_0 d)
\end{gather}
Using Eq.~\eqref{beta_j_in_terms_of_beta_c}, the above simplifies to
\begin{gather}
\tr(T_\text{cell})
\approx -2
-4\beta_{N_\text{u}}\sum_{j=1}^{N_\text{u}-1}\beta_\text{c},
\end{gather}
which is the same as Eq.~\eqref{even_N_u_tr_M_cell}.

\bibliography{references}

\begin{thebibliography}{38}%
\makeatletter
\providecommand \@ifxundefined [1]{%
 \@ifx{#1\undefined}
}%
\providecommand \@ifnum [1]{%
 \ifnum #1\expandafter \@firstoftwo
 \else \expandafter \@secondoftwo
 \fi
}%
\providecommand \@ifx [1]{%
 \ifx #1\expandafter \@firstoftwo
 \else \expandafter \@secondoftwo
 \fi
}%
\providecommand \natexlab [1]{#1}%
\providecommand \enquote  [1]{``#1''}%
\providecommand \bibnamefont  [1]{#1}%
\providecommand \bibfnamefont [1]{#1}%
\providecommand \citenamefont [1]{#1}%
\providecommand \href@noop [0]{\@secondoftwo}%
\providecommand \href [0]{\begingroup \@sanitize@url \@href}%
\providecommand \@href[1]{\@@startlink{#1}\@@href}%
\providecommand \@@href[1]{\endgroup#1\@@endlink}%
\providecommand \@sanitize@url [0]{\catcode `\\12\catcode `\$12\catcode
  `\&12\catcode `\#12\catcode `\^12\catcode `\_12\catcode `\%12\relax}%
\providecommand \@@startlink[1]{}%
\providecommand \@@endlink[0]{}%
\providecommand \url  [0]{\begingroup\@sanitize@url \@url }%
\providecommand \@url [1]{\endgroup\@href {#1}{\urlprefix }}%
\providecommand \urlprefix  [0]{URL }%
\providecommand \Eprint [0]{\href }%
\providecommand \doibase [0]{http://dx.doi.org/}%
\providecommand \selectlanguage [0]{\@gobble}%
\providecommand \bibinfo  [0]{\@secondoftwo}%
\providecommand \bibfield  [0]{\@secondoftwo}%
\providecommand \translation [1]{[#1]}%
\providecommand \BibitemOpen [0]{}%
\providecommand \bibitemStop [0]{}%
\providecommand \bibitemNoStop [0]{.\EOS\space}%
\providecommand \EOS [0]{\spacefactor3000\relax}%
\providecommand \BibitemShut  [1]{\csname bibitem#1\endcsname}%
\let\auto@bib@innerbib\@empty
\bibitem [{\citenamefont {Hau}\ \emph {et~al.}(1999)\citenamefont {Hau},
  \citenamefont {Harris}, \citenamefont {Dutton},\ and\ \citenamefont
  {Behroozi}}]{hau_nature1999}%
  \BibitemOpen
  \bibfield  {author} {\bibinfo {author} {\bibfnamefont {L.~V.}\ \bibnamefont
  {Hau}}, \bibinfo {author} {\bibfnamefont {S.~E.}\ \bibnamefont {Harris}},
  \bibinfo {author} {\bibfnamefont {Z.}~\bibnamefont {Dutton}}, \ and\ \bibinfo
  {author} {\bibfnamefont {C.~H.}\ \bibnamefont {Behroozi}},\ }\href {\doibase
  10.1038/17561} {\bibfield  {journal} {\bibinfo  {journal} {Nature}\ }\textbf
  {\bibinfo {volume} {397}},\ \bibinfo {pages} {594} (\bibinfo {year}
  {1999})}\BibitemShut {NoStop}%
\bibitem [{\citenamefont {Lukin}(2003)}]{lukin_rmp03}%
  \BibitemOpen
  \bibfield  {author} {\bibinfo {author} {\bibfnamefont {M.~D.}\ \bibnamefont
  {Lukin}},\ }\href {\doibase 10.1103/RevModPhys.75.457} {\bibfield  {journal}
  {\bibinfo  {journal} {Rev. Mod. Phys.}\ }\textbf {\bibinfo {volume} {75}},\
  \bibinfo {pages} {457} (\bibinfo {year} {2003})}\BibitemShut {NoStop}%
\bibitem [{\citenamefont {Liu}\ \emph {et~al.}(2001)\citenamefont {Liu},
  \citenamefont {Dutton}, \citenamefont {Behroozi},\ and\ \citenamefont
  {Hau}}]{liu_nature2001}%
  \BibitemOpen
  \bibfield  {author} {\bibinfo {author} {\bibfnamefont {C.}~\bibnamefont
  {Liu}}, \bibinfo {author} {\bibfnamefont {Z.}~\bibnamefont {Dutton}},
  \bibinfo {author} {\bibfnamefont {C.~H.}\ \bibnamefont {Behroozi}}, \ and\
  \bibinfo {author} {\bibfnamefont {L.~V.}\ \bibnamefont {Hau}},\ }\href
  {\doibase 10.1038/35054017} {\bibfield  {journal} {\bibinfo  {journal}
  {Nature}\ }\textbf {\bibinfo {volume} {409}},\ \bibinfo {pages} {490}
  (\bibinfo {year} {2001})}\BibitemShut {NoStop}%
\bibitem [{\citenamefont {Phillips}\ \emph {et~al.}(2001)\citenamefont
  {Phillips}, \citenamefont {Fleischhauer}, \citenamefont {Mair}, \citenamefont
  {Walsworth},\ and\ \citenamefont {Lukin}}]{phillips_prl01}%
  \BibitemOpen
  \bibfield  {author} {\bibinfo {author} {\bibfnamefont {D.~F.}\ \bibnamefont
  {Phillips}}, \bibinfo {author} {\bibfnamefont {A.}~\bibnamefont
  {Fleischhauer}}, \bibinfo {author} {\bibfnamefont {A.}~\bibnamefont {Mair}},
  \bibinfo {author} {\bibfnamefont {R.~L.}\ \bibnamefont {Walsworth}}, \ and\
  \bibinfo {author} {\bibfnamefont {M.~D.}\ \bibnamefont {Lukin}},\ }\href
  {\doibase 10.1103/PhysRevLett.86.783} {\bibfield  {journal} {\bibinfo
  {journal} {Phys. Rev. Lett.}\ }\textbf {\bibinfo {volume} {86}},\ \bibinfo
  {pages} {783} (\bibinfo {year} {2001})}\BibitemShut {NoStop}%
\bibitem [{\citenamefont {Harris}\ and\ \citenamefont
  {Hau}(1999)}]{harris_prl99}%
  \BibitemOpen
  \bibfield  {author} {\bibinfo {author} {\bibfnamefont {S.~E.}\ \bibnamefont
  {Harris}}\ and\ \bibinfo {author} {\bibfnamefont {L.~V.}\ \bibnamefont
  {Hau}},\ }\href {\doibase 10.1103/PhysRevLett.82.4611} {\bibfield  {journal}
  {\bibinfo  {journal} {Phys. Rev. Lett.}\ }\textbf {\bibinfo {volume} {82}},\
  \bibinfo {pages} {4611} (\bibinfo {year} {1999})}\BibitemShut {NoStop}%
\bibitem [{\citenamefont {Bajcsy}\ \emph {et~al.}(2009)\citenamefont {Bajcsy},
  \citenamefont {Hofferberth}, \citenamefont {Balic}, \citenamefont {Peyronel},
  \citenamefont {Hafezi}, \citenamefont {Zibrov}, \citenamefont {Vuletic},\
  and\ \citenamefont {Lukin}}]{bajcsy_prl09}%
  \BibitemOpen
  \bibfield  {author} {\bibinfo {author} {\bibfnamefont {M.}~\bibnamefont
  {Bajcsy}}, \bibinfo {author} {\bibfnamefont {S.}~\bibnamefont {Hofferberth}},
  \bibinfo {author} {\bibfnamefont {V.}~\bibnamefont {Balic}}, \bibinfo
  {author} {\bibfnamefont {T.}~\bibnamefont {Peyronel}}, \bibinfo {author}
  {\bibfnamefont {M.}~\bibnamefont {Hafezi}}, \bibinfo {author} {\bibfnamefont
  {A.~S.}\ \bibnamefont {Zibrov}}, \bibinfo {author} {\bibfnamefont
  {V.}~\bibnamefont {Vuletic}}, \ and\ \bibinfo {author} {\bibfnamefont
  {M.~D.}\ \bibnamefont {Lukin}},\ }\href {\doibase
  10.1103/PhysRevLett.102.203902} {\bibfield  {journal} {\bibinfo  {journal}
  {Phys. Rev. Lett.}\ }\textbf {\bibinfo {volume} {102}},\ \bibinfo {pages}
  {203902} (\bibinfo {year} {2009})}\BibitemShut {NoStop}%
\bibitem [{\citenamefont {Gorshkov}\ \emph {et~al.}(2011)\citenamefont
  {Gorshkov}, \citenamefont {Otterbach}, \citenamefont {Fleischhauer},
  \citenamefont {Pohl},\ and\ \citenamefont {Lukin}}]{gorshkov_prl11}%
  \BibitemOpen
  \bibfield  {author} {\bibinfo {author} {\bibfnamefont {A.~V.}\ \bibnamefont
  {Gorshkov}}, \bibinfo {author} {\bibfnamefont {J.}~\bibnamefont {Otterbach}},
  \bibinfo {author} {\bibfnamefont {M.}~\bibnamefont {Fleischhauer}}, \bibinfo
  {author} {\bibfnamefont {T.}~\bibnamefont {Pohl}}, \ and\ \bibinfo {author}
  {\bibfnamefont {M.~D.}\ \bibnamefont {Lukin}},\ }\href {\doibase
  10.1103/PhysRevLett.107.133602} {\bibfield  {journal} {\bibinfo  {journal}
  {Phys. Rev. Lett.}\ }\textbf {\bibinfo {volume} {107}},\ \bibinfo {pages}
  {133602} (\bibinfo {year} {2011})}\BibitemShut {NoStop}%
\bibitem [{\citenamefont {Chen}\ \emph {et~al.}(2013)\citenamefont {Chen},
  \citenamefont {Beck}, \citenamefont {B{\"u}cker}, \citenamefont {Gullans},
  \citenamefont {Lukin}, \citenamefont {Tanji-Suzuki},\ and\ \citenamefont
  {Vuleti{\'c}}}]{chen_science13}%
  \BibitemOpen
  \bibfield  {author} {\bibinfo {author} {\bibfnamefont {W.}~\bibnamefont
  {Chen}}, \bibinfo {author} {\bibfnamefont {K.~M.}\ \bibnamefont {Beck}},
  \bibinfo {author} {\bibfnamefont {R.}~\bibnamefont {B{\"u}cker}}, \bibinfo
  {author} {\bibfnamefont {M.}~\bibnamefont {Gullans}}, \bibinfo {author}
  {\bibfnamefont {M.~D.}\ \bibnamefont {Lukin}}, \bibinfo {author}
  {\bibfnamefont {H.}~\bibnamefont {Tanji-Suzuki}}, \ and\ \bibinfo {author}
  {\bibfnamefont {V.}~\bibnamefont {Vuleti{\'c}}},\ }\href {\doibase
  10.1126/science.1238169} {\bibfield  {journal} {\bibinfo  {journal}
  {Science}\ }\textbf {\bibinfo {volume} {341}},\ \bibinfo {pages} {768}
  (\bibinfo {year} {2013})}\BibitemShut {NoStop}%
\bibitem [{\citenamefont {Fleischhauer}\ and\ \citenamefont
  {Lukin}(2000)}]{fleischhauer_prl2000}%
  \BibitemOpen
  \bibfield  {author} {\bibinfo {author} {\bibfnamefont {M.}~\bibnamefont
  {Fleischhauer}}\ and\ \bibinfo {author} {\bibfnamefont {M.~D.}\ \bibnamefont
  {Lukin}},\ }\href {\doibase 10.1103/PhysRevLett.84.5094} {\bibfield
  {journal} {\bibinfo  {journal} {Phys. Rev. Lett.}\ }\textbf {\bibinfo
  {volume} {84}},\ \bibinfo {pages} {5094} (\bibinfo {year}
  {2000})}\BibitemShut {NoStop}%
\bibitem [{\citenamefont {Andr\'e}\ and\ \citenamefont
  {Lukin}(2002)}]{andre_prl02a}%
  \BibitemOpen
  \bibfield  {author} {\bibinfo {author} {\bibfnamefont {A.}~\bibnamefont
  {Andr\'e}}\ and\ \bibinfo {author} {\bibfnamefont {M.~D.}\ \bibnamefont
  {Lukin}},\ }\href {\doibase 10.1103/PhysRevLett.89.143602} {\bibfield
  {journal} {\bibinfo  {journal} {Phys. Rev. Lett.}\ }\textbf {\bibinfo
  {volume} {89}},\ \bibinfo {pages} {143602} (\bibinfo {year}
  {2002})}\BibitemShut {NoStop}%
\bibitem [{\citenamefont {Bajcsy}\ \emph {et~al.}(2003)\citenamefont {Bajcsy},
  \citenamefont {Zibrov},\ and\ \citenamefont {Lukin}}]{bajcsy_nature03a}%
  \BibitemOpen
  \bibfield  {author} {\bibinfo {author} {\bibfnamefont {M.}~\bibnamefont
  {Bajcsy}}, \bibinfo {author} {\bibfnamefont {A.~S.}\ \bibnamefont {Zibrov}},
  \ and\ \bibinfo {author} {\bibfnamefont {M.~D.}\ \bibnamefont {Lukin}},\
  }\href {\doibase 10.1038/nature02176} {\bibfield  {journal} {\bibinfo
  {journal} {Nature}\ }\textbf {\bibinfo {volume} {426}},\ \bibinfo {pages}
  {638} (\bibinfo {year} {2003})}\BibitemShut {NoStop}%
\bibitem [{\citenamefont {Chang}\ \emph {et~al.}(2008)\citenamefont {Chang},
  \citenamefont {Gritsev}, \citenamefont {Morigi}, \citenamefont {Vuleti\'{c}},
  \citenamefont {Lukin},\ and\ \citenamefont {Demler}}]{chang_naturephys08a}%
  \BibitemOpen
  \bibfield  {author} {\bibinfo {author} {\bibfnamefont {D.~E.}\ \bibnamefont
  {Chang}}, \bibinfo {author} {\bibfnamefont {V.}~\bibnamefont {Gritsev}},
  \bibinfo {author} {\bibfnamefont {G.}~\bibnamefont {Morigi}}, \bibinfo
  {author} {\bibfnamefont {V.}~\bibnamefont {Vuleti\'{c}}}, \bibinfo {author}
  {\bibfnamefont {M.~D.}\ \bibnamefont {Lukin}}, \ and\ \bibinfo {author}
  {\bibfnamefont {E.~A.}\ \bibnamefont {Demler}},\ }\href {\doibase
  10.1038/nphys1074} {\bibfield  {journal} {\bibinfo  {journal} {Nat Phys}\
  }\textbf {\bibinfo {volume} {4}},\ \bibinfo {pages} {884} (\bibinfo {year}
  {2008})}\BibitemShut {NoStop}%
\bibitem [{\citenamefont {Hafezi}\ \emph {et~al.}(2012)\citenamefont {Hafezi},
  \citenamefont {Chang}, \citenamefont {Gritsev}, \citenamefont {Demler},\ and\
  \citenamefont {Lukin}}]{hafezi_pra12a}%
  \BibitemOpen
  \bibfield  {author} {\bibinfo {author} {\bibfnamefont {M.}~\bibnamefont
  {Hafezi}}, \bibinfo {author} {\bibfnamefont {D.~E.}\ \bibnamefont {Chang}},
  \bibinfo {author} {\bibfnamefont {V.}~\bibnamefont {Gritsev}}, \bibinfo
  {author} {\bibfnamefont {E.}~\bibnamefont {Demler}}, \ and\ \bibinfo {author}
  {\bibfnamefont {M.~D.}\ \bibnamefont {Lukin}},\ }\href {\doibase
  10.1103/PhysRevA.85.013822} {\bibfield  {journal} {\bibinfo  {journal} {Phys.
  Rev. A}\ }\textbf {\bibinfo {volume} {85}},\ \bibinfo {pages} {013822}
  (\bibinfo {year} {2012})}\BibitemShut {NoStop}%
\bibitem [{\citenamefont {Le~Kien}\ \emph {et~al.}(2004)\citenamefont
  {Le~Kien}, \citenamefont {Balykin},\ and\ \citenamefont
  {Hakuta}}]{le_kien_pra04}%
  \BibitemOpen
  \bibfield  {author} {\bibinfo {author} {\bibfnamefont {F.}~\bibnamefont
  {Le~Kien}}, \bibinfo {author} {\bibfnamefont {V.~I.}\ \bibnamefont
  {Balykin}}, \ and\ \bibinfo {author} {\bibfnamefont {K.}~\bibnamefont
  {Hakuta}},\ }\href {\doibase 10.1103/PhysRevA.70.063403} {\bibfield
  {journal} {\bibinfo  {journal} {Phys. Rev. A}\ }\textbf {\bibinfo {volume}
  {70}},\ \bibinfo {pages} {063403} (\bibinfo {year} {2004})}\BibitemShut
  {NoStop}%
\bibitem [{\citenamefont {Vetsch}\ \emph {et~al.}(2010)\citenamefont {Vetsch},
  \citenamefont {Reitz}, \citenamefont {Sagu\'e}, \citenamefont {Schmidt},
  \citenamefont {Dawkins},\ and\ \citenamefont
  {Rauschenbeutel}}]{vetsch_prl10}%
  \BibitemOpen
  \bibfield  {author} {\bibinfo {author} {\bibfnamefont {E.}~\bibnamefont
  {Vetsch}}, \bibinfo {author} {\bibfnamefont {D.}~\bibnamefont {Reitz}},
  \bibinfo {author} {\bibfnamefont {G.}~\bibnamefont {Sagu\'e}}, \bibinfo
  {author} {\bibfnamefont {R.}~\bibnamefont {Schmidt}}, \bibinfo {author}
  {\bibfnamefont {S.~T.}\ \bibnamefont {Dawkins}}, \ and\ \bibinfo {author}
  {\bibfnamefont {A.}~\bibnamefont {Rauschenbeutel}},\ }\href {\doibase
  10.1103/PhysRevLett.104.203603} {\bibfield  {journal} {\bibinfo  {journal}
  {Phys. Rev. Lett.}\ }\textbf {\bibinfo {volume} {104}},\ \bibinfo {pages}
  {203603} (\bibinfo {year} {2010})}\BibitemShut {NoStop}%
\bibitem [{\citenamefont {Goban}\ \emph {et~al.}(2012)\citenamefont {Goban},
  \citenamefont {Choi}, \citenamefont {Alton}, \citenamefont {Ding},
  \citenamefont {Lacro\^ute}, \citenamefont {Pototschnig}, \citenamefont
  {Thiele}, \citenamefont {Stern},\ and\ \citenamefont {Kimble}}]{goban_prl12}%
  \BibitemOpen
  \bibfield  {author} {\bibinfo {author} {\bibfnamefont {A.}~\bibnamefont
  {Goban}}, \bibinfo {author} {\bibfnamefont {K.~S.}\ \bibnamefont {Choi}},
  \bibinfo {author} {\bibfnamefont {D.~J.}\ \bibnamefont {Alton}}, \bibinfo
  {author} {\bibfnamefont {D.}~\bibnamefont {Ding}}, \bibinfo {author}
  {\bibfnamefont {C.}~\bibnamefont {Lacro\^ute}}, \bibinfo {author}
  {\bibfnamefont {M.}~\bibnamefont {Pototschnig}}, \bibinfo {author}
  {\bibfnamefont {T.}~\bibnamefont {Thiele}}, \bibinfo {author} {\bibfnamefont
  {N.~P.}\ \bibnamefont {Stern}}, \ and\ \bibinfo {author} {\bibfnamefont
  {H.~J.}\ \bibnamefont {Kimble}},\ }\href {\doibase
  10.1103/PhysRevLett.109.033603} {\bibfield  {journal} {\bibinfo  {journal}
  {Phys. Rev. Lett.}\ }\textbf {\bibinfo {volume} {109}},\ \bibinfo {pages}
  {033603} (\bibinfo {year} {2012})}\BibitemShut {NoStop}%
\bibitem [{\citenamefont {Yu}\ \emph {et~al.}(2014)\citenamefont {Yu},
  \citenamefont {Hood}, \citenamefont {Muniz}, \citenamefont {Martin},
  \citenamefont {Norte}, \citenamefont {Hung}, \citenamefont {Meenehan},
  \citenamefont {Cohen}, \citenamefont {Painter},\ and\ \citenamefont
  {Kimble}}]{yu_apl14}%
  \BibitemOpen
  \bibfield  {author} {\bibinfo {author} {\bibfnamefont {S.-P.}\ \bibnamefont
  {Yu}}, \bibinfo {author} {\bibfnamefont {J.~D.}\ \bibnamefont {Hood}},
  \bibinfo {author} {\bibfnamefont {J.~A.}\ \bibnamefont {Muniz}}, \bibinfo
  {author} {\bibfnamefont {M.~J.}\ \bibnamefont {Martin}}, \bibinfo {author}
  {\bibfnamefont {R.}~\bibnamefont {Norte}}, \bibinfo {author} {\bibfnamefont
  {C.-L.}\ \bibnamefont {Hung}}, \bibinfo {author} {\bibfnamefont {S.~M.}\
  \bibnamefont {Meenehan}}, \bibinfo {author} {\bibfnamefont {J.~D.}\
  \bibnamefont {Cohen}}, \bibinfo {author} {\bibfnamefont {O.}~\bibnamefont
  {Painter}}, \ and\ \bibinfo {author} {\bibfnamefont {H.~J.}\ \bibnamefont
  {Kimble}},\ }\href {\doibase 10.1063/1.4868975} {\bibfield  {journal}
  {\bibinfo  {journal} {Applied Physics Letters}\ }\textbf {\bibinfo {volume}
  {104}},\ \bibinfo {eid} {111103} (\bibinfo {year} {2014})}\BibitemShut
  {NoStop}%
\bibitem [{\citenamefont {Goban}\ \emph {et~al.}(2014)\citenamefont {Goban},
  \citenamefont {Hung}, \citenamefont {Yu}, \citenamefont {Hood}, \citenamefont
  {Muniz}, \citenamefont {Lee}, \citenamefont {Martin}, \citenamefont
  {McClung}, \citenamefont {Choi}, \citenamefont {Chang}, \citenamefont
  {Painter},\ and\ \citenamefont {Kimble}}]{goban_ncomms2014}%
  \BibitemOpen
  \bibfield  {author} {\bibinfo {author} {\bibfnamefont {A.}~\bibnamefont
  {Goban}}, \bibinfo {author} {\bibfnamefont {C.-L.}\ \bibnamefont {Hung}},
  \bibinfo {author} {\bibfnamefont {S.-P.}\ \bibnamefont {Yu}}, \bibinfo
  {author} {\bibfnamefont {J.~D.}\ \bibnamefont {Hood}}, \bibinfo {author}
  {\bibfnamefont {J.~A.}\ \bibnamefont {Muniz}}, \bibinfo {author}
  {\bibfnamefont {J.~H.}\ \bibnamefont {Lee}}, \bibinfo {author} {\bibfnamefont
  {M.~J.}\ \bibnamefont {Martin}}, \bibinfo {author} {\bibfnamefont {A.~C.}\
  \bibnamefont {McClung}}, \bibinfo {author} {\bibfnamefont {K.~S.}\
  \bibnamefont {Choi}}, \bibinfo {author} {\bibfnamefont {D.~E.}\ \bibnamefont
  {Chang}}, \bibinfo {author} {\bibfnamefont {O.}~\bibnamefont {Painter}}, \
  and\ \bibinfo {author} {\bibfnamefont {H.~J.}\ \bibnamefont {Kimble}},\
  }\href {http://dx.doi.org/10.1038/ncomms4808} {\bibfield  {journal} {\bibinfo
   {journal} {Nat Commun}\ }\textbf {\bibinfo {volume} {5}} (\bibinfo {year}
  {2014})}\BibitemShut {NoStop}%
\bibitem [{\citenamefont {Witthaut}\ and\ \citenamefont
  {Sørensen}(2010)}]{witthaut_njp2010}%
  \BibitemOpen
  \bibfield  {author} {\bibinfo {author} {\bibfnamefont {D.}~\bibnamefont
  {Witthaut}}\ and\ \bibinfo {author} {\bibfnamefont {A.~S.}\ \bibnamefont
  {Sørensen}},\ }\href {http://stacks.iop.org/1367-2630/12/i=4/a=043052}
  {\bibfield  {journal} {\bibinfo  {journal} {New Journal of Physics}\ }\textbf
  {\bibinfo {volume} {12}},\ \bibinfo {pages} {043052} (\bibinfo {year}
  {2010})}\BibitemShut {NoStop}%
\bibitem [{\citenamefont {Chang}\ \emph {et~al.}(2011)\citenamefont {Chang},
  \citenamefont {Safavi-Naeini}, \citenamefont {Hafezi},\ and\ \citenamefont
  {Painter}}]{chang_njp11a}%
  \BibitemOpen
  \bibfield  {author} {\bibinfo {author} {\bibfnamefont {D.~E.}\ \bibnamefont
  {Chang}}, \bibinfo {author} {\bibfnamefont {A.~H.}\ \bibnamefont
  {Safavi-Naeini}}, \bibinfo {author} {\bibfnamefont {M.}~\bibnamefont
  {Hafezi}}, \ and\ \bibinfo {author} {\bibfnamefont {O.}~\bibnamefont
  {Painter}},\ }\href {http://stacks.iop.org/1367-2630/13/i=2/a=023003}
  {\bibfield  {journal} {\bibinfo  {journal} {New Journal of Physics}\ }\textbf
  {\bibinfo {volume} {13}},\ \bibinfo {pages} {023003} (\bibinfo {year}
  {2011})}\BibitemShut {NoStop}%
\bibitem [{\citenamefont {Moiseev}\ and\ \citenamefont
  {Ham}(2006)}]{moiseev_pra2006}%
  \BibitemOpen
  \bibfield  {author} {\bibinfo {author} {\bibfnamefont {S.~A.}\ \bibnamefont
  {Moiseev}}\ and\ \bibinfo {author} {\bibfnamefont {B.~S.}\ \bibnamefont
  {Ham}},\ }\href {\doibase 10.1103/PhysRevA.73.033812} {\bibfield  {journal}
  {\bibinfo  {journal} {Phys. Rev. A}\ }\textbf {\bibinfo {volume} {73}},\
  \bibinfo {pages} {033812} (\bibinfo {year} {2006})}\BibitemShut {NoStop}%
\bibitem [{\citenamefont {Zimmer}\ \emph {et~al.}(2008)\citenamefont {Zimmer},
  \citenamefont {Otterbach}, \citenamefont {Unanyan}, \citenamefont {Shore},\
  and\ \citenamefont {Fleischhauer}}]{zimmer_pra08}%
  \BibitemOpen
  \bibfield  {author} {\bibinfo {author} {\bibfnamefont {F.~E.}\ \bibnamefont
  {Zimmer}}, \bibinfo {author} {\bibfnamefont {J.}~\bibnamefont {Otterbach}},
  \bibinfo {author} {\bibfnamefont {R.~G.}\ \bibnamefont {Unanyan}}, \bibinfo
  {author} {\bibfnamefont {B.~W.}\ \bibnamefont {Shore}}, \ and\ \bibinfo
  {author} {\bibfnamefont {M.}~\bibnamefont {Fleischhauer}},\ }\href {\doibase
  10.1103/PhysRevA.77.063823} {\bibfield  {journal} {\bibinfo  {journal} {Phys.
  Rev. A}\ }\textbf {\bibinfo {volume} {77}},\ \bibinfo {pages} {063823}
  (\bibinfo {year} {2008})}\BibitemShut {NoStop}%
\bibitem [{\citenamefont {Moiseev}\ \emph {et~al.}(2014)\citenamefont
  {Moiseev}, \citenamefont {Sidorova},\ and\ \citenamefont
  {Ham}}]{moiseev_pra2014}%
  \BibitemOpen
  \bibfield  {author} {\bibinfo {author} {\bibfnamefont {S.~A.}\ \bibnamefont
  {Moiseev}}, \bibinfo {author} {\bibfnamefont {A.~I.}\ \bibnamefont
  {Sidorova}}, \ and\ \bibinfo {author} {\bibfnamefont {B.~S.}\ \bibnamefont
  {Ham}},\ }\href {\doibase 10.1103/PhysRevA.89.043802} {\bibfield  {journal}
  {\bibinfo  {journal} {Phys. Rev. A}\ }\textbf {\bibinfo {volume} {89}},\
  \bibinfo {pages} {043802} (\bibinfo {year} {2014})}\BibitemShut {NoStop}%
\bibitem [{\citenamefont {Gorshkov}\ \emph
  {et~al.}(2007{\natexlab{a}})\citenamefont {Gorshkov}, \citenamefont
  {Andr\'e}, \citenamefont {Lukin},\ and\ \citenamefont
  {S\o{}rensen}}]{gorshkov_pra07_2}%
  \BibitemOpen
  \bibfield  {author} {\bibinfo {author} {\bibfnamefont {A.~V.}\ \bibnamefont
  {Gorshkov}}, \bibinfo {author} {\bibfnamefont {A.}~\bibnamefont {Andr\'e}},
  \bibinfo {author} {\bibfnamefont {M.~D.}\ \bibnamefont {Lukin}}, \ and\
  \bibinfo {author} {\bibfnamefont {A.~S.}\ \bibnamefont {S\o{}rensen}},\
  }\href {\doibase 10.1103/PhysRevA.76.033805} {\bibfield  {journal} {\bibinfo
  {journal} {Phys. Rev. A}\ }\textbf {\bibinfo {volume} {76}},\ \bibinfo
  {pages} {033805} (\bibinfo {year} {2007}{\natexlab{a}})}\BibitemShut
  {NoStop}%
\bibitem [{\citenamefont {Hansen}\ and\ \citenamefont
  {M\o{}lmer}(2007)}]{hansen_pra07a}%
  \BibitemOpen
  \bibfield  {author} {\bibinfo {author} {\bibfnamefont {K.~R.}\ \bibnamefont
  {Hansen}}\ and\ \bibinfo {author} {\bibfnamefont {K.}~\bibnamefont
  {M\o{}lmer}},\ }\href {\doibase 10.1103/PhysRevA.75.065804} {\bibfield
  {journal} {\bibinfo  {journal} {Phys. Rev. A}\ }\textbf {\bibinfo {volume}
  {75}},\ \bibinfo {pages} {065804} (\bibinfo {year} {2007})}\BibitemShut
  {NoStop}%
\bibitem [{\citenamefont {Nikoghosyan}\ and\ \citenamefont
  {Fleischhauer}(2009)}]{nikoghosyan_pra09a}%
  \BibitemOpen
  \bibfield  {author} {\bibinfo {author} {\bibfnamefont {G.}~\bibnamefont
  {Nikoghosyan}}\ and\ \bibinfo {author} {\bibfnamefont {M.}~\bibnamefont
  {Fleischhauer}},\ }\href {\doibase 10.1103/PhysRevA.80.013818} {\bibfield
  {journal} {\bibinfo  {journal} {Phys. Rev. A}\ }\textbf {\bibinfo {volume}
  {80}},\ \bibinfo {pages} {013818} (\bibinfo {year} {2009})}\BibitemShut
  {NoStop}%
\bibitem [{\citenamefont {Lin}\ \emph {et~al.}(2009)\citenamefont {Lin},
  \citenamefont {Liao}, \citenamefont {Peters}, \citenamefont {Chou},
  \citenamefont {Wang}, \citenamefont {Cho}, \citenamefont {Kuan},\ and\
  \citenamefont {Yu}}]{lin_prl09}%
  \BibitemOpen
  \bibfield  {author} {\bibinfo {author} {\bibfnamefont {Y.-W.}\ \bibnamefont
  {Lin}}, \bibinfo {author} {\bibfnamefont {W.-T.}\ \bibnamefont {Liao}},
  \bibinfo {author} {\bibfnamefont {T.}~\bibnamefont {Peters}}, \bibinfo
  {author} {\bibfnamefont {H.-C.}\ \bibnamefont {Chou}}, \bibinfo {author}
  {\bibfnamefont {J.-S.}\ \bibnamefont {Wang}}, \bibinfo {author}
  {\bibfnamefont {H.-W.}\ \bibnamefont {Cho}}, \bibinfo {author} {\bibfnamefont
  {P.-C.}\ \bibnamefont {Kuan}}, \ and\ \bibinfo {author} {\bibfnamefont
  {I.~A.}\ \bibnamefont {Yu}},\ }\href {\doibase
  10.1103/PhysRevLett.102.213601} {\bibfield  {journal} {\bibinfo  {journal}
  {Phys. Rev. Lett.}\ }\textbf {\bibinfo {volume} {102}},\ \bibinfo {pages}
  {213601} (\bibinfo {year} {2009})}\BibitemShut {NoStop}%
\bibitem [{\citenamefont {Wu}\ \emph {et~al.}(2010{\natexlab{a}})\citenamefont
  {Wu}, \citenamefont {Artoni},\ and\ \citenamefont {La~Rocca}}]{wu_pra10}%
  \BibitemOpen
  \bibfield  {author} {\bibinfo {author} {\bibfnamefont {J.-H.}\ \bibnamefont
  {Wu}}, \bibinfo {author} {\bibfnamefont {M.}~\bibnamefont {Artoni}}, \ and\
  \bibinfo {author} {\bibfnamefont {G.~C.}\ \bibnamefont {La~Rocca}},\ }\href
  {\doibase 10.1103/PhysRevA.81.033822} {\bibfield  {journal} {\bibinfo
  {journal} {Phys. Rev. A}\ }\textbf {\bibinfo {volume} {81}},\ \bibinfo
  {pages} {033822} (\bibinfo {year} {2010}{\natexlab{a}})}\BibitemShut
  {NoStop}%
\bibitem [{\citenamefont {Wu}\ \emph {et~al.}(2010{\natexlab{b}})\citenamefont
  {Wu}, \citenamefont {Artoni},\ and\ \citenamefont {La~Rocca}}]{wu_pra10_2}%
  \BibitemOpen
  \bibfield  {author} {\bibinfo {author} {\bibfnamefont {J.-H.}\ \bibnamefont
  {Wu}}, \bibinfo {author} {\bibfnamefont {M.}~\bibnamefont {Artoni}}, \ and\
  \bibinfo {author} {\bibfnamefont {G.~C.}\ \bibnamefont {La~Rocca}},\ }\href
  {\doibase 10.1103/PhysRevA.82.013807} {\bibfield  {journal} {\bibinfo
  {journal} {Phys. Rev. A}\ }\textbf {\bibinfo {volume} {82}},\ \bibinfo
  {pages} {013807} (\bibinfo {year} {2010}{\natexlab{b}})}\BibitemShut
  {NoStop}%
\bibitem [{\citenamefont {Peters}\ \emph {et~al.}(2012)\citenamefont {Peters},
  \citenamefont {Su}, \citenamefont {Chen}, \citenamefont {Wang}, \citenamefont
  {Gou},\ and\ \citenamefont {Yu}}]{peters_pra12}%
  \BibitemOpen
  \bibfield  {author} {\bibinfo {author} {\bibfnamefont {T.}~\bibnamefont
  {Peters}}, \bibinfo {author} {\bibfnamefont {S.-W.}\ \bibnamefont {Su}},
  \bibinfo {author} {\bibfnamefont {Y.-H.}\ \bibnamefont {Chen}}, \bibinfo
  {author} {\bibfnamefont {J.-S.}\ \bibnamefont {Wang}}, \bibinfo {author}
  {\bibfnamefont {S.-C.}\ \bibnamefont {Gou}}, \ and\ \bibinfo {author}
  {\bibfnamefont {I.~A.}\ \bibnamefont {Yu}},\ }\href {\doibase
  10.1103/PhysRevA.85.023838} {\bibfield  {journal} {\bibinfo  {journal} {Phys.
  Rev. A}\ }\textbf {\bibinfo {volume} {85}},\ \bibinfo {pages} {023838}
  (\bibinfo {year} {2012})}\BibitemShut {NoStop}%
\bibitem [{\citenamefont {Shen}\ and\ \citenamefont {Fan}(2005)}]{shen_ol05}%
  \BibitemOpen
  \bibfield  {author} {\bibinfo {author} {\bibfnamefont {J.~T.}\ \bibnamefont
  {Shen}}\ and\ \bibinfo {author} {\bibfnamefont {S.}~\bibnamefont {Fan}},\
  }\href {\doibase 10.1364/OL.30.002001} {\bibfield  {journal} {\bibinfo
  {journal} {Opt. Lett.}\ }\textbf {\bibinfo {volume} {30}},\ \bibinfo {pages}
  {2001} (\bibinfo {year} {2005})}\BibitemShut {NoStop}%
\bibitem [{\citenamefont {Gorshkov}\ \emph
  {et~al.}(2007{\natexlab{b}})\citenamefont {Gorshkov}, \citenamefont
  {Andr\'e}, \citenamefont {Lukin},\ and\ \citenamefont
  {S\o{}rensen}}]{gorshkov_pra07_1}%
  \BibitemOpen
  \bibfield  {author} {\bibinfo {author} {\bibfnamefont {A.~V.}\ \bibnamefont
  {Gorshkov}}, \bibinfo {author} {\bibfnamefont {A.}~\bibnamefont {Andr\'e}},
  \bibinfo {author} {\bibfnamefont {M.~D.}\ \bibnamefont {Lukin}}, \ and\
  \bibinfo {author} {\bibfnamefont {A.~S.}\ \bibnamefont {S\o{}rensen}},\
  }\href {\doibase 10.1103/PhysRevA.76.033804} {\bibfield  {journal} {\bibinfo
  {journal} {Phys. Rev. A}\ }\textbf {\bibinfo {volume} {76}},\ \bibinfo
  {pages} {033804} (\bibinfo {year} {2007}{\natexlab{b}})}\BibitemShut
  {NoStop}%
\bibitem [{\citenamefont {S\o{}rensen}\ and\ \citenamefont
  {S\o{}rensen}(2008)}]{soerensen_mw_pra08}%
  \BibitemOpen
  \bibfield  {author} {\bibinfo {author} {\bibfnamefont {M.~W.}\ \bibnamefont
  {S\o{}rensen}}\ and\ \bibinfo {author} {\bibfnamefont {A.~S.}\ \bibnamefont
  {S\o{}rensen}},\ }\href {\doibase 10.1103/PhysRevA.77.013826} {\bibfield
  {journal} {\bibinfo  {journal} {Phys. Rev. A}\ }\textbf {\bibinfo {volume}
  {77}},\ \bibinfo {pages} {013826} (\bibinfo {year} {2008})}\BibitemShut
  {NoStop}%
\bibitem [{\citenamefont {Arlinghaus}\ and\ \citenamefont
  {Holthaus}(2011)}]{holthaus_prb2011}%
  \BibitemOpen
  \bibfield  {author} {\bibinfo {author} {\bibfnamefont {S.}~\bibnamefont
  {Arlinghaus}}\ and\ \bibinfo {author} {\bibfnamefont {M.}~\bibnamefont
  {Holthaus}},\ }\href {\doibase 10.1103/PhysRevB.84.054301} {\bibfield
  {journal} {\bibinfo  {journal} {Phys. Rev. B}\ }\textbf {\bibinfo {volume}
  {84}},\ \bibinfo {pages} {054301} (\bibinfo {year} {2011})}\BibitemShut
  {NoStop}%
\bibitem [{\citenamefont {Holthaus}(2016)}]{holthaus_floquet_tutorial}%
  \BibitemOpen
  \bibfield  {author} {\bibinfo {author} {\bibfnamefont {M.}~\bibnamefont
  {Holthaus}},\ }\href {http://stacks.iop.org/0953-4075/49/i=1/a=013001}
  {\bibfield  {journal} {\bibinfo  {journal} {Journal of Physics B: Atomic,
  Molecular and Optical Physics}\ }\textbf {\bibinfo {volume} {49}},\ \bibinfo
  {pages} {013001} (\bibinfo {year} {2016})}\BibitemShut {NoStop}%
\bibitem [{\citenamefont {Chang}\ \emph {et~al.}(2012)\citenamefont {Chang},
  \citenamefont {Jiang}, \citenamefont {Gorshkov},\ and\ \citenamefont
  {Kimble}}]{chang_njp12}%
  \BibitemOpen
  \bibfield  {author} {\bibinfo {author} {\bibfnamefont {D.~E.}\ \bibnamefont
  {Chang}}, \bibinfo {author} {\bibfnamefont {L.}~\bibnamefont {Jiang}},
  \bibinfo {author} {\bibfnamefont {A.~V.}\ \bibnamefont {Gorshkov}}, \ and\
  \bibinfo {author} {\bibfnamefont {H.~J.}\ \bibnamefont {Kimble}},\ }\href
  {http://stacks.iop.org/1367-2630/14/i=6/a=063003} {\bibfield  {journal}
  {\bibinfo  {journal} {New Journal of Physics}\ }\textbf {\bibinfo {volume}
  {14}},\ \bibinfo {pages} {063003} (\bibinfo {year} {2012})}\BibitemShut
  {NoStop}%
\bibitem [{\citenamefont {Stephen}(2006)}]{stephen_2006}%
  \BibitemOpen
  \bibfield  {author} {\bibinfo {author} {\bibfnamefont {N.}~\bibnamefont
  {Stephen}},\ }\href {\doibase 10.1098/rspa.2006.1669} {\bibfield  {journal}
  {\bibinfo  {journal} {Proceedings of the Royal Society of London A:
  Mathematical, Physical and Engineering Sciences}\ }\textbf {\bibinfo {volume}
  {462}},\ \bibinfo {pages} {2245} (\bibinfo {year} {2006})}\BibitemShut
  {NoStop}%
\bibitem [{\citenamefont {Deutsch}\ \emph {et~al.}(1995)\citenamefont
  {Deutsch}, \citenamefont {Spreeuw}, \citenamefont {Rolston},\ and\
  \citenamefont {Phillips}}]{deutsch_pra95a}%
  \BibitemOpen
  \bibfield  {author} {\bibinfo {author} {\bibfnamefont {I.~H.}\ \bibnamefont
  {Deutsch}}, \bibinfo {author} {\bibfnamefont {R.~J.~C.}\ \bibnamefont
  {Spreeuw}}, \bibinfo {author} {\bibfnamefont {S.~L.}\ \bibnamefont
  {Rolston}}, \ and\ \bibinfo {author} {\bibfnamefont {W.~D.}\ \bibnamefont
  {Phillips}},\ }\href {\doibase 10.1103/PhysRevA.52.1394} {\bibfield
  {journal} {\bibinfo  {journal} {Phys. Rev. A}\ }\textbf {\bibinfo {volume}
  {52}},\ \bibinfo {pages} {1394} (\bibinfo {year} {1995})}\BibitemShut
  {NoStop}%
\end{thebibliography}%

\end{document}